\documentclass{article}

\usepackage[english]{babel}

\usepackage[letterpaper,top=2cm,bottom=2cm,left=3cm,right=3cm,marginparwidth=1.75cm]{geometry}
\usepackage{natbib}
\usepackage{multirow}
 
\usepackage{amsmath}
\usepackage{authblk}
\usepackage{authblk}
\usepackage[normalem]{ulem}
\usepackage[dvipsnames]{xcolor}
\usepackage{graphicx}
\usepackage{amsfonts} 
\usepackage{bm}
\usepackage{subcaption}
\usepackage{graphicx}
\usepackage{xspace}
\usepackage{float}
\usepackage{empheq}
\usepackage[utf8]{inputenc}
\usepackage{mathtools}
\usepackage{array}
\usepackage[normalem]{ulem}
\newcolumntype{P}[1]{>{\centering\arraybackslash}p{#1}}
\usepackage[colorlinks=true, allcolors=blue]{hyperref}
\usepackage[format=plain,textfont=it]{caption}

\date{}

\title{A consensus-constrained parsimonious Gaussian mixture model for clustering hyperspectral images}

\author[1]{Ganesh Babu\thanks{ganesh.babu@ucdconnect.ie}}
\author[2]{Aoife Gowen\thanks{aoife.gowen@ucd.ie}}
\author[1]{Michael Fop\thanks{michael.fop@ucd.ie}}
\author[1]{Isobel Claire Gormley\thanks{claire.gormley@ucd.ie\\ Isobel Claire Gormley and Michael Fop contributed equally to this work.}}
\affil[1]{School of Mathematics and Statistics,}
\affil[2]{School of Biosystems and Food Engineering\protect \\ University College Dublin, Ireland.}

\begin{document}
\maketitle

\begin{abstract}
The use of hyperspectral imaging to investigate food samples has grown due to the improved performance and lower cost of instrumentation. Food engineers use hyperspectral images to classify the type and quality of a food sample, typically using classification methods. In order to train these methods, every pixel in each training image needs to be labelled. Typically, computationally cheap threshold-based approaches are used to label the pixels, and classification methods are trained based on those labels. However, threshold-based approaches are subjective and cannot be generalized across hyperspectral images taken in different conditions and of different foods.
Here a consensus-constrained parsimonious Gaussian mixture model (ccPGMM) is proposed to label pixels in hyperspectral images using a model-based clustering approach. The ccPGMM utilizes information that is available on some pixels and specifies constraints on those pixels belonging to the
same or different clusters 
while clustering the rest of the pixels in the image. A latent variable model is used to represent the high-dimensional data in terms of a small number of underlying latent factors. To ensure computational feasibility, a consensus clustering approach is employed, where the data are divided into multiple randomly selected subsets of variables and constrained clustering is applied to each data subset; the clustering results are then consolidated across all data subsets to provide a consensus clustering solution. The ccPGMM approach is applied to simulated datasets and real hyperspectral images of three types of puffed cereal, corn, rice, and wheat. Improved clustering performance and computational efficiency are demonstrated when compared to other current state-of-the-art approaches. 
\end{abstract}

\section{Introduction}
\label{Introduction}

Hyperspectral imaging is a spectroscopic method combining digital imaging with spectroscopy \citep{gowen_et_al_2019}. Hyperspectral imaging collects an image as a function of the light, in which each pixel is a vector of reflectance or fluorescence information of the sample under study in the continuous spectrum. The idea behind producing hyperspectral images lies in the interaction between the photons emitted by a light source and the physical and chemical properties of the sample \citep{amigo_et_al_2020}. The interaction allows hyperspectral images to capture critical information and traits of a sample not visible to the naked eye. The critical information captured on the continuous spectrum can be affected by spatial interference and spectral and redundant noise \citep{amigo_jos_2020}. These issues are effectively handled with computationally intensive spectral preprocessing techniques \citep{amigo_jos_2010} like denoising and scatter correction \citep{rinnan_et_al_2009,thomas_et_al_2011,geladi_et_al_1985}. The preprocessed information is then used to detect chemical contamination, adulteration, and presence of foreign components in the food samples under study \citep{feng_et_al_2012}. Thanks to the increasing availability of statistical methods for handling complex data, food engineers are routinely employing classification methods for segmentation, food sample detection, and analysis of large volumes of hyperspectral food images.

To train a classification method, each pixel of a hyperspectral image needs to be appropriately labelled as e.g., either the background or the food sample. Usually, the images are captured on appropriate backgrounds that facilitate segmentation. In such cases, to select the regions of interest or the food sample under study, a process called masking is performed to remove the background pixels from the image \citep{amigo_et_al_2015}. To create the mask, multiple techniques are employed \citep{pal_et_al_1993}, with the most widely used being $k$-means clustering \citep{amigo_et_al_2015} and threshold-based approaches \citep{pal_et_al_1993,gowen_et_al_2019,xu_et_al_2020}. In \cite{amigo_et_al_2015}, $k$-means clustering with 2 clusters is applied to a set of hyperspectral images to separate the background from the food sample. \cite{abdi_et_al_2010}, \cite{gowen_et_al_2019}, and \cite{xu_et_al_2020} highlight some of the prominent threshold-based approaches used to label the different food samples and background pixels in a given hyperspectral image. In threshold-based approaches, the threshold for the segmentation of the pixels is selected by producing a histogram of some useful pixel-related summary of the entire spectrum, like the mean or the principal component scores of the wavelength intensities for each pixel. The selection of which summary to employ in the histogram depends on the type of background and the food sample under study. In the histogram, typically a bimodal distribution is observed and a threshold is subjectively chosen as the point that separates the two modes.  The quality of the threshold inferred labels in each hyperspectral image depends on several decisions like the summary metric used and which variables are selected for computing the summary. Furthermore, with the threshold approach, only hard classification of pixels as background or food type are possible, ignoring the inherent uncertainty in the labelling process, particularly in certain areas of the image, like the pixels around the edges of the food sample \citep{amigo_et_al_2020}.

A major challenge in labelling the pixels of the hyperspectral images is that each image has a large number of pixels and for each pixel, a large amount of highly correlated reflectance information is captured in a continuous $p$-dimensional spectrum. Further, often threshold and $k$-means clustering approaches are applied to each hyperspectral image individually, which is impractical as the number of images increases and does not allow for the borrowing of information across images. Also, the threshold approach and $k$-means clustering with 2 clusters often only work well when there is exactly one type of food sample in a hyperspectral image. As the number of food samples in a given image increases, multiple thresholds need to be selected from the histogram, which is even more challenging. Overall, threshold-based approaches are subjective and are not easily generalisable across hyperspectral images.

Approaches to clustering pixels in hyperspectral images are receiving increasing attention \citep{zhai:2021}.
The density based DBSCAN clustering approach \citep{khan_et_al_2014,hahsler_et_al_2019} has been widely used in imaging. Further, model-based clustering approaches have proved effective, for example via 
Gaussian mixture models \citep[GMM,][]{mclust_2016,mcnicholas_et_al_2016,charles_et_al_2019} and parsimonious versions 
\citep{mcnicholas_et_al_2008}. \cite{bouveyron:2014} propose a dimension reduction-based approach to segmenting hyperspectral images of Mars. Further, \cite{jacques:2016} consider co-clustering of pixels and wavelengths in the context of  hyperspectral imaging of an oil-in-water
emulsion using a latent block model.

Here, we propose a novel model-based clustering approach, a consensus-constrained parsimonious Gaussian mixture model (ccPGMM), to label the pixels of multiple hyperspectral images simultaneously in a computationally feasible manner while accounting for uncertainty. A key aspect of ccPGMM is that it involves a constrained parsimonious Gaussian mixture model (constrained-PGMM) in which information regarding the types of food samples present in the hyperspectral images along with information about some pixels is incorporated as constraints to perform informed clustering. Further, as hyperspectral images have a large number of pixels, each of which has an associated high-dimensional spectrum of $p$ strongly correlated variables, ccPGMM employs a parsimonious Gaussian mixture model (PGMM) which describes the high-dimensional data using a small number of latent variables. To enable computationally efficient implementation, ccPGMM adopts a consensus approach, whereby the high-dimensional spectrum is divided into randomly selected subsets of $d < p$ variables, and a constrained-PGMM is fitted to each subset of variables on all pixels. The clustering solutions of the constrained-PGMM fitted to each subset are consolidated to provide a final consensus clustering allocation for each pixel. The performance of the proposed ccPGMM approach is demonstrated, and compared to other state-of-the-art approaches, through thorough simulation studies and application to real hyperspectral images of puffed cereals \citep{gowen_et_al_2019}.

In what follows, Section \ref{sec:data} gives details of the motivating dataset containing hyperspectral images of puffed cereals. Section \ref{ccPGMM_Section} outlines the ccPGMM approach and discusses its inference and implementation details. Section \ref{simstudy} delineates thorough simulation studies that explore the performance of ccPGMM under different settings. Section \ref{application} discusses the application of ccPGMM to the motivating puffed cereal images and Section \ref{discussion} summarises the contributions of ccPGMM and gives an overview of future research directions.  The R \citep{rsoftware} code to implement ccPGMM is available at \href{https://github.com/GaneshBabu1604/ccPGMM.git}{github.com/GaneshBabu1604/ccPGMM.git}.

\section{Hyperspectral images of puffed cereals}
\label{sec:data}

The ccPGMM approach is motivated by the need to simultaneously label pixels in nine hyperspectral images of three types of puffed cereal: wheat, corn, and rice \citep{gowen_et_al_2019}. Each pixel in an image records the near-infrared (NIR) spectrum, with $p = 101$ equally spaced variables spanning the wavelengths in the 880-1720nm interval. Each image $l = 1, \ldots, L$, where $L = 9$ contains only one puffed cereal and is a three-dimensional tensor of dimension $S_l \times T_l \times p$ where $S_l$ and $T_l$ are the numbers of row and column pixels respectively, which vary across the images as they are of different dimensions (details are given in Appendix \ref{appendix_ImageDimensions}). Figure \ref{cereals_grey} shows a greyscale image of one hyperspectral image of each puffed cereal type, where, for each pixel, the intensity of the colour is given by the average of the reflectance information captured in the NIR spectrum. 

For the purpose of ccPGMM, the pixels are assumed to be independent and so the three-dimensional tensor of each image $l$ is represented by a rectangular dataset of dimension $(S_l T_l) \times p$. These $L = 9$ rectangular datasets are then collated to form a single dataset of dimension $N \times p$, where $N = \sum_{l=1}^L S_l T_l$. Here, the final dataset of nine images results in $N = 28039$ pixels, each measured over $p = 101$ wavelengths which are highly correlated and contain spectral noise. 

The objective is to simultaneously label all pixels in all nine images as either background or a cereal type, taking into account available information on the images (i.e., it is known which cereal type is in each image) and on some pixels (e.g., some corner pixels are likely to be background and should be clustered together), in a computationally feasible manner.
To demonstrate the utility of the proposed ccPGMM approach, the method is used in a supervised manner to classify the pixels of an additional hyperspectral image that contains multiple puffed cereal grains of potentially different type. Details of the dimensions of this multi-grain image are given in Appendix \ref{appendix_ImageDimensions}.

\begin{figure*}[tb]
\begin{subfigure}{0.3\textwidth}
\centering
\includegraphics[width=0.9\textwidth,height=5cm]{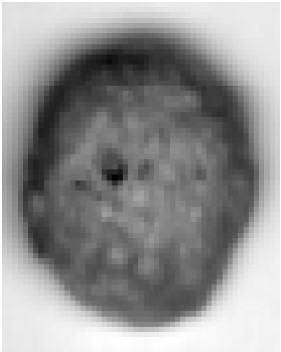} 
\caption{Wheat}
\label{wheat_grey}
\end{subfigure}
\begin{subfigure}{0.3\textwidth}
\centering
\includegraphics[width=0.9\linewidth,height=5cm]{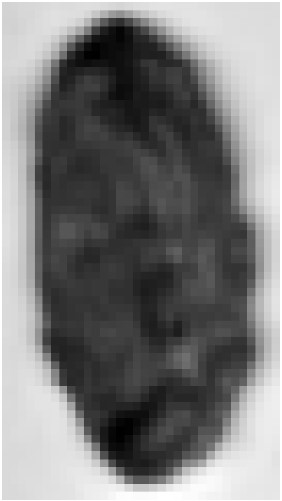}
\caption{Corn}
\label{corn_grey}
\end{subfigure}
\begin{subfigure}{0.3\textwidth}
\centering
\includegraphics[width=0.9\linewidth,height=5cm]{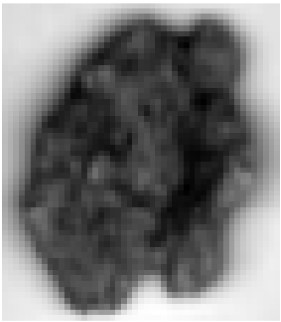}
\caption{Rice}
\label{rice_grey}
\end{subfigure}
\centering
\caption{Greyscale images of one hyperspectral image of each puffed cereal type. For each pixel, the intensity of the color corresponds to the average of the reflectance information captured in the NIR spectrum.}
\label{cereals_grey}
\end{figure*}

\section{The ccPGMM and its inference} \label{ccPGMM_Section}

\subsection{Parsimonious Gaussian mixture models (PGMM)}\label{PGMM}
Factor analysis is a latent variable model that represents a large number $p$ of correlated variables using a smaller number $q \ll p$ of underlying latent factors. In a single factor analysis model \citep{mclachlan_et_al_2000}, each observation $\mathbf{x}_n \in \mathbb{R}^p$ for $n = 1, \ldots, N$ in data $\mathbf{X} = \{\mathbf{x}_1, \ldots, \mathbf{x}_N\}$ of dimension $N \times p$ is modelled as $\mathbf{x}_n = \bm{\mu} + \bm{\Lambda} \mathbf{u}_n + \bm{\varepsilon}_n
$,
where $\bm{\mu}$ is the mean, $\bm{\Lambda}$ is the $p \times q$ loadings matrix and $\mathbf{u}_n \sim \mathcal{N}_q(\mathbf{0},\mathbf{I})$ and $\bm{\varepsilon}_n \sim \mathcal{N}_p(\mathbf{0},\mathbf{\Psi})$ are the latent factor score and specific factor, respectively, for observation $n$. 
Therefore, $\mathbf{x}_n \sim \mathcal{N}_p(\bm{\mu},\bm{\Sigma})$, where ${\bm{\Sigma}} = \bm{\Lambda\Lambda^{\top} + \Psi}$ provides a parsimonious covariance structure.

The parsimonious Gaussian mixture model (PGMM)  \citep{mclachlan_et_al_2004, mcnicholas_et_al_2008,mcnicholas_et_al_2010} is a finite mixture of $G$ factor analysis models such that each mixture component $g$ has a multivariate Gaussian distribution with mean $\bm{\mu}_{g}$ and covariance $\bm{\Sigma}_g = \bm{\Lambda}_g\bm{\Lambda}_g^{\top}+\bm{\Psi}_g$. Denoting the probability of membership of mixture component $g$ as $\tau_g$, under PGMM, we have
$
{f}(\mathbf{x}_n) = \sum_{g=1}^{G} \tau_{g}\: \mathcal{N}_p(\mathbf{x}_{n};\bm{\mu}_{g},\bm{\Lambda}_{g},\bm{\Psi}_{g}).
$
By constraining $\bm{\Lambda}_g$ and $\bm{\Psi}_g$ to be the same or different across components, a family of PGMMs has been developed \citep{mcnicholas_et_al_2008} with inference via an alternating expectation-conditional maximization\citep[AECM,][]{ghahramani_et_al_1996, mclachlan_et_al_2003}. These PGMMs facilitate model-based clustering of high-dimensional data.

\subsection{Constrained parsimonious Gaussian mixture models} \label{constrained-PGMM}
\begin{figure*}[tb]
\begin{subfigure}{0.33\textwidth}
\centering
\includegraphics[width=0.9\textwidth,height=5cm]{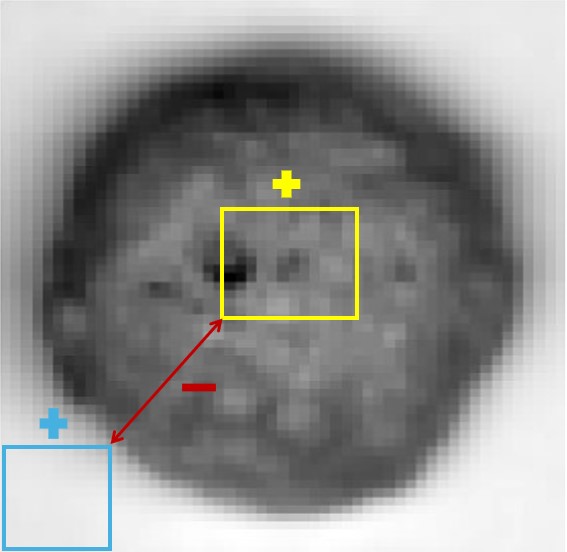}
\caption{Wheat}
\label{wheat_constraints}
\end{subfigure}
\begin{subfigure}{0.33\textwidth}
\centering
\includegraphics[width=0.9\linewidth,height=5cm]{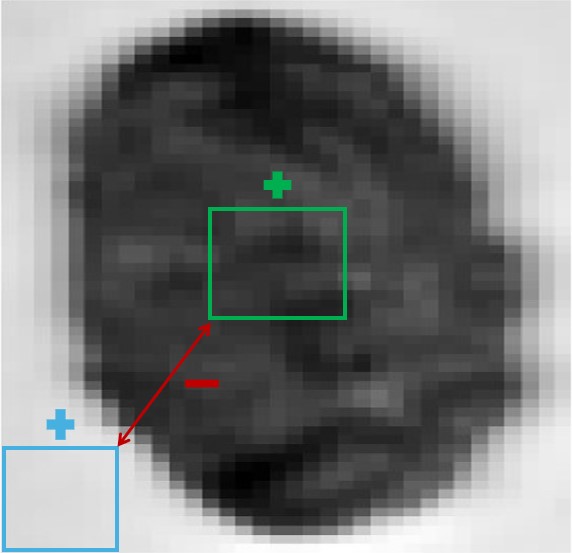} 
\caption{Corn}
\label{corn_constraints}
\end{subfigure}
\centering
\caption{An illustration of positive ($+$) and negative ($-$) constraints. Four blocks of three types of pixels are highlighted. Pixels in the blue blocks must be clustered together, as must the pixels in the yellow block and the pixels in the green block. However, pixels in the blue blocks must not be clustered with those in the yellow block or with those in the green block, and vice versa.}
\label{cereal_constraints}
\end{figure*}

  When labelling pixels in a set of hyperspectral images, there is often information about the type of food sample(s) present in some of the hyperspectral images, along with indirect information on the likely labels of some pixels in the images. For example, in Figure \ref{cereal_constraints}, four blocks of pixels are highlighted in three different colours in the two images. The pixels in the blue blocks located at the corner of the images are assumed to belong to the same cluster, corresponding to the background. On the other hand, the pixels in the yellow and green blocks are located in the center of their images, where the corresponding cereal is observed to be situated. Pixels in the blue blocks should be clustered together, as must the pixels in the yellow block and the pixels in the green block; this information imposes what are known as positive constraints. Also, the pixels in the blue blocks should not be placed in the same cluster as pixels in the yellow or green blocks, and vice versa; this information imposes what are known as negative constraints. It is not necessary to specify constraints in all images if the user prefers or is uncertain, e.g., as may occur in a case where multiple puffed cereal grains of potentially different types are present in a single training image. As highlighted by \cite{melnykov_et_al_2016}, the positive and negative constraints relate to pixels belonging to the same or different clusters, rather than directly specifying pixel labels; unless in a trivial case, assigning labels to pixels prior to model fitting is questionable since the components are yet to be formed and there is therefore inherent uncertainty.

  \cite{melnykov_et_al_2016} proposed a constrained Gaussian mixture model that uses such additional information about some observations as constraints to perform informed model-based clustering. However, the high-dimensionality of the hyperspectral imaging data poses computational challenges for this approach. Here, we build on \cite{melnykov_et_al_2016} and allow for constraints to be incorporated when clustering using a PGMM. The resulting constrained parsimonious Gaussian mixture model (constrained-PGMM) facilitates clustering and therefore labelling of the pixels of the hyperspectral images. Denoting the parameter vector $\bm{\Theta} = \{\tau_1, \ldots, \tau_G, \bm{\mu}_1, \ldots, \bm{\mu}_G, \bm{\Lambda}_1, \ldots, \bm{\Lambda}_G,\bm{\Psi}_1, \ldots, \bm{\Psi}_G\}$, under the constrained-PGMM, the observed data log-likelihood is 

\begin{equation}\label{obs_log_lik}
    \ell_o(\bm{\Theta}; \mathbf{X}) = \sum_{n=1}^{N} \log \sum_{g=1}^G \tau_g\: \mathcal{N}_p(\mathbf{x}_n ; \bm{\mu}_g, \bm{\Lambda}_g, \bm{\Psi}_g).
\end{equation}
As (\ref{obs_log_lik}) is difficult to maximize directly, an AECM algorithm is used to fit the constrained-PGMM. This requires the introduction of a latent component indicator $\mathbf{z}_n = (z_{n1}, \ldots, z_{nG})'$, for $n = 1, \ldots, N$, where $z_{ng} = 1$ if observation $n$ belongs to cluster $g$ or 
 0 otherwise. The AECM algorithm works in two cycles and allows for a different definition of the complete data in each cycle. In the first cycle, the component indicators $\mathbf{Z}$ = \{$\mathbf{z}_1$,\ldots, $\mathbf{z}_N$\} are assumed to be the missing data. Further, $\bm{\Theta}$ is partitioned as $(\bm{\Theta}_1, \bm{\Theta}_2)$ where $\bm{\Theta}_1$ denotes the mixing proportions $\tau_g$ and the mean parameters $\bm{\mu}_g$ ($g=1, \ldots, G$), and $\bm{\Theta}_2$ denotes the loadings $\bm{\Lambda}_g$ and variances $\bm{\Psi}_g$  for $g=1, \ldots, G$. The complete data log-likelihood in the first cycle is then
\begin{equation}   \ell_c(\bm{\Theta}; \mathbf{X},\mathbf{Z}) = \sum_{n=1}^{N}\sum_{g=1}^{G} {z}_{ng} [\log \tau_g + \log \: \mathcal{N}_p(\mathbf{x}_n;\bm{\mu}_g,\bm{\Lambda}_g,\bm{\Psi}_g)]
\end{equation}
and the expected complete data log-likelihood is
\begin{eqnarray*}
    Q_{1}(\bm{\Theta}) \!\! & = &  \!\! \sum_{n=1}^{N}\sum_{g=1}^{G} E[{z}_{ng};\mathbf{x}_n]\Bigg[\log\tau_g - \frac{p}{2}\log(2\pi) - \frac{1}{2}\log(|\bm{\Lambda}_g\:\bm{\Lambda}_g^{\top}\:+\:\bm{\Psi}_g|) \\
     & ~ & -\frac{1}{2}[(\mathbf{x}_{n}-\bm{\mu}_{g})^{\top}(\bm{\Lambda}_g\:\bm{\Lambda}_g^{\top}\:+\:\bm{\Psi}_g)^{-1}(\mathbf{x}_{n}-\bm{\mu}_{g})]\Bigg].   
\end{eqnarray*}
The posterior probability ${\hat{z}_{ng}}$ of observation $n$ belonging to cluster $g$ is
\begin{eqnarray}\label{pos_prob_noconstraints}
E[{z}_{ng};\mathbf{x}_{n}]  & =  & \hat{{z}}_{ng} \: = \: \Pr({z}_{ng} = 1 ; \mathbf{x}_{n},\hat{\tau}_{g},\hat{\bm{\mu}}_{g},\hat{\bm{\Lambda}}_g,\hat{\bm{\Psi}}_g) 
    = 
\frac{\hat{\tau}_g\:\mathcal{N}_p(\mathbf{x}_{n};\hat{\bm{\mu}}_g,\hat{\bm{\Lambda}}_g\:\hat{\bm{\Lambda}}_g^{\top}\:+\:\hat{\bm{\Psi}}_g)}{\sum_{g^{'}=1}^{G}\hat{\tau}_{g^{'}}\:\mathcal{N}_p(\mathbf{x}_{n};\hat{\bm{\mu}}_{g^{'}},\hat{\bm{\Lambda}}_{g^{'}}\:\hat{\bm{\Lambda}}_{g^{'}}^{\top}\:+\:\hat{\bm{\Psi}}_{g^{'}})}
\end{eqnarray}
where $\hat{\cdot}$ denotes an initial or current parameter estimate, as relevant. Let $B_1$ be the set of $J$ pixels (indexed by $j$) which must be clustered together, shown in the blue blocks in Figure \ref{cereal_constraints}. This is a positive constraint (denoted $+$). The posterior probability $\hat{z}_{{B_1}g}^{+}$ of the pixels in $B_1$ belonging to cluster $g$ given the positive constraints ($+$) is
\begin{equation}\nonumber
\centering
    \indent\indent \hat{{z}}_{{B_{1}},g}^{+} = \frac{\prod\limits_{\substack{j \in {B_{1}} \\ j = 1}}^J \hat{\tau}_g \mathcal{N}_p(\mathbf{x}_j;\hat{\bm{\mu}}_g,\hat{\bm{\Lambda}}_g\:\hat{\bm{\Lambda}}_g^{\top}\:+\:\hat{\bm{\Psi}}_g)}{\displaystyle\sum_{g^{'}=1}^{G} \prod\limits_{\substack{j \in {B_{1}} \\ j = 1}}^J \hat{\tau}_{g^{'}} \mathcal{N}_p(\mathbf{x}_j;\hat{\bm{\mu}}_{g^{'}},\hat{\bm{\Lambda}}_{g^{'}}\:\hat{\bm{\Lambda}}_{g^{'}}^{\top}\:+\:\hat{\bm{\Psi}}_{g^{'}})}.
\end{equation}
Let $B_2$ be the set of $K$ pixels (indexed by $k$) which must be clustered together, shown in the yellow block of Figure \ref{cereal_constraints}; these are positive constraints (denoted $+$). However, the pixels in $B_1$ should not be clustered together with the pixels in $B_2$; this is a negative constraint (denoted $-$). 
Therefore, the joint posterior probability $\hat{{z}}_{{B_{1}},g}^{+ \atop -}$ of the pixels in $B_1$ belonging to cluster $g$ and the pixels in $B_2$ not belonging to cluster $g$ (i.e., given the positive and negative constraints $+ \atop -$) is
\begin{equation}\label{posterior_prob_pos_neg}
\centering
    \indent\indent \hat{{z}}_{{B_{1}},g}^{+ \atop -} = \frac{\prod\limits_{\substack{j \in {B_{1}} \\ j = 1}}^J \hat{\tau}_g \mathcal{N}_p(\mathbf{x}_j;\hat{\bm{\mu}}_g,\hat{\bm{\Lambda}}_g\:\hat{\bm{\Lambda}}_g^{\top}\:+\:\hat{\bm{\Psi}}_g) \sum\limits_{\substack{f=1 \\ f \neq g}}^G\prod\limits_{\substack{k \in {B_{2}} \\ k = 1}}^K \hat{\tau}_f \mathcal{N}_p(\mathbf{x}_k;\hat{\bm{\mu}}_f,\hat{\bm{\Lambda}}_f\:\hat{\bm{\Lambda}}_f^{\top}\:+\:\hat{\bm{\Psi}}_f)}{\displaystyle\sum_{g^{'}=1}^{G} \prod\limits_{\substack{j \in {B_{1}} \\ j = 1}}^J \hat{\tau}_{g^{'}} \mathcal{N}_p(\mathbf{x}_j;\hat{\bm{\mu}}_{g^{'}},\hat{\bm{\Lambda}}_{g^{'}}\:\hat{\bm{\Lambda}}_{g^{'}}^{\top}\:+\:\hat{\bm{\Psi}}_{g^{'}})\sum\limits_{\substack{f=1 \\ f \neq g^{'}}}^G\prod\limits_{\substack{k \in {B_{2}} \\ k = 1}}^K \hat{\tau}_f \mathcal{N}_p(\mathbf{x}_k;\hat{\bm{\mu}}_f,\hat{\bm{\Lambda}}_f\:\hat{\bm{\Lambda}}_f^{\top}\:+\:\hat{\bm{\Psi}}_f)}.
\end{equation}
Similarly, the joint posterior probability $\hat{{z}}_{{B_{2}},g}^{+ \atop -}$ of the pixels in $B_2$ belonging to cluster $g$ and the pixels in $B_1$ not belonging to cluster $g$ is
\begin{equation}\nonumber
\centering
    \indent\indent \hat{{z}}_{{B_{2}},g}^{+ \atop -} = \frac{\prod\limits_{\substack{k \in {B_{2}} \\ k = 1}}^K \hat{\tau}_g \mathcal{N}_p(\mathbf{x}_k;\hat{\bm{\mu}}_g,\hat{\bm{\Lambda}}_g\:\hat{\bm{\Lambda}}_g^{\top}\:+\:\hat{\bm{\Psi}}_g) \sum\limits_{\substack{f=1 \\ f \neq g}}^G\prod\limits_{\substack{j \in {B_{1}} \\ j = 1}}^J \hat{\tau}_f \mathcal{N}_p(\mathbf{x}_j;\hat{\bm{\mu}}_f,\hat{\bm{\Lambda}}_f\:\hat{\bm{\Lambda}}_f^{\top}\:+\:\hat{\bm{\Psi}}_f)}{\displaystyle\sum_{g^{'}=1}^{G} \prod\limits_{\substack{k \in {B_{2}} \\ k = 1}}^K \hat{\tau}_{g^{'}} \mathcal{N}_p(\mathbf{x}_k;\hat{\bm{\mu}}_{g^{'}},\hat{\bm{\Lambda}}_{g^{'}}\:\hat{\bm{\Lambda}}_{g^{'}}^{\top}\:+\:\hat{\bm{\Psi}}_{g^{'}})\sum\limits_{\substack{f=1 \\ f \neq g^{'}}}^G\prod\limits_{\substack{j \in {B_{1}} \\ j = 1}}^J \hat{\tau}_f \mathcal{N}_p(\mathbf{x}_j;\hat{\bm{\mu}}_f,\hat{\bm{\Lambda}}_f\:\hat{\bm{\Lambda}}_f^{\top}\:+\:\hat{\bm{\Psi}}_f)}.
\end{equation}
In (\ref{posterior_prob_pos_neg}), only one negative constraint is considered, however (\ref{posterior_prob_pos_neg}) can easily be extended to accommodate multiple such negative constraints. The remaining pixels in Figure \ref{cereal_constraints} which are not in the highlighted blocks have no constraints. To complete the first cycle of the AECM, the model parameters $\bm{\Theta}_1$ are updated in the conditional maximization (CM) step with $\bm{\Theta}_2$ held fixed as in PGMM \citep[see][]{mcnicholas_et_al_2008}; further details are given in Appendix~\ref{appendix_AECM}.

In the second cycle, both component indicators $\mathbf{Z} = \{\mathbf{z}_1, \ldots, \mathbf{z}_n\}$ and latent factors $\mathbf{U} = \{\mathbf{u}_1, \ldots, \mathbf{u}_n\}$ are assumed to be the missing data. In the E-step of the second cycle, the expected value of $\hat{{z}}_{B_1,g}^{+ \atop -}$ is computed as in (\ref{posterior_prob_pos_neg}), with $\bm{\Theta}_1$ updated to the estimates from the CM-step of the first cycle. The expected value of $\mathbf{u}_n (n = 1, \ldots, N$) 
and estimates of $\bm{\Theta}_2$ are computed as in PGMM; details are reported in Appendix~\ref{appendix_AECM}.

The AECM algorithm moves between these two cycles, iteratively updating the model parameters, until convergence. The values of $\hat{\mathbf{Z}}$ at convergence are the posterior probabilities of cluster membership for each pixel; labels can be obtained based, for example, on the \emph{maximum a posteriori} values of $\hat{\mathbf{Z}}$. However, constrained-PGMM exhibits very long run times when fitted to the hyperspectral image data and could not be fitted to the entire $p = 101$ variables of the motivating dataset in a computationally efficient manner.

\subsection{Consensus-constrained parsimonious Gaussian mixture models} \label{ccPGMM}
To improve the computational efficiency of constrained-PGMM, we propose a consensus-constrained parsimonious Gaussian mixture model (ccPGMM), inspired by \cite{russell_et_al_2015}. \cite{russell_et_al_2015} proposed a consensus approach to avoid the selection of a single best model based on the Bayesian information criterion \citep[BIC,][]{neath_et_al_2012,schwarz_1978}. In \cite{russell_et_al_2015}, the clustering solutions of multiple models fitted on the same data are consolidated to provide the final consensus clustering solution. Previously, \cite{strehl_et_al_2002} proposed three ensemble clustering methods namely a cluster-based similarity partitioning algorithm, a hyper-graph partitioning algorithm, and a meta-clustering algorithm to combine multiple clustering solutions into one, but the uncertainty in cluster membership was not accounted for. \cite{fern_et_al_2003} and \cite{punera_et_al_2008} extended these approaches, incorporating the posterior probabilities into the ensemble. 

Unlike \cite{russell_et_al_2015}, the proposed ccPGMM takes a divide-and-conquer approach. In ccPGMM, a constrained-PGMM is fitted to $M$ subsets (indexed by $m$) of $d < p$ randomly selected variables from the high-dimensional NIR spectrum. The $M$ posterior probabilities which account for the uncertainty of the cluster memberships of the $N$ pixels are then consolidated to provide the final consensus clustering solution. An $N \times N$ similarity matrix $\mathbf{S}^{m}$ for $m = 1, \ldots, M$ is computed based on the estimated posterior probabilities $\hat{\mathbf{Z}}^m$ available on convergence of the AECM after fitting the constrained-PGMM to the subset $m$. Each entry of $\mathbf{S}^{m}$ is a similarity score $S_{ij}^m$ computed as
\begin{equation}\nonumber
    S_{ij}^{m}= 
\begin{cases}
    \hat{\mathbf{z}}_{i}^{m}(\hat{\mathbf{z}}_{j}^{m})^{\top}& \text{if } i\neq j,\\
    1              & \text{if } i = j,
\end{cases}    
\end{equation}
where $\hat{\mathbf{z}}_{i}^{m}$ and $\hat{\mathbf{z}}_{j}^{m}$ are the posterior probabilities of cluster membership for observations $i$ and $j$ respectively. The $M$ similarity matrices are then averaged to compute the final similarity matrix $\mathbf{S}$, whose entries $S_{ij}$ are given by 

\begin{equation}\nonumber
    {S}_{ij} = \frac{1}{M} \sum_{m=1}^{M} \displaystyle \sum_{g=1}^{G} \hat{z}_{ig}^{m}\hat{z}_{jg}^{m},
\end{equation}
where $\hat{z}_{ig}^{m}$ and $\hat{z}_{jg}^{m}$ are the posterior probabilities of observation $i$ and $j$ belonging to group $g$, inferred from data subset $m$. A dissimilarity matrix $\mathbf{D}$ is then computed as
$\mathbf{D} = 1 - \mathbf{S}$ and hierarchical clustering with complete linkage \citep{sokal_1963} is performed employing $\mathbf{D}$. As the number of clusters $G$ is known based on the application at hand, the resulting dendrogram is cut to give $G$ clusters and the final consensus clustering solution.

\subsection{Technical details}\label{tech_detail}
In the proposed ccPGMM, a single constrained-PGMM with a fixed number of clusters $G$ and a fixed number of factors $q$ is fitted to $M$ data subsets with $d$ variables per subset; the values of $G$, $q$, $M$, and $d$ require specification. The number of clusters $G$ is known and fixed in advance according to the particular application. For example, in the motivating application here (see Section \ref{sec:data}) where there are three cereal types and the background, $G = 4$. The selection of the number of factors $q$ is based on monitoring the proportion of variance explained by the first $q$ principal components obtained from the application of principal component analysis (PCA) to the entire dataset. There is a trade-off between the values of $M$, the number of subsets of variables, and $d$, the number of variables per subset. Typically, $d$ is smaller than $p$ but too small values may provide little clustering information. Further, very large values of $d$ hamper the computational feasibility. Also, the choice of $M$ must be cognisant of the choice of $d$ to ensure good coverage of all the variables in the $M$ data subsets. These choices are explored further in Sections \ref{simstudy} and \ref{application}.

In terms of specifying the constraints, there are technically no limits on the number of pixels that can be used. The size of the blocks of pixels (constraints) highlighted in Figure \ref{cereal_constraints} can vary and the number of pixels in one block need not be the same as the number of pixels in the other blocks. Usually, the hyperspectral images of food samples are plotted as greyscale images, and the maximum region of pixels that are distinguishable are selected as constraints. Increasing the number of positive constraints and the number of pixels within the positive constraints does not significantly increase the computational cost of fitting the ccPGMM, as in \cite{melnykov_et_al_2016}. However, an increase in the number of negative constraints will result in more complex posterior probability calculations in (\ref{posterior_prob_pos_neg}), negatively impacting the computational cost. Investigations on the choice of constrained pixels are in Sections \ref{simstudy} and \ref{application}.

The AECM algorithm, used to fit the constrained-PGMM to each data subset, is initialized with the cluster labels obtained from applying $k$-means clustering on the respective data subset. The initial values of $\bm{\Lambda}_g$ and $\bm{\Psi}_g$ are computed based on the eigenvalue decomposition of $\bm{\Sigma}_g$ \citep{mcnicholas_et_al_2008}. The relative change in the log-likelihood is used to assess the convergence of the AECM with a tolerance of \textbf{$1e{-}6$}.

\section{Simulation studies}
\label{simstudy}
Simulation studies are conducted to assess the performance of ccPGMM and compare it to the performance of the threshold approach and existing state-of-the-art methods such as DBSCAN \citep{khan_et_al_2014,hahsler_et_al_2019}, Gaussian mixture models \citep[GMM,][]{mclust_2016,mcnicholas_et_al_2016,charles_et_al_2019}, PGMM 
\citep{mcnicholas_et_al_2008} and consensus PGMM (cPGMM) i.e., ccPGMM with no constraints. For DBSCAN \citep{hahsler_et_al_2019}, the minimum points required in the neighborhood for core points was set to $2p$, and the size of the neighborhood was decided from the $k$-nearest neighbours distance plot  \citep{peterson_et_al_2009}. For the GMM, $G=4$ as determined by the setting and the full suite of models available in the \texttt{mclust} \citep{mclust_2016} R package were fitted. For PGMM, the unconstrained model of \cite{mcnicholas_et_al_2023} was fitted with $G=4$ and $q$ set by considering the proportion of variance in the dataset explained by its principal components. 
Convergence in the GMM and PGMM approaches was assessed using the default settings detailed in their respective R packages. Consensus-based approaches were implemented with $G=4$ and $q$ set by monitoring the proportion of variance explained by the data principal components, considering $M = \{10, 25, 50, 100\}$ randomly selected subsets with $d = \{10, 20\}$ variables.

\begin{figure*}[tb]
\begin{subfigure}{0.3\textwidth}
\centering
\includegraphics[width=0.9\textwidth,height=5cm]{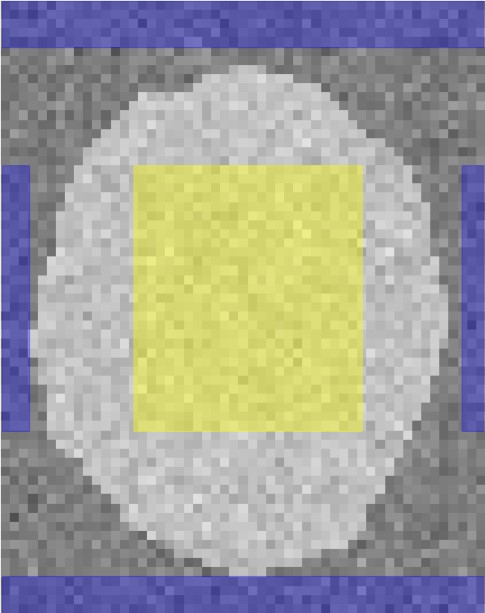} 
\caption{Wheat}
\label{wheat_subset_constraints}
\end{subfigure}
\begin{subfigure}{0.3\textwidth}
\centering
\includegraphics[width=0.9\linewidth,height=5cm]{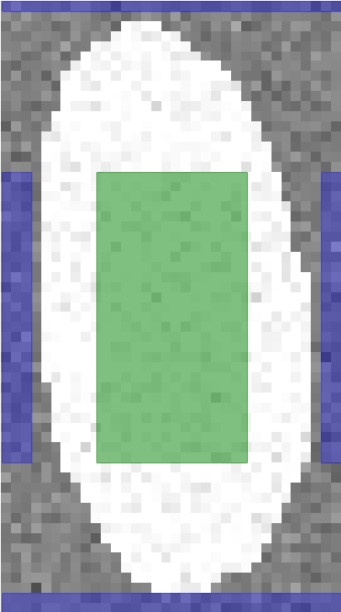}
\caption{Corn}
\label{corn_subset_constraints}
\end{subfigure}
\begin{subfigure}{0.3\textwidth}
\centering
\includegraphics[width=0.9\linewidth,height=5cm]{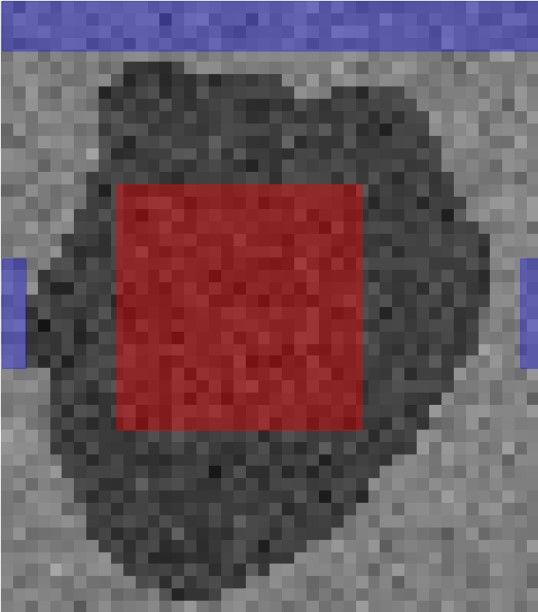}
\caption{Rice}
\label{rice_subset_constraints}
\end{subfigure}
\centering
\caption{Greyscale image of three simulated hyperspectral images for one of the five well-separated cluster datasets. The pixels in the blue blocks must be clustered together,  as must the pixels in the yellow, green, and red blocks. The pixels in one coloured block must not be clustered with the pixels in other coloured blocks.}
\label{simulation_constraints}
\end{figure*}

Four scenarios are considered: low overlap clusters (scenario 1), mild overlap clusters (scenario 2), high overlap clusters (scenario 3), and synthetic puffed cereal data (scenario 4). For each scenario, five datasets are simulated. In scenarios 1--3, each simulated dataset contains three simulated hyperspectral images (one for each cereal type), for a total of $N = 7825$ pixels, where each pixel has an associated spectrum of $p = 101$ strongly correlated variables. The three images are simulated according to labels resulting from applying the threshold approach to three of the hyperspectral images in the motivating dataset detailed in Section \ref{sec:data}. The $p$ variables for each of the three types of cereal and the background pixels are generated from different factor analysis models with $q = 3$ and parameters $\bm{\mu}_g$, $\bm{\Lambda}_g$, and $\bm{\Psi}_g$ for $g = 1, \ldots, G$ where in this setting $G = 4$. To emulate the real data, a strong correlation is induced between the $p$ variables at each pixel by, for every set of $5$ consecutive variables, generating the associated values in the loadings matrix from a $\mathcal{N}_q(\mathcal{U}(0.3,0.9),0.03)$ and the values of $\mbox{diag}(\bm{\Psi}_g)$ from a $\mathrm{Ga}(\mathcal{U}(0,0.1),1)$ where 
 $\mathcal{U}$ and $\mathrm{Ga}$ denote the uniform and gamma distributions, respectively.  The values of the means $\bm{\mu}_g$ for $g = 1, \ldots, G$ are used to control the degree of overlap of the clusters in each scenario; details are given in the respective subsections of Section \ref{simstudy}. To determine the constraints, the greyscale images of the simulated hyperspectral images are examined to decide on the sets of pixels to be used as constraints, as shown in Figure \ref{simulation_constraints}. Here, this resulted in $2,956$ pixels of the $N = 7825$ pixels $(37.7\%)$ being used to inform the constraints. The $1,356$ pixels in the blue blocks must be clustered together, as must the $750$ pixels in the yellow block, the $450$ pixels in the green block, and the $400$ pixels in the red block. These are the positive constraints. The pixels in one coloured block must not be clustered together with the pixels in other coloured blocks; these are the negative constraints.

In scenario 4, a synthetic cereal dataset is considered which mirrors the properties of the motivating hyperspectral images. Each dataset contains $L = 9$ simulated hyperspectral images (three for each cereal type) with $N = 28,039$ and where each pixel has an associated $p = 101$ length spectrum. The $p$ variables for each cereal type and background pixels are generated from the relevant factor analysis model, fitted to the real data; details are discussed in Section \ref{synthetic_cereal}. In this scenario, $12,215$ pixels of the $N = 28,039$ pixels $(43.5\%)$ are selected as constraints by plotting the greyscale images of the synthetic puffed cereals. Of the $12,215$ pixels selected as constraints, $5,900$ are in the blue blocks, $3,130$ are in the yellow blocks, $1,535$ are in the green blocks, and $1,650$ are in the red blocks. Further, to assess the impact of the proportion of pixels used as constraints on the clustering performance, a smaller set of $6,951$ pixels ($24.7\%$) are selected as constraints. Of the $6,951$ pixels, $3,801$ are in the blue blocks, $1,540$ is in the yellow blocks, $783$ are in the green blocks and $827$ are in the red blocks. The set of the $43.5\%$ of pixels and the $24.7\%$ of pixels used as constraints are illustrated in Figures \ref{fig:synthetic_cereal_constraints_large} and \ref{fig:synthetic_cereal_constraints_small} respectively in Appendix \ref{appendix_selectedConstraints}. 

The simulation study is conducted on an Intel(R) core(TM) i7-10850H CPU @ 2.7 GHz processor with 16 GB RAM and 64-bit operating system. The code used to generate the datasets for each scenario and to conduct the simulation studies is available on \href{https://github.com/GaneshBabu1604/ccPGMM.git}{GitHub}.

\subsection{Scenario 1: low overlap clusters}\label{wellsep}
To generate the low overlap clusters, the cluster mean parameters are generated as $\bm{\mu}_g \sim \mathcal{N}_p(a, 0.5)$ for $g = 1, \ldots, 4$, where $a  \in \{-5, 0, 5, 10\}$. The $p$ variables of the background, wheat, corn, and rice pixels in each dataset are then generated from the four low overlapping factor analysis models.

\begin{figure}[tb]
\centering
\includegraphics[width=\textwidth,height=6cm]{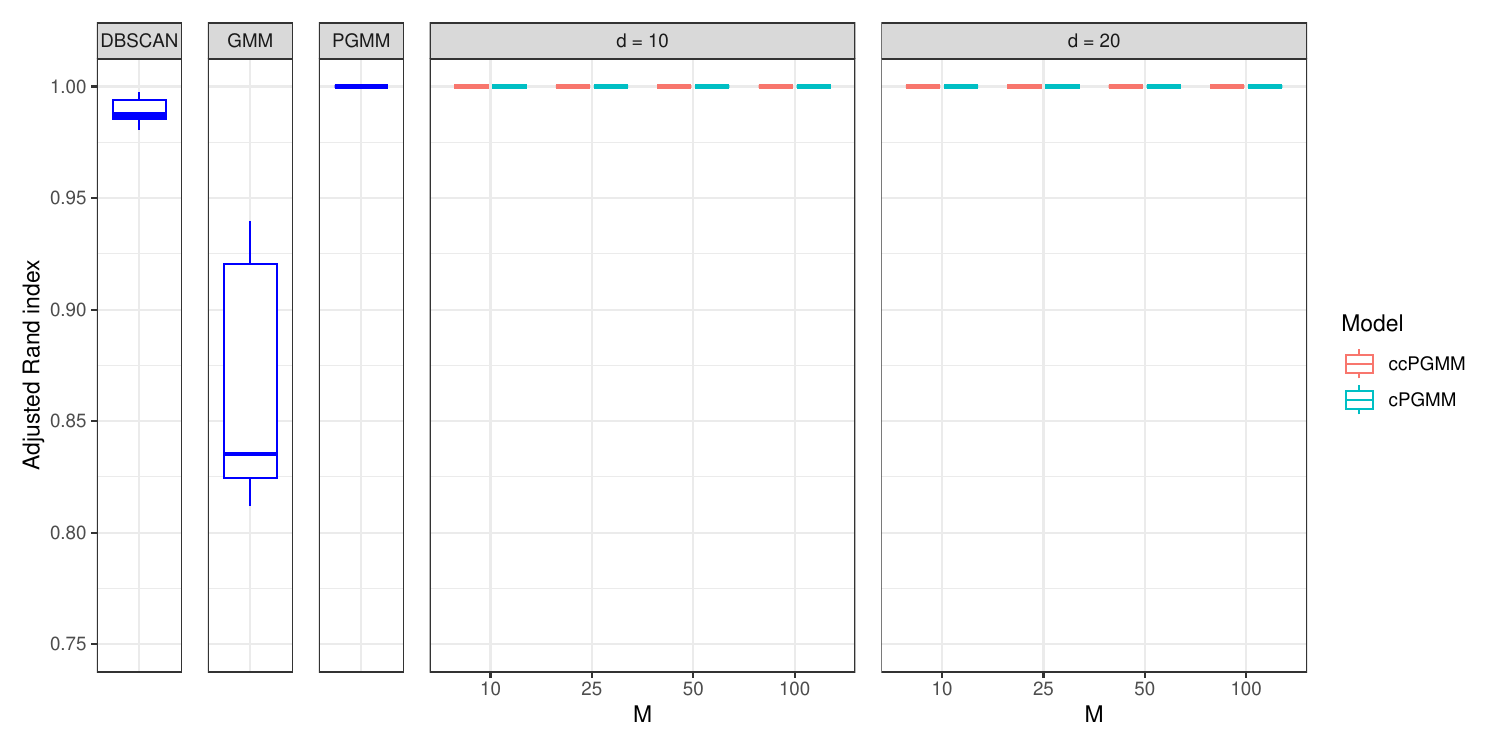} 
\centering
\caption{For each of the five simulated
datasets with low overlap clusters, the ARI between the known labels and the clustering solutions of DBSCAN, GMM, and PGMM fitted on $p = 101$ variables (first three panels from the left) and cPGMM and ccPGMM (last two panels on the right) fitted with different settings of $M$ and $d$.}
\label{Lowoverlapcluster_simulation_ARI}
\end{figure}

\begin{figure}[tb]
\centering
\includegraphics[width=\textwidth,height=6cm]{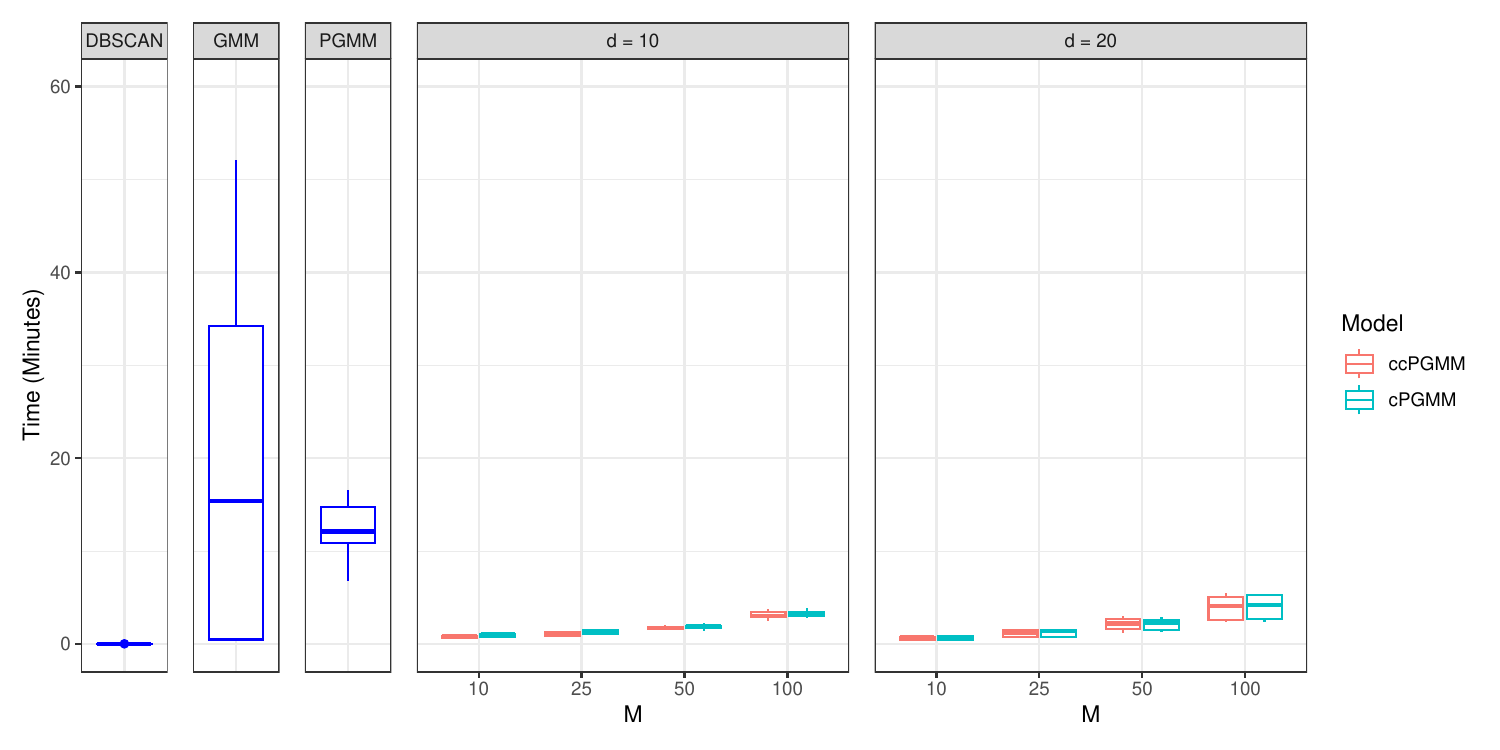} 
\centering
\caption{For each of the five simulated datasets with low overlap clusters, the time taken to fit DBSCAN, GMM, and PGMM on $p = 101$ variables (first three panels from the left) and cPGMM and ccPGMM (last two panels on the right) fitted with different settings of $M$ and $d$.}
\label{Lowoverlapcluster_simulation_time}
\end{figure}

Figure 
 \ref{Lowoverlapcluster_simulation_ARI} shows the adjusted Rand index \citep[ARI,][]{hubert_et_al_1985} between the known labels and the clustering solutions for the five simulated datasets in scenario 1 under the DBSCAN, GMM, PGMM, cPGMM, and ccPGMM approaches. For the parsimonious models, PCA suggested $q=1$. Figure \ref{Lowoverlapcluster_simulation_time} shows the time taken in minutes to fit the methods. All approaches demonstrate strong clustering performance. In terms of computational cost, DBSCAN clusters each dataset in less than a minute while GMM and PGMM take an average of $21$ minutes and $12$ minutes respectively. The cPGMM and ccPGMM fitted with different settings of $M$ and $d$ have lower computational costs than GMM and PGMM, but similar to DBSCAN, with no loss of accuracy.

\subsection{Scenario 2: mild overlap clusters}
\begin{figure}[tb]
\centering
\includegraphics[width=\textwidth,height=6cm]{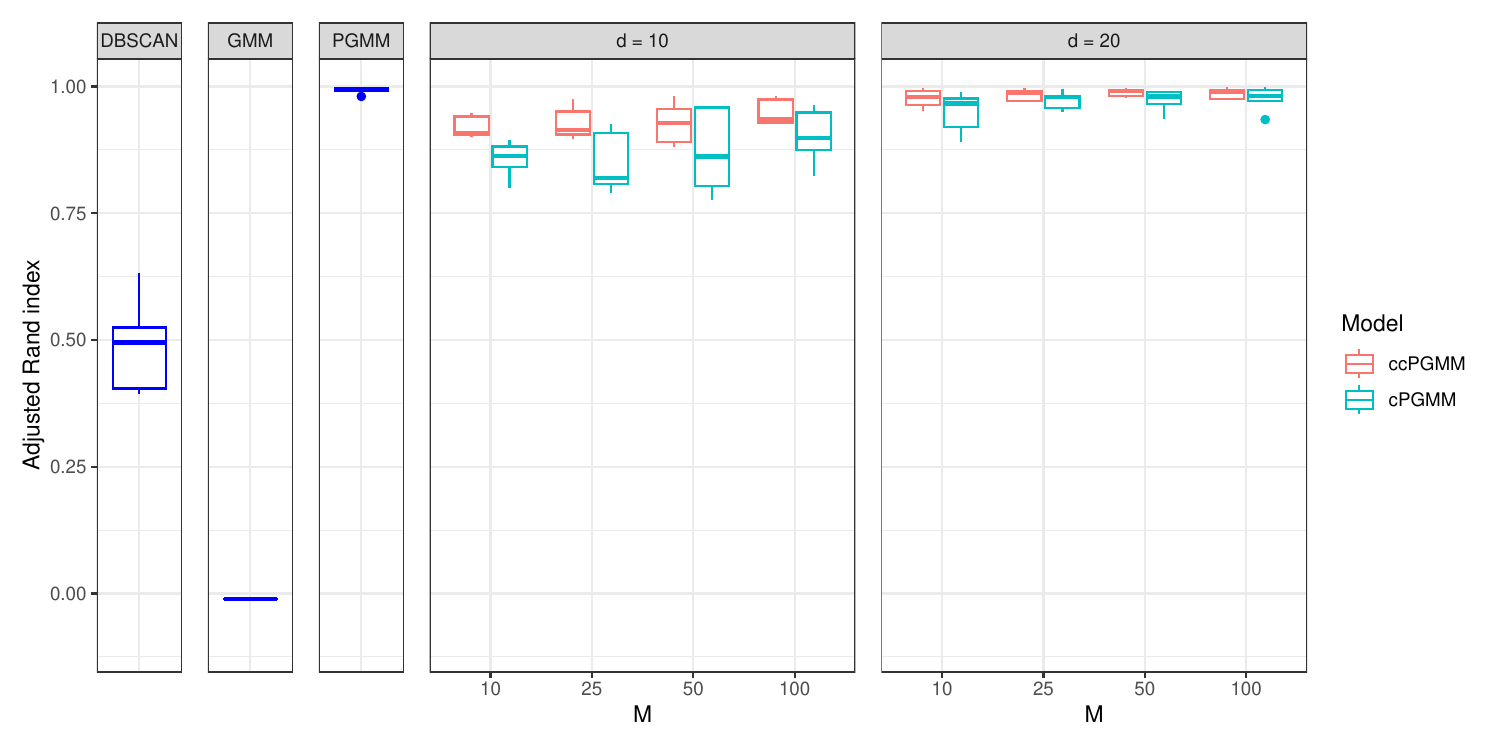} 
\centering
\caption{For each of the five datasets with mild overlap clusters, the ARI of DBSCAN, GMM, and PGMM fitted on $p = 101$ variables (first three panels from the left) and cPGMM and ccPGMM (last two panels on the right) fitted with different settings of $M$ and $d$.}
\label{Mildoverlapcluster_simulation_ARI}
\end{figure}

\begin{figure}[tb]
\centering
\includegraphics[width=\textwidth,height=6cm]{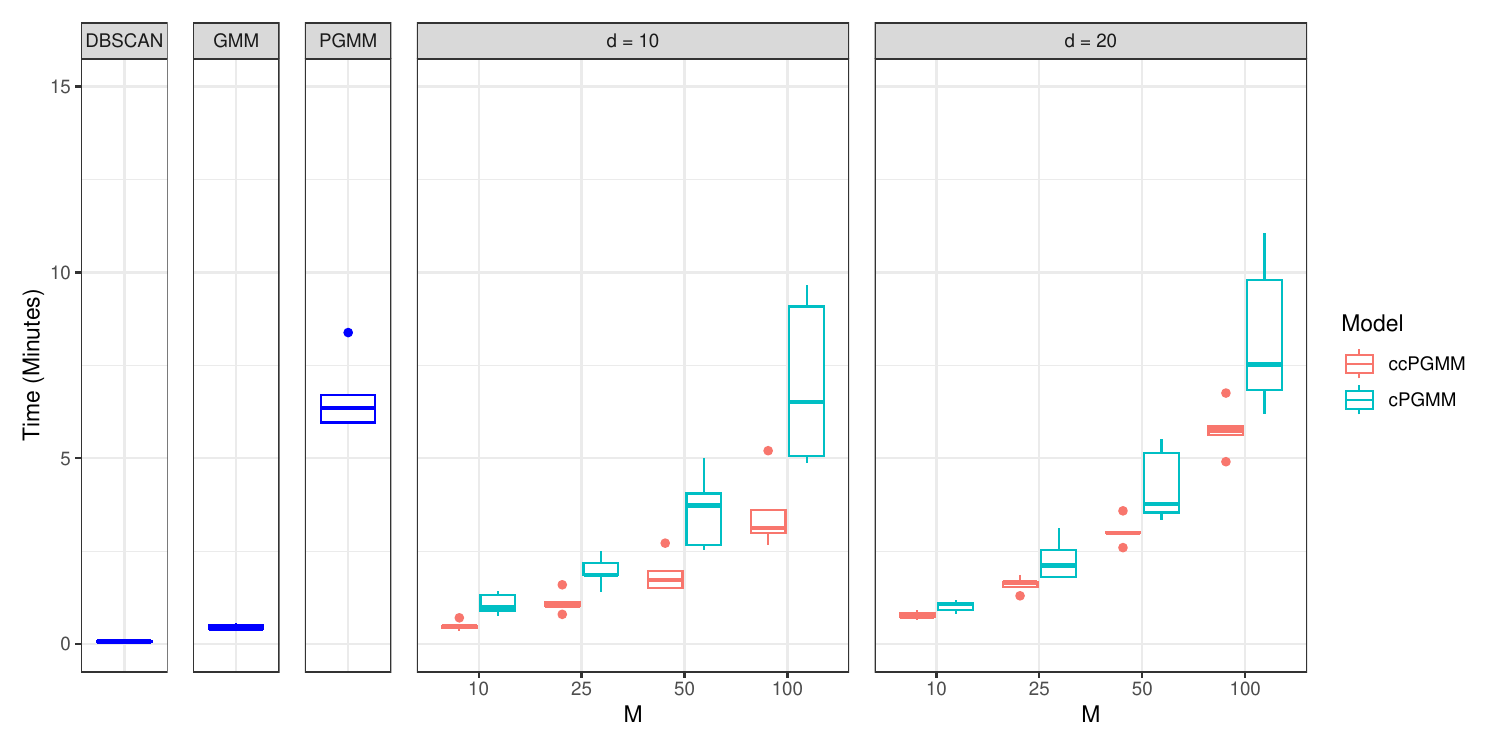} 
\centering
\caption{For each of the five datasets with mild overlap clusters, the time taken to fit DBSCAN, GMM, and PGMM on $p = 101$ variables (first three panels from the left) and cPGMM and ccPGMM (last two panels on the right) fitted with different settings of $M$ and $d$.}
\label{Mildoverlapcluster_simulation_time}
\end{figure}

The $p$ variables of the background, wheat, corn and rice pixels in each dataset are generated from four mildly overlapping factor analysis models, with the cluster mean parameters generated as $\bm{\mu}_g \sim \mathcal{N}_p(a, 1)$ where $a = \{-2.5, 0, 2.5, 5\}$ for $g = 1, \ldots, 4$. Figures \ref{Mildoverlapcluster_simulation_ARI} and \ref{Mildoverlapcluster_simulation_time} show the ARI of the cluster solutions of DBSCAN, GMM, PGMM, cPGMM, and ccPGMM fitted on the five datasets with mild overlap clusters and the time taken in minutes to fit them, respectively. For the parsimonious models, PCA suggested $q=1$. While GMM struggles to uncover the clustering structure, DBSCAN, and PGMM fitted to all $p$ variables clustered the pixels with an average ARI of $0.48$ and $0.99$, respectively. In terms of cPGMM and ccPGMM, Figure \ref{Mildoverlapcluster_simulation_ARI} shows that increasing $d$ from $10$ to $20$ improves the clustering performance of both methods, with the performance of ccPGMM being preferable. Run times are comparable overall, with cPGMM with the highest values of $M$ and $d$ intuitively taking the longest. The ccPGMM approach with $M \geq 10$ and $d = 20$ is competitively computationally efficient with an average ARI $> 0.97$, a similar performance to PGMM fitted to all $p$ variables, but at a reduced computational cost.

\subsection{Scenario 3: high overlap clusters} 
\label{highoverlap}
\begin{figure}[tb]
\centering
\includegraphics[width=\textwidth,height=6cm]{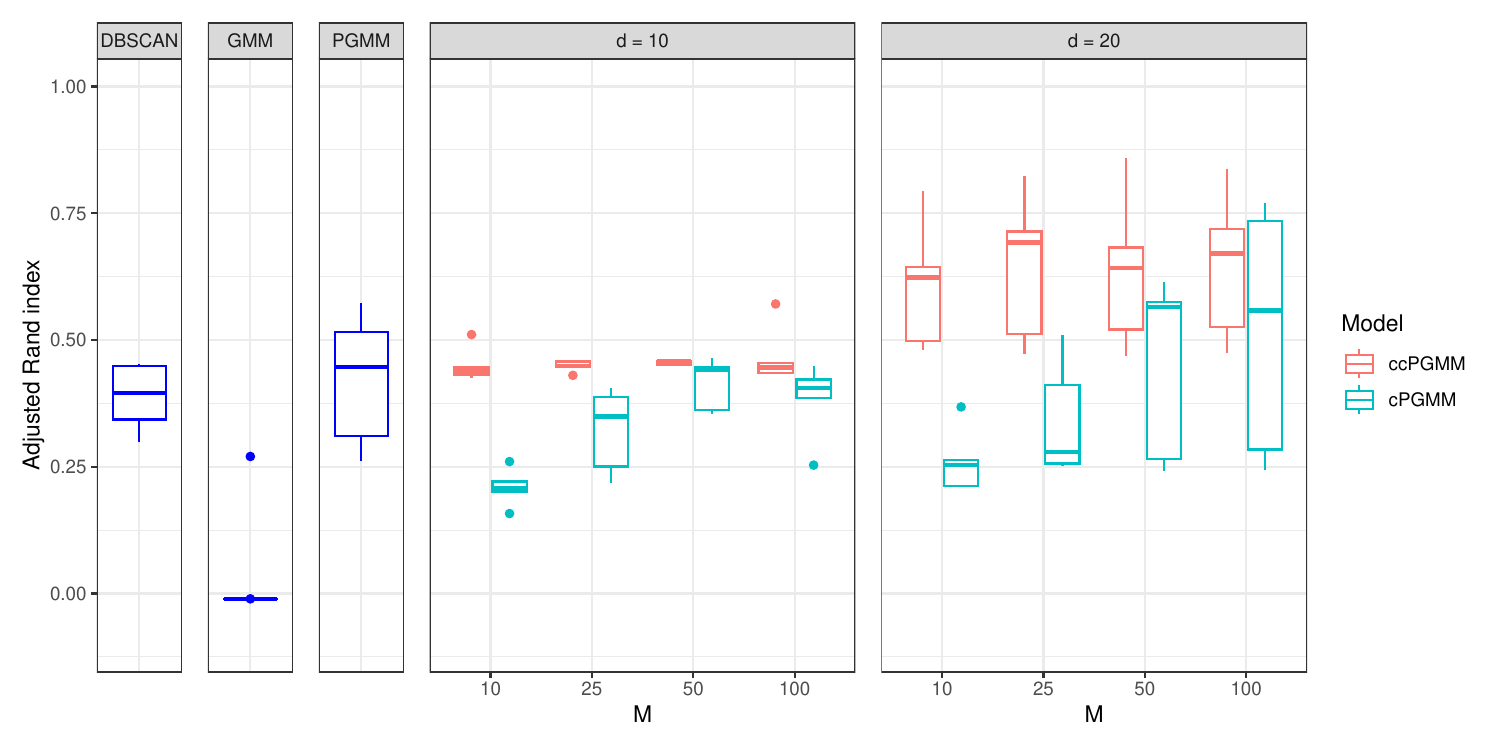} 
\centering
\caption{For each of the five datasets with high overlap clusters, the ARI of DBSCAN, GMM, and PGMM fitted on $p = 101$ variables (first three panels from the left) and cPGMM and ccPGMM (last two panels on the right) with different settings of $M$ and $d$.}
\label{Highoverlapcluster_simulation_ARI}
\end{figure}

\begin{figure}[tb]
\centering
\includegraphics[width=\textwidth,height=6cm]{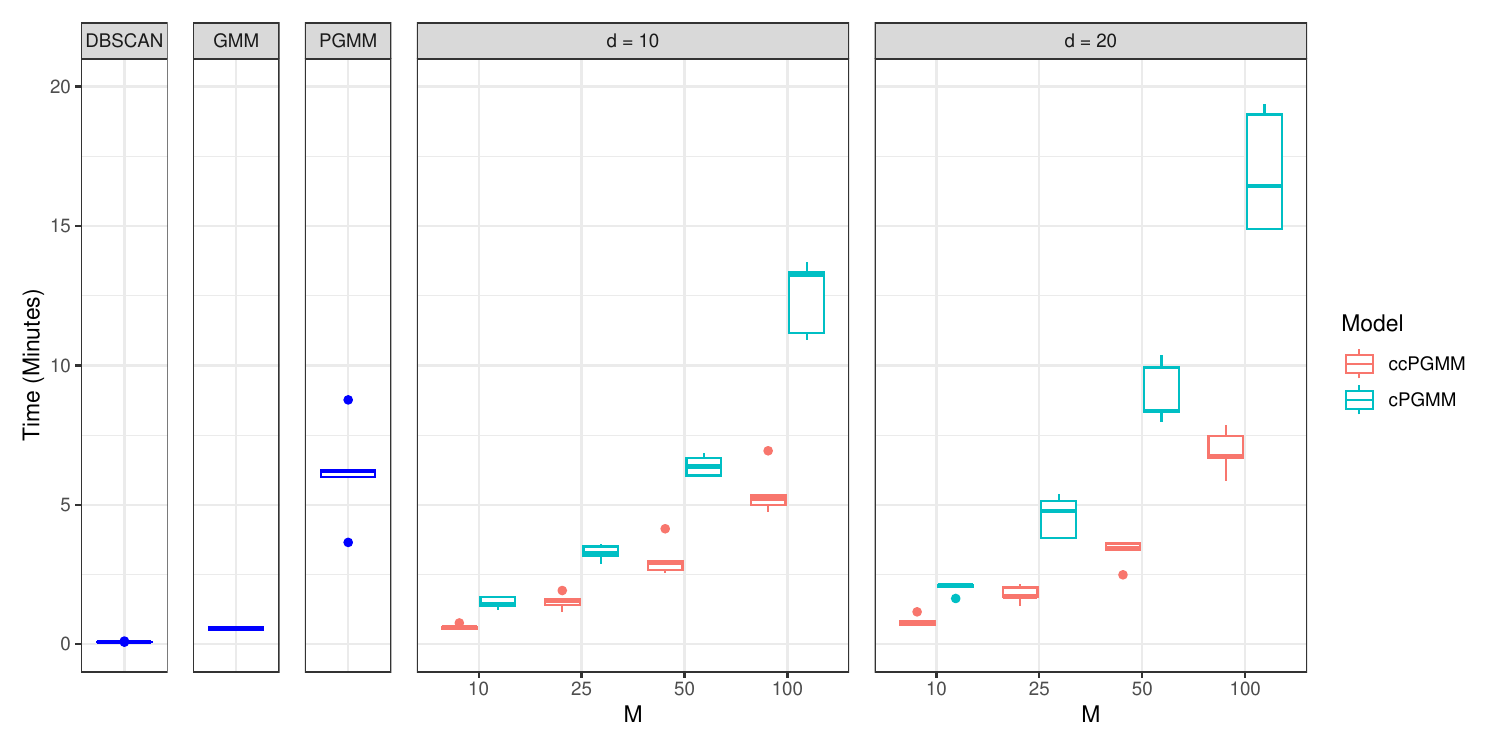} 
\centering
\caption{For each of the five datasets with high overlap clusters, the time taken to fit DBSCAN, GMM, and PGMM on $p = 101$ variables (first three panels from the left) and cPGMM and ccPGMM (last two panels on the right) with different settings of $M$ and $d$.}
\label{Highoverlapcluster_simulation_time}
\end{figure}

To generate data with high overlap clusters, the cluster mean parameters are generated as $\bm{\mu}_g \sim \mathcal{N}_p(a, 1.5)$, where $a \in \{-1.25, 0, 1.25, 2.5\}$ for $g = 1, \ldots, 4$ respectively. Figures \ref{Highoverlapcluster_simulation_ARI} and \ref{Highoverlapcluster_simulation_time} show the ARI between the known pixel labels and the clustering solutions of DBSCAN, GMM, PGMM, cPGMM, and ccPGMM and the time taken in minutes by each method, respectively. For the parsimonious models, PCA suggested $q=1$. DBSCAN clusters the datasets with an average ARI of $0.38$ in less than a minute on average. The GMM struggles to uncover the clustering structure. The PGMM approach performs competitively, achieving an average ARI of $0.42$ in $6$ minutes on average. The ccPGMM approach performs well in this scenario: with $M \geq 25$ and $d = 20$ an average ARI of $0.64$ is achieved, which outperforms PGMM fitted to all $p$ variables. In terms of clustering performance, ccPGMM performs better than cPGMM across all settings of $M$ and $d$, in comparable time.

\subsection{Scenario 4: synthetic puffed cereal data}
\label{synthetic_cereal}
Here a synthetic cereal dataset is considered which mirrors the properties of the motivating hyperspectral images. To generate the data, PCA is fitted to the real puffed cereal images and based on the cumulative proportion of variance explained (Table \ref{pca_table} in Appendix \ref{appendix_pcatable}), $q = 2$ is selected. Threshold labels, generated as proposed by \cite{xu_et_al_2020}, are used to identify the pixels belonging to the 3 cereal types and the background in the real puffed cereal images. Four individual factor analysis models with $q = 2$ are then fitted to pixels highly likely to be from each grain type in the motivating dataset.  Finally, to generate the synthetic cereal images, for each pixel type, $p$ variables are generated from the resulting factor analysis models fitted to the background, wheat, corn, and rice pixels in the real data.

\begin{figure}[tb]
\centering
\includegraphics[width=\textwidth,height=6cm]{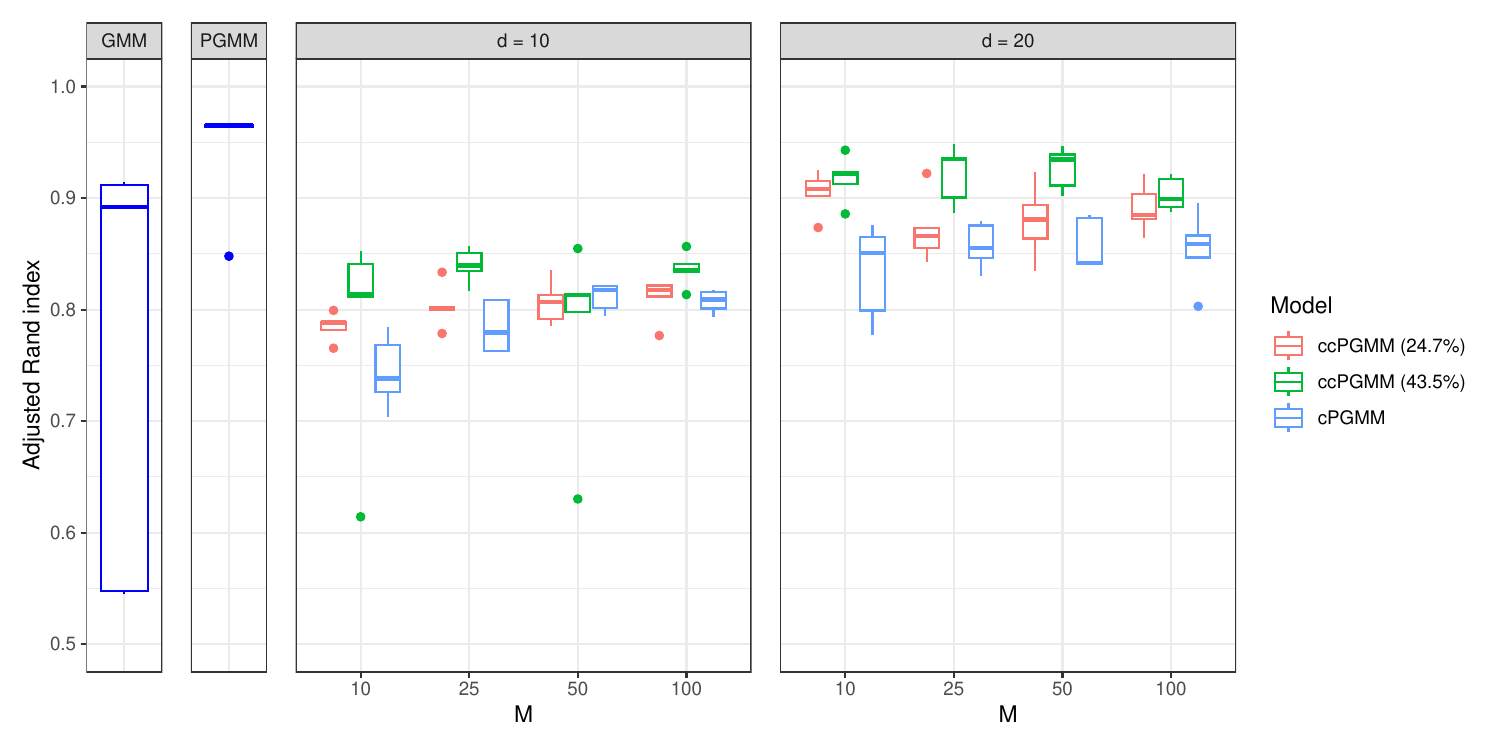} 
\centering
\caption{For each of the five synthetic cereal datasets, the ARI of GMM, and PGMM fitted on $p = 101$ variables (first two panels from the left) and cPGMM and ccPGMM (last two panels on the right) fitted with different settings of $M$ and $d$. For ccPGMM, two settings with $24.7\%$ and $43.5\%$ of pixels used as constraints are shown.}
\label{Syntheticcereal_simulation_ARI}
\end{figure}

\begin{figure}[tb]
\centering
\includegraphics[width=\textwidth,height=6cm]{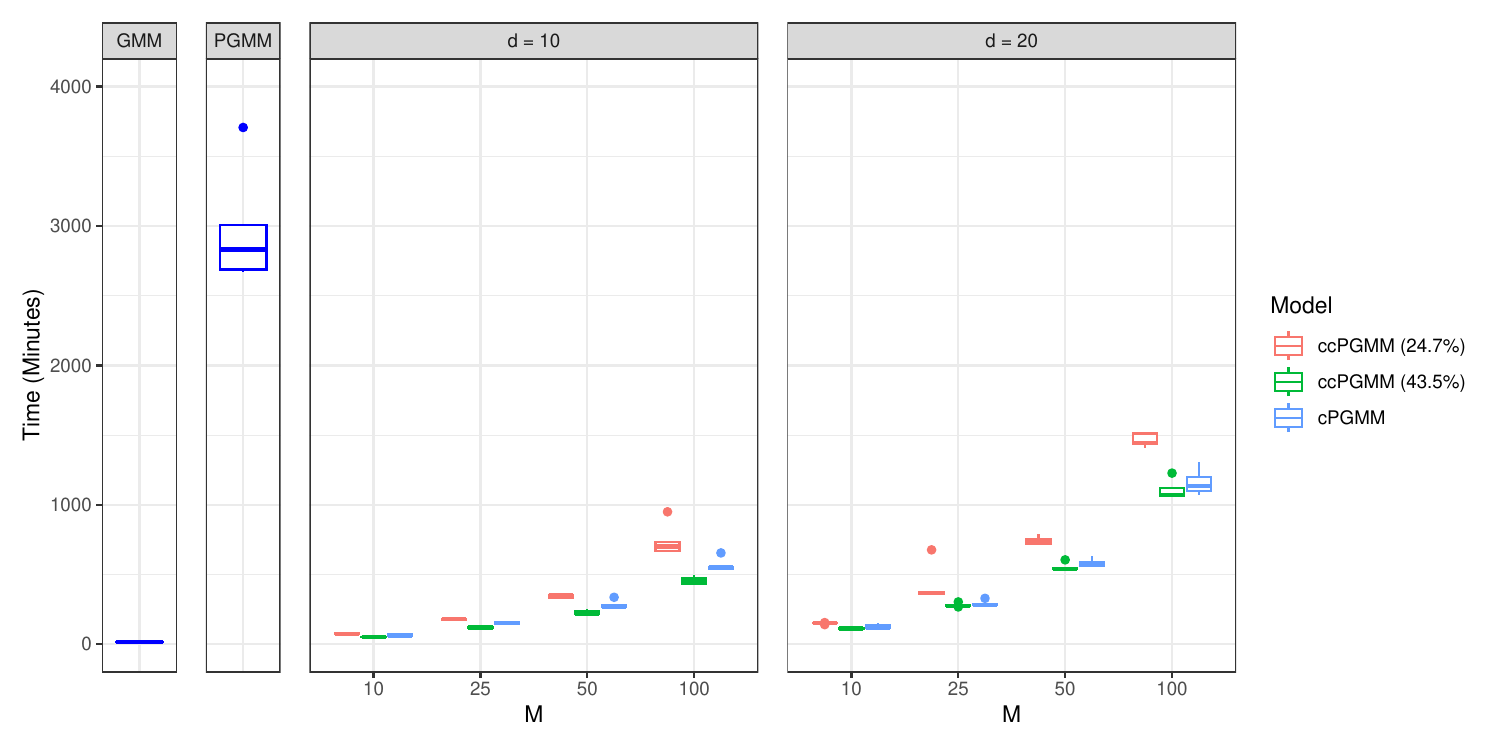} 
\centering
\caption{For each of the five synthetic cereal datasets, the time taken to fit GMM, and PGMM on $p = 101$ variables (first two panels from the left) and cPGMM and ccPGMM (last two panels on the right) fitted with different settings of $M$ and $d$. For ccPGMM, two settings with $24.7\%$ and $43.5\%$ of pixels used as constraints are shown.}
\label{Syntheticcereal_simulation_time}
\end{figure}

Figures \ref{Syntheticcereal_simulation_ARI} and \ref{Syntheticcereal_simulation_time} show, respectively, the ARI of the cluster solutions of the different clustering methods fitted to the five synthetic cereal datasets, and the time taken in minutes for estimation. Intuitively, since the synthetic cereal datasets are generated from a mixture of factor analyzers, PGMM fitted to all $p$ variables shows strong clustering performance (average ARI of $0.94$). However, it takes an average of $50$ hours to cluster each synthetic cereal dataset. The GMM approach has a much lower computational cost than PGMM but the quality of the clustering solutions is inconsistent. For DBSCAN (not illustrated), while computationally cheap, clustering performance is poor with an average ARI of $0.32$ with a variance of $0.11$. The ccPGMM with $43.5\%$ of pixels as constraints and ccPGMM with $24.7\%$ of pixels as constraints fitted with $M \geq 10$ and $d = 20$ shows strong clustering performances. Intuitively, ccPGMM with larger set of constraints ($43.5\%$) performs slightly better than ccPGMM with smaller set of constraints ($24.7\%$) and cPGMM across different settings of $M$ and $d$. For the ccPGMM approach with $43.5\%$ pixels as constraints, fitted with $M \geq 25$ and $d = 20$ has an average ARI of $0.92$. Overall, ccPGMM shows comparable clustering performance to PGMM fitted to all $p$ variables, in less than one-tenth of the computational time.


\section{Application to hyperspectral images of puffed cereals}\label{application}
To simultaneously cluster the pixels of the $9$ hyperspectral images in the motivating dataset, we apply ccPGMM with $G = 4$, as there are $3$ puffed cereal types (wheat, corn, and rice) and the background, and $q = 2$ based on Table \ref{pca_table} in Appendix \ref{appendix_pcatable}.  Based on the performance observed in simulation study scenario 4 (Section \ref{synthetic_cereal}) which mirrored the real data, $M = 25$ subsets and $d = 20$  variables per subset were employed. As detailed in Section \ref{simstudy}, the greyscale version of the hyperspectral images was considered and, based on pixels for which the labelling was clear,  $12,215$ pixels out of the $N = 28,039$ pixels were selected $(43.5\%)$; these pixels informed the constraints. Also, a smaller set of $6,951$ pixels out of the $N = 28,039$ pixels ($24.7\%$) are selected as constraints to assess the impact of the proportion of pixels used as constraints on the clustering performance. For comparison purposes, DBSCAN (with settings as in Section \ref{simstudy}), GMM with $G = 4$ (as implemented in Section \ref{simstudy}, considering the full suite of covariance models and using BIC to select the best one), and the unconstrained PGMM with $G = 4$ and $q = 2$ are fitted to the motivating dataset with all $p$ variables. The threshold approach of \cite{xu_et_al_2020} is also applied, individually, to each hyperspectral image to label the pixels.

\begin{figure}
     \centering
    \begin{tabular}{c}
    \begin{subfigure}[b]{0.32\textwidth}
         \centering
         \includegraphics[width=\textwidth,height=2.5cm]{Figures/Wheat_1_Grey.jpg}
         \caption{}
         \label{fig:wheat_grey}
     \end{subfigure}
     \hfill
     \begin{subfigure}[b]{0.32\textwidth}
         \centering
         \includegraphics[width=\textwidth,height=2.5cm]{Figures/Corn_1_Grey.jpg}
         \caption{}
         \label{fig:corn_grey}
     \end{subfigure}
     \hfill
     \begin{subfigure}[b]{0.32\textwidth}
         \centering
         \includegraphics[width=\textwidth,height=2.5cm]{Figures/Rice_1_Grey.jpg}
         \caption{}
         \label{fig:rice_grey}
     \end{subfigure}\\
    \begin{subfigure}[b]{0.32\textwidth}
         \centering
         \includegraphics[width=\textwidth,height=2.5cm]{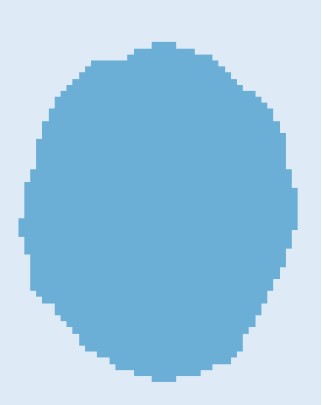}
         \caption{}
         \label{fig:wheat_thresh}
     \end{subfigure}
     \hfill
     \begin{subfigure}[b]{0.32\textwidth}
         \centering
         \includegraphics[width=\textwidth,height=2.5cm]{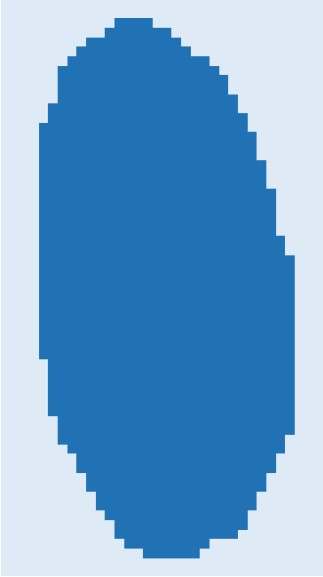}
         \caption{}
         \label{fig:corn_thresh}
     \end{subfigure}
     \hfill
     \begin{subfigure}[b]{0.32\textwidth}
         \centering
         \includegraphics[width=\textwidth,height=2.5cm]{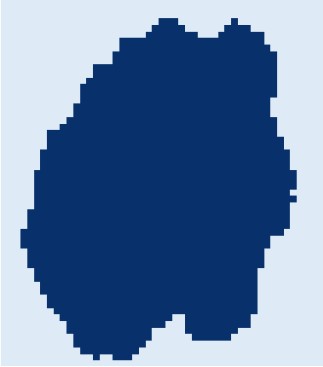}
         \caption{}
         \label{fig:rice_thresh}
     \end{subfigure}\\
    \begin{subfigure}[b]{0.32\textwidth}
         \centering
         \includegraphics[width=\textwidth,height=2.5cm]{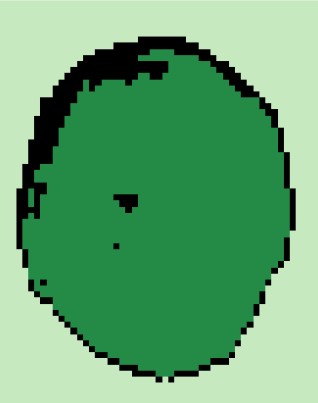}
         \caption{}
         \label{fig:wheat_dbscan}
     \end{subfigure}
     \hfill
     \begin{subfigure}[b]{0.32\textwidth}
         \centering
         \includegraphics[width=\textwidth,height=2.5cm]{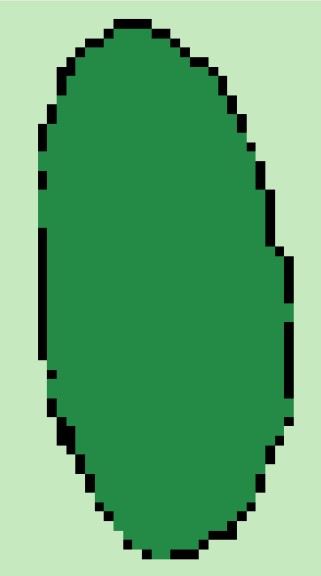}
         \caption{}
         \label{fig:corn_dbscan}
     \end{subfigure}
     \hfill
     \begin{subfigure}[b]{0.32\textwidth}
         \centering
         \includegraphics[width=\textwidth,height=2.5cm]{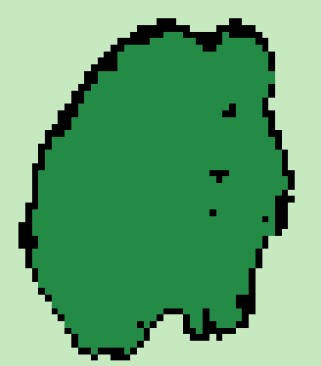}
         \caption{}
         \label{fig:rice_dbscan}
     \end{subfigure}\\
    \begin{subfigure}[b]{0.32\textwidth}
         \centering
         \includegraphics[width=\textwidth,height=2.5cm]{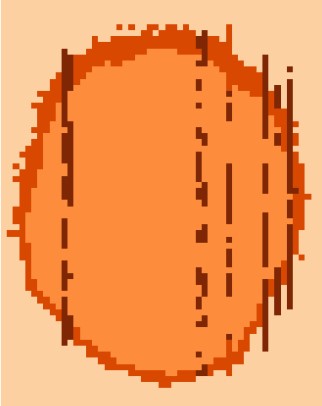}
         \caption{}
         \label{fig:wheat_gmm}
     \end{subfigure}
     \hfill
     \begin{subfigure}[b]{0.32\textwidth}
         \centering
         \includegraphics[width=\textwidth,height=2.5cm]{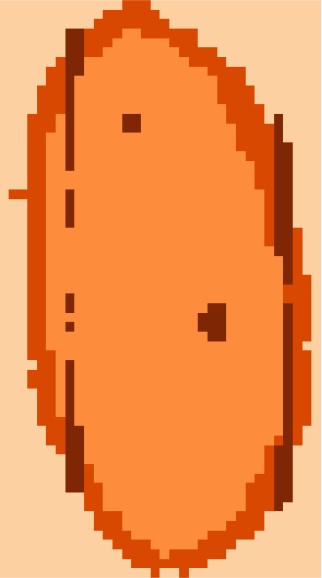}
         \caption{}
         \label{fig:corn_gmm}
     \end{subfigure}
     \hfill
     \begin{subfigure}[b]{0.32\textwidth}
         \centering
         \includegraphics[width=\textwidth,height=2.5cm]{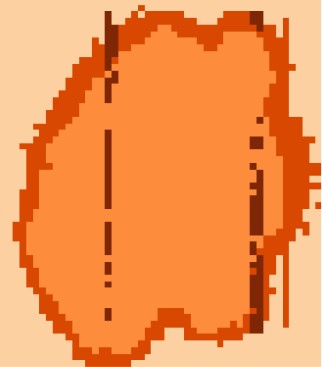}
         \caption{}
         \label{fig:rice_gmm}
     \end{subfigure}\\
    \begin{subfigure}[b]{0.32\textwidth}
         \centering
         \includegraphics[width=\textwidth,height=2.5cm]{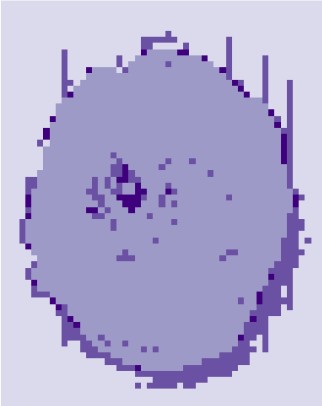}
         \caption{}
         \label{fig:wheat_pgmm}
     \end{subfigure}
     \hfill
     \begin{subfigure}[b]{0.32\textwidth}
         \centering
         \includegraphics[width=\textwidth,height=2.5cm]{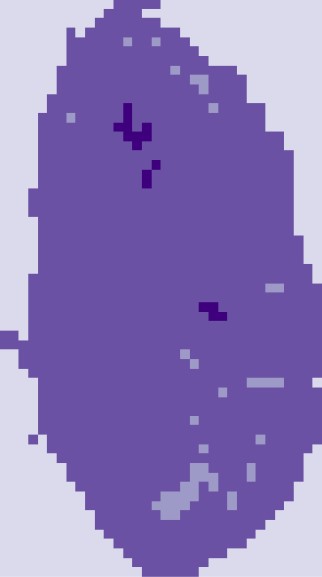}
         \caption{}
         \label{fig:corn_pgmm}
     \end{subfigure}
     \hfill
     \begin{subfigure}[b]{0.32\textwidth}
         \centering
         \includegraphics[width=\textwidth,height=2.5cm]{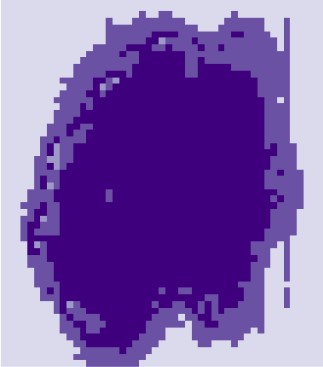}
         \caption{}
         \label{fig:rice_pgmm}
     \end{subfigure}\\
    \begin{subfigure}[b]{0.32\textwidth}
         \centering
         \includegraphics[width=\textwidth,height=2.5cm]{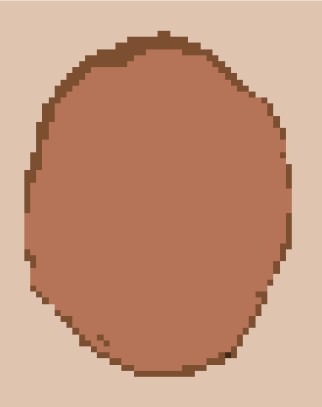}
         \caption{}
         \label{fig:wheat_ccpgmm}
     \end{subfigure}
     \hfill
     \begin{subfigure}[b]{0.32\textwidth}
         \centering
         \includegraphics[width=\textwidth,height=2.5cm]{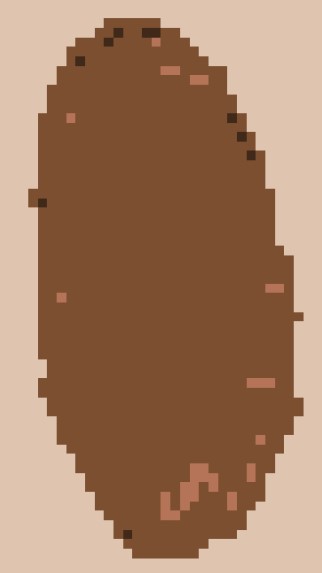}
         \caption{}
         \label{fig:corn_ccpgmm}
     \end{subfigure}
     \hfill
     \begin{subfigure}[b]{0.32\textwidth}
         \centering
         \includegraphics[width=\textwidth,height=2.5cm]{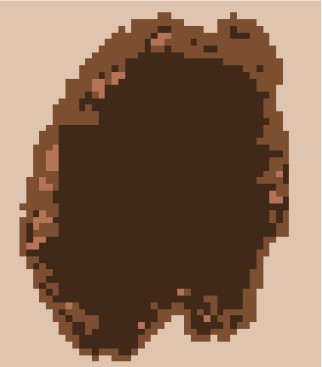}
         \caption{}
         \label{fig:rice_ccpgmm}
     \end{subfigure}\\
     \begin{subfigure}[b]{0.32\textwidth}
         \centering         \includegraphics[width=\textwidth,height=2.5cm]{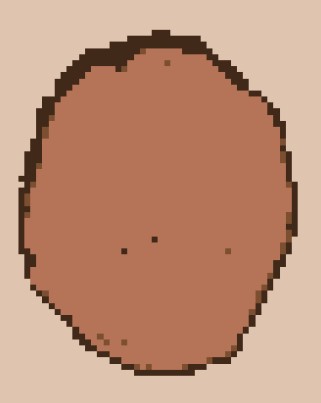}
         \caption{}
         \label{fig:wheat_ccpgmm_small}
     \end{subfigure}
     \hfill
     \begin{subfigure}[b]{0.32\textwidth}
         \centering
         \includegraphics[width=\textwidth,height=2.5cm]{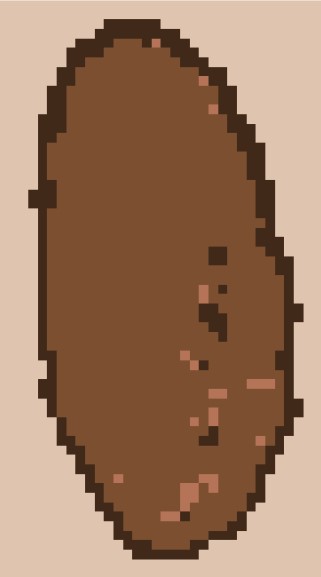}
         \caption{}
         \label{fig:corn_ccpgmm_small}
     \end{subfigure}
     \hfill
     \begin{subfigure}[b]{0.32\textwidth}
         \centering
         \includegraphics[width=\textwidth,height=2.5cm]{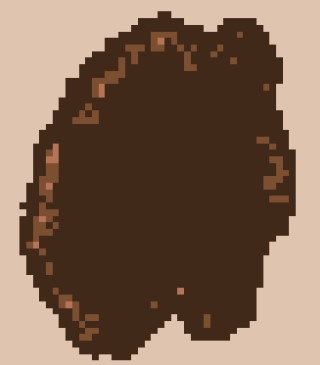}
         \caption{}
         \label{fig:rice_ccpgmm_small}
     \end{subfigure}\\
     \end{tabular}
        \caption{Greyscale images and pixel labels for three of the nine hyperspectral images ($l = \{1,4,7\}$) of the cereals using the threshold approach, DBSCAN, GMM, PGMM, ccPGMM with $43.5\%$pixels as constraints, and ccPGMM with $24.7\%$ pixels as constraints respectively for wheat (\ref{fig:wheat_grey}, \ref{fig:wheat_thresh}, \ref{fig:wheat_dbscan}, \ref{fig:wheat_gmm}, \ref{fig:wheat_pgmm}, \ref{fig:wheat_ccpgmm}, \ref{fig:wheat_ccpgmm_small}), corn (\ref{fig:corn_grey}, \ref{fig:corn_thresh}, \ref{fig:corn_dbscan}, \ref{fig:corn_gmm}, \ref{fig:corn_pgmm}, \ref{fig:corn_ccpgmm}, \ref{fig:corn_ccpgmm_small}) and rice (\ref{fig:rice_grey}, \ref{fig:rice_thresh}, \ref{fig:rice_dbscan}, \ref{fig:rice_gmm}, \ref{fig:rice_pgmm}, \ref{fig:rice_ccpgmm}, \ref{fig:rice_ccpgmm_small}).}
        \label{fig:results_cereal}
\end{figure}

Figure \ref{fig:results_cereal} shows the greyscale images for three of the nine hyperspectral images, and the pixel labels as generated using the threshold approach, DBSCAN, GMM, PGMM, ccPGMM with $43.5\%$ pixels as constraints and ccPGMM with $24.7\%$ pixels as constraints; Figures \ref{fig:results_cereal_2} and \ref{fig:results_cereal_3} in Appendix \ref{appendix_additionalclusterresults} illustrate the same for the remaining six hyperspectral images. Figures \ref{fig:wheat_thresh}, \ref{fig:corn_thresh}, and \ref{fig:rice_thresh} show the labels generated for each hyperspectral image by the threshold approach. Each type of pixel is coloured in a different shade of blue according to its label; the resulting labels are very close to the respective greyscale images (\ref{fig:wheat_grey}, \ref{fig:corn_grey}, and \ref{fig:rice_grey}).  
However, the threshold approach is ad-hoc involving subjective choice,  
and does not provide any clustering uncertainties. Figures \ref{fig:wheat_dbscan}, \ref{fig:corn_dbscan}, and \ref{fig:rice_dbscan} show the cluster solutions of DBSCAN where each pixel is coloured in a different shade of green depending on its label. On these data, it took less than a minute to run the DBSCAN approach. However, while DBSCAN correctly distinguishes the background pixels from the cereal pixels in each hyperspectral image, the pixels of the three different cereals are incorrectly allocated to the same cluster. Similarly, GMM clusters the pixels of the three types of cereals into a single cluster as shown in Figures \ref{fig:wheat_gmm}, \ref{fig:corn_gmm}, and \ref{fig:rice_gmm}. In addition, GMM allocates the uncertain edge pixels of the cereal grains to a separate cluster and allocates some pixels that likely correspond to spectral noise into another separate cluster. It takes nearly $16$ minutes to fit all the GMM models and the ``EVE" model \citep{mclust_2016} is selected based on the BIC. Figures \ref{fig:wheat_pgmm}, \ref{fig:corn_pgmm}, and \ref{fig:rice_pgmm} illustrate the pixel labels based on the application of PGMM to all $p$ variables. Each label obtained from the fitted PGMM is coloured in a different shade of purple. Though PGMM has correctly allocated several background and cereal pixels to their respective separate clusters, it also appears to have incorrectly clustered some together, giving non-contiguous images. In Figure \ref{fig:corn_pgmm}, which shows a corn cereal, many background pixels are allocated to the corn cluster. In Figure \ref{fig:wheat_pgmm} and \ref{fig:rice_pgmm}, corresponding to wheat and rice cereals respectively, several edge pixels of the cereals are incorrectly allocated to the corn cluster. In addition, PGMM takes nearly $72$ hours to cluster the motivating dataset.

 Figures \ref{fig:wheat_ccpgmm}, \ref{fig:corn_ccpgmm}, and \ref{fig:rice_ccpgmm} show the labels of the pixels, in different shades of brown, obtained by applying ccPGMM with $43.5\%$ of pixels as constraints. Applying ccPGMM to all nine hyperspectral images simultaneously took $6$ hours and the cluster labels are close to the labels obtained by applying the threshold approach to each image individually. Compared to the other clustering solutions considered, ccPGMM distinguishes the cereals well from the background while, overall, pixels are allocated to different clusters corresponding to corn, wheat, and rice. However, some pixels are still incorrectly mixed across clusters, particularly around the edges of the wheat and rice cereals. Figures \ref{fig:wheat_ccpgmm_small}, \ref{fig:corn_ccpgmm_small}, and \ref{fig:rice_ccpgmm_small} show the cluster labels obtained by applying ccPGMM with $24.7\%$ pixels as constraints in different shades of brown. The ccPGMM fitted with $24.7\%$ of pixels as constraints clusters the pixels as effectively as the ccPGMM with $43.5\%$ of pixels as constraints. However, some pixels around the edges of the wheat and corn cereal are incorrectly clustered as rice. 
 
 Importantly, the uncertainties associated with the cluster membership for each pixel are available given the model-based approach to clustering. Under ccPGMM, let $\hat{z}^m_{ng}$ be the maximum estimated posterior probability (corresponding to the \emph{maximum a posteriori} cluster $g$) for pixel $n$ and variable subset $m$, obtained on convergence of constrained-PGMM. Then, the uncertainty for pixel $n$ is $\frac{1}{M}\sum_{m=1}^M (1 - \hat{z}^m_{ng})$. For GMM and PGMM cluster labels, the associated uncertainty for each pixel is calculated as $1$ minus the value of the corresponding maximum estimated posterior probability.
 The uncertainty plots are given in Figures \ref{fig:results_cereal_1_uncertainty}, \ref{fig:results_cereal_2_uncertainty}, and \ref{fig:results_cereal_3_uncertainty} in Appendix \ref{appendix_clusteruncert}; intuitively the pixels around the edges of the cereals have the highest clustering uncertainties.

\subsection{Classification of pixels in a hyperspectral image with multiple puffed cereal grains}\label{Mixture_puffed_cereal_classification}

\begin{figure}
     \centering
    \begin{tabular}{c}
    \begin{subfigure}[b]{0.48\textwidth}
         \centering         \includegraphics[width=\textwidth,height=4cm]{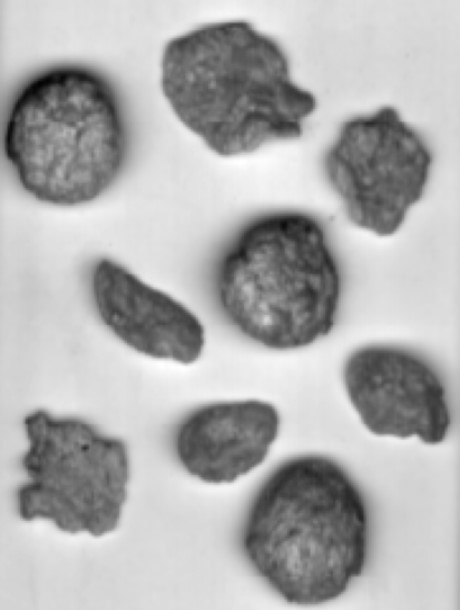}
         \caption{}         \label{fig:Mixture_Puffed_Cereal_Grey}
     \end{subfigure}
     \hfill
     \begin{subfigure}[b]{0.48\textwidth}
         \centering         \includegraphics[width=\textwidth,height=4cm]{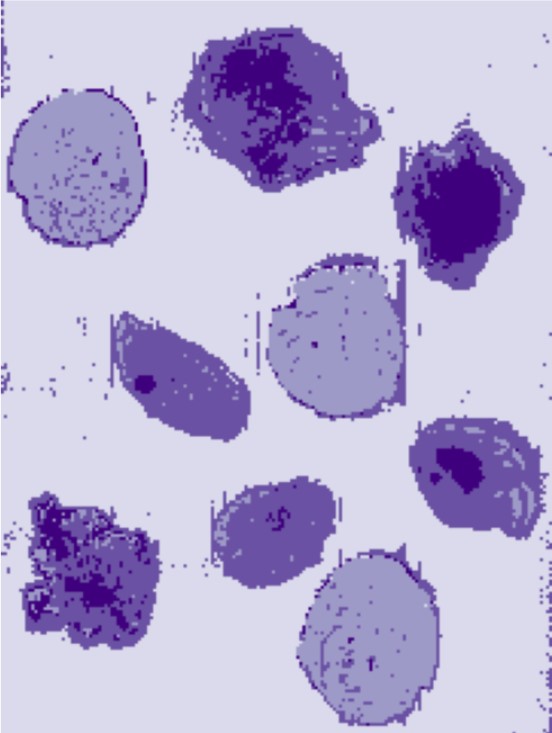}
         \caption{}         \label{fig:Mixture_Puffed_Cereal_PGMM_Classification}
     \end{subfigure}\\
    \begin{subfigure}[b]{0.48\textwidth}
         \centering         \includegraphics[width=\textwidth,height=4cm]{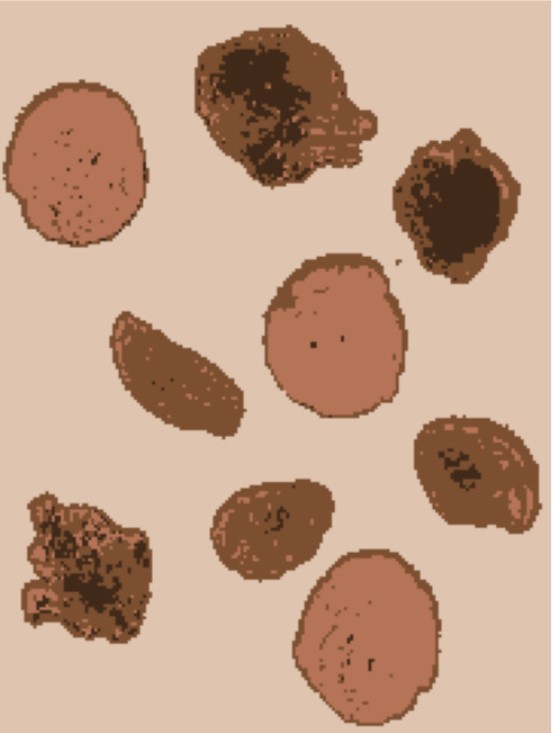}
         \caption{}         \label{fig:Mixture_Puffed_Cereal_ccPGMM_LargeConstraints_Classification}
     \end{subfigure}
     \hfill
     \begin{subfigure}[b]{0.48\textwidth}
         \centering         \includegraphics[width=\textwidth,height=4cm]{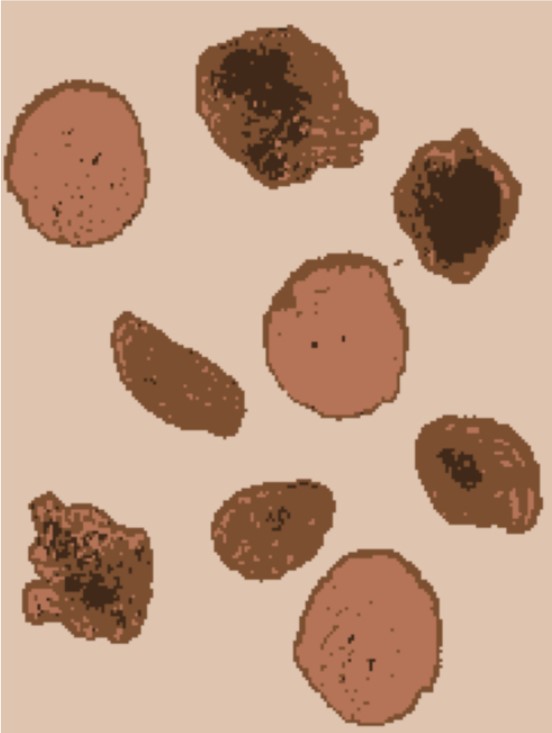}
         \caption{}         \label{fig:Mixture_Puffed_Cereal_ccPGMM_SmallConstraints_Classification}
     \end{subfigure}\\
     \end{tabular}
        \caption{The greyscale image of the multi-grain hyperspectral image which contains three grains of each of wheat, corn, and rice. (\ref{fig:Mixture_Puffed_Cereal_Grey}); for each pixel, the intensity of the color corresponds to the average of its NIR spectrum. Classification labels for each pixel using the fitted PGMM (\ref{fig:Mixture_Puffed_Cereal_PGMM_Classification}), ccPGMM with $43.5\%$ of pixels as constraints (\ref{fig:Mixture_Puffed_Cereal_ccPGMM_LargeConstraints_Classification}), and ccPGMM with $24.7\%$ of pixels as constraints (\ref{fig:Mixture_Puffed_Cereal_ccPGMM_SmallConstraints_Classification}).}
        \label{fig:Mixture_Puffed_Cereal_Classification}
\end{figure}

To demonstrate the utility of ccPGMM, it is used here in a supervised manner to classify the pixels of a hyperspectral image that contains multiple puffed cereal grains of  different types. In this multi-grain hyperspectral image $\mathbf{Y} = \{\mathbf{y}_1, \ldots, \mathbf{y}_N\}$ with $N = 43,621$ pixels, each pixel $\mathbf{y}_n$ contains the reflectance information captured at the $p = 101$ wavelengths of an NIR spectrum. Figure \ref{fig:Mixture_Puffed_Cereal_Grey} shows the greyscale image of this multi-grain hyperspectral image; for each pixel, the intensity of the colour corresponds to the average of the reflectance information captured in the NIR spectrum.

To classify the pixels of this multi-grain image, ccPGMM with $G = 4$ and $q = 2$ is first fitted to the $L = 9$ single-grain hyperspectral images, with $M = 25$ and $d = 20$. The resulting parameter estimates 
$(\hat{\tau}_g^{m}, \bm{\hat{\mu}}_g^{m}, \hat{\bm{\Lambda}}^{m}_g, \hat{\bm{\Psi}}^{m}_g)$ for $g = 1, \ldots, G$ and $m = 1, \ldots, M$ are used to compute the posterior probability of membership of cluster $g$ for each pixel $n$ i.e., $\hat{z}^m_{ng} \propto \hat{\tau}^m_g \mathcal{N}_d(\mathbf{y}^{m}_n;\hat{\bm{\mu}}^{m}_g,\hat{\bm{\Lambda}}^{m}_g\:(\hat{\bm{\Lambda}}^{m}_g)^{\top}\:+\:\hat{\bm{\Psi}}^{m}_g)$.
Pixel $n$'s classification label in subset $m$, $c^m_{n}$, is the cluster for which it has maximum probability of membership.
Thus for each pixel $n$, a vector of classification labels $\mathbf{c}_n = \{c^1_{n}, \ldots,c^M_{n}\}$ is then available. Label switching in the classification labels across the $M$ subsets is corrected, taking as reference the cluster label of the pixels constrained while fitting ccPGMM, to ensure consistency. The final consensus classification label $c_n^*$ for pixel $n$ is the mode of $\mathbf{c}_n$, with associated uncertainty of $1 - \frac{1}{M} \#(c^*_n)$, where $\#(c^*_n)$ is the frequency of $c_n^*$ in $\mathbf{c}_n$. For comparison, the estimated parameters of PGMM fitted to the $L=9$ hyperspectral images (as in Section \ref{application}) are used to compute $\hat{z}_{ng}$ as in (\ref{pos_prob_noconstraints}); labels, and associated uncertainties, are obtained based on \emph{maximum a posteriori} values.

Figures \ref{fig:Mixture_Puffed_Cereal_PGMM_Classification}, \ref{fig:Mixture_Puffed_Cereal_ccPGMM_LargeConstraints_Classification} and \ref{fig:Mixture_Puffed_Cereal_ccPGMM_SmallConstraints_Classification} show the classification labels for the pixels using the fitted PGMM, ccPGMM with $43.5\%$ of pixels constrained, and ccPGMM with $24.7\%$ of pixels constrained, respectively. As true labels are unavailable, performance is assessed by visual inspection. The PGMM approach finds it difficult to discern between background and grain pixels, and also between grain pixels. Both ccPGMM approaches do well in discerning background from grain pixels and show more homogeneity in labels within grains, but there is still confusion between grains. Figure \ref{fig:Mixture_Puffed_Cereal_Classification_Uncertainty} in Appendix \ref{appendix_clusteruncert} shows the pixels' classification uncertainties; intuitively uncertainty is higher in the consensus-based approaches and, within the consensus-based approaches, classification uncertainties are higher when fewer pixels are constrained.

\section{Discussion}\label{discussion}
A novel model-based approach to simultaneously labelling pixels of hyperspectral images while availing of additional information about the relationship between groups of pixels within the images is proposed. The proposed ccPGMM demonstrates strong clustering performance when compared to existing approaches both in terms of quality of the clustering solution, and the computational time required to cluster large datasets. 

While the ccPGMM approach is less subjective than the currently used threshold approach, user decisions must still be made. For example, the selection of the number of factors $q$, while well established, is subjective. Examining alternative approaches to choosing $q$ either using model selection criteria or using shrinkage priors \citep{bhattacharya_et_al_2011} in a Bayesian context would address this, but increase the computational cost. Indeed, under the ccPGMM approach, allowing for different values of $q$ when fitting to the $M$ different subsets of variables could bring improved performance but would require further exploration. Further, the choice of $M$ and $d$ are currently data-driven and therefore subjective. A very large $p$ will likely require large values of $d$ and $M$ for appropriate coverage of the available set of variables and to ensure that informative variables are employed for clustering, consequently increasing the computational cost of ccPGMM. Performing variable selection before or as part of the clustering algorithm \citep{DeanRaftery2006, mcparland_et_al_2017,fop_et_al_2018} would help to reduce the set of variables to be considered. Further, considering co-clustering in the context of ccPGMM, similar to \cite{jacques:2016}, could tackle the high-dimensionality by facilitating the simultaneous clustering of pixels and variables.

Though ccPGMM is computationally efficient, an increase in the number of observations $N$ leads to scalability issues as generating the multiple $N \times N$ similarity matrices is computationally and memory intensive. The efficient use of accelerators and GPUs \citep{arefin_et_al_2012} could resolve this but at a higher equipment cost.  Additionally, discarding the non-informative clustering partitions in the consensus approach could help to reduce the number of similarity matrices required for the consensus solution, thereby reducing memory requirements in parallel computation. An approach based on the entropy of the clustering allocations \citep{Gray_2011} in the ensemble is currently under exploration, and further investigation is needed to fully evaluate its effectiveness.

In ccPGMM, the consensus approach has been explored to handle
high-dimensional data of the same type. However, the consensus approach could intuitively be employed to cluster data of multiple types and/or from multiple sources. For example, the consensus approach could be used to merge clustering solutions from both NIR and visible spectrum hyperspectral images of same sample.

Overall, the proposed ccPGMM model works well when clustering a large number of pixels in a set of hyperspectral images, and provides associated clustering uncertainty, in a computationally efficient manner. As such, the approach should have a broader application to other hyperspectral images where labelling of the pixels is required.

\section*{Acknowledgements}
This publication has emanated from research conducted with the financial support of Science Foundation Ireland under Grant number 18/CRT/6049. For the purpose of Open Access, the author has applied a CC BY public copyright licence to any Author Accepted Manuscript version arising from this submission. The authors are grateful to members of the Working Group in Model-based Clustering for discussions that improved this work.

{\small
\bibliographystyle{apalike}
\bibliography{main.bib}

\begin{thebibliography}{}

\bibitem[Abdi and Williams, 2010]{abdi_et_al_2010}
Abdi, H. and Williams, L.~J. (2010).
\newblock Principal component analysis.
\newblock {\em Wiley Interdisciplinary Reviews: Computational Statistics}, 2(4):433--459.

\bibitem[Amigo, 2010]{amigo_jos_2010}
Amigo, J.~M. (2010).
\newblock Practical issues of hyperspectral imaging analysis of solid dosage forms.
\newblock {\em Analytical and Bioanalytical Chemistry}, 398(1):93--109.

\bibitem[Amigo, 2020]{amigo_jos_2020}
Amigo, J.~M. (2020).
\newblock Hyperspectral and multispectral imaging: setting the scene.
\newblock In {\em Data Handling in Science and Technology}, volume~32, pages 3--16. Elsevier.

\bibitem[Amigo et~al., 2015]{amigo_et_al_2015}
Amigo, J.~M., Babamoradi, H., and Elcoroaristizabal, S. (2015).
\newblock Hyperspectral image analysis. {A} tutorial.
\newblock {\em Analytica Chimica Acta}, 896:34--51.

\bibitem[Amigo and Santos, 2020]{amigo_et_al_2020}
Amigo, J.~M. and Santos, C. (2020).
\newblock Preprocessing of hyperspectral and multispectral images.
\newblock In {\em Data Handling in Science and Technology}, volume~32, pages 37--53. Elsevier.

\bibitem[Arefin et~al., 2012]{arefin_et_al_2012}
Arefin, A.~S., Riveros, C., Berretta, R., and Moscato, P. (2012).
\newblock Computing large-scale distance matrices on {GPU}.
\newblock In {\em The 7th International Conference on Computer Science \& Education (ICCSE 2012)}, pages 576--580. IEEE.

\bibitem[Bhattacharya and Dunson, 2011]{bhattacharya_et_al_2011}
Bhattacharya, A. and Dunson, D.~B. (2011).
\newblock Sparse {B}ayesian infinite factor models.
\newblock {\em Biometrika}, 98(2):291--306.

\bibitem[Bocklitz et~al., 2011]{thomas_et_al_2011}
Bocklitz, T., Walter, A., Hartmann, K., R{\"o}sch, P., and Popp, J. (2011).
\newblock How to pre-process {R}aman spectra for reliable and stable models?
\newblock {\em Analytica Chimica Acta}, 704(1-2):47--56.

\bibitem[Bouveyron and Brunet-Saumard, 2014]{bouveyron:2014}
Bouveyron, C. and Brunet-Saumard, C. (2014).
\newblock Discriminative variable selection for clustering with the sparse {F}isher-{EM} algorithm.
\newblock {\em Computational Statistics}, 29(3):489--513.

\bibitem[Bouveyron et~al., 2019]{charles_et_al_2019}
Bouveyron, C., Celeux, G., Murphy, T.~B., and Raftery, A.~E. (2019).
\newblock {\em Model-based clustering and classification for data science: with applications in {R}}, volume~50.
\newblock Cambridge University Press.

\bibitem[Feng and Sun, 2012]{feng_et_al_2012}
Feng, Y.-Z. and Sun, D.-W. (2012).
\newblock Application of hyperspectral imaging in food safety inspection and control: a review.
\newblock {\em Critical Reviews in Food Science and Nutrition}, 52(11):1039--1058.

\bibitem[Fern and Brodley, 2003]{fern_et_al_2003}
Fern, X.~Z. and Brodley, C.~E. (2003).
\newblock Random projection for high dimensional data clustering: {A} cluster ensemble approach.
\newblock In {\em Proceedings of the Twentieth International Conference on Machine Learning (ICML-2003)}, pages 186--193.

\bibitem[Fop and Murphy, 2018]{fop_et_al_2018}
Fop, M. and Murphy, T.~B. (2018).
\newblock Variable selection methods for model-based clustering.
\newblock {\em Statistics Surveys}, 12:18--65.

\bibitem[Geladi et~al., 1985]{geladi_et_al_1985}
Geladi, P., MacDougall, D., and Martens, H. (1985).
\newblock Linearization and scatter-correction for near-infrared reflectance spectra of meat.
\newblock {\em Applied Spectroscopy}, 39(3):491--500.

\bibitem[Ghahramani et~al., 1996]{ghahramani_et_al_1996}
Ghahramani, Z., Hinton, G.~E., et~al. (1996).
\newblock The {EM} algorithm for mixtures of factor analyzers.
\newblock Technical report, Technical Report CRG-TR-96-1, University of Toronto.

\bibitem[Gowen et~al., 2019]{gowen_et_al_2019}
Gowen, A.~A., Xu, J.-L., and Herrero-Langreo, A. (2019).
\newblock Comparison of spectral selection methods in the development of classification models from visible near infrared hyperspectral imaging data.
\newblock {\em Journal of Spectral Imaging}, 8.

\bibitem[Gray, 2011]{Gray_2011}
Gray, R.~M. (2011).
\newblock {\em Entropy and information theory}.
\newblock Springer Science \& Business Media.

\bibitem[Hahsler et~al., 2019]{hahsler_et_al_2019}
Hahsler, M., Piekenbrock, M., and Doran, D. (2019).
\newblock dbscan: {F}ast density-based clustering with {R}.
\newblock {\em Journal of Statistical Software}, 91:1--30.

\bibitem[Hubert and Arabie, 1985]{hubert_et_al_1985}
Hubert, L. and Arabie, P. (1985).
\newblock Comparing partitions.
\newblock {\em Journal of Classification}, 2:193--218.

\bibitem[Jacques and Ruckebusch, 2016]{jacques:2016}
Jacques, J. and Ruckebusch, C. (2016).
\newblock Model-based co-clustering for hyperspectral images.
\newblock {\em Journal of Spectral Imaging}.

\bibitem[Khan et~al., 2014]{khan_et_al_2014}
Khan, K., Rehman, S.~U., Aziz, K., Fong, S., and Sarasvady, S. (2014).
\newblock {DBSCAN}: {P}ast, present and future.
\newblock In {\em The Fifth International Conference on the Applications of Digital Information and Web Technologies (ICADIWT 2014)}, pages 232--238. IEEE.

\bibitem[McLachlan and Peel, 2000]{mclachlan_et_al_2000}
McLachlan, G. and Peel, D. (2000).
\newblock Mixtures of factor analyzers.
\newblock In {\em Proceedings of the Seventeenth International Conference on Machine Learning}. Citeseer.

\bibitem[McLachlan and Peel, 2004]{mclachlan_et_al_2004}
McLachlan, G.~J. and Peel, D. (2004).
\newblock {\em Finite mixture models}.
\newblock John Wiley \& Sons.

\bibitem[McLachlan et~al., 2003]{mclachlan_et_al_2003}
McLachlan, G.~J., Peel, D., and Bean, R.~W. (2003).
\newblock Modelling high-dimensional data by mixtures of factor analyzers.
\newblock {\em Computational Statistics \& Data Analysis}, 41(3-4):379--388.

\bibitem[McNicholas, 2016]{mcnicholas_et_al_2016}
McNicholas, P.~D. (2016).
\newblock Model-based clustering.
\newblock {\em Journal of Classification}, 33(3):331--373.

\bibitem[McNicholas et~al., 2023]{mcnicholas_et_al_2023}
McNicholas, P.~D., ElSherbiny, A., McDaid, A.~F., and Murphy, T.~B. (2023).
\newblock {\em pgmm: Parsimonious Gaussian mixture models}.
\newblock R package version 1.2.7.

\bibitem[McNicholas and Murphy, 2008]{mcnicholas_et_al_2008}
McNicholas, P.~D. and Murphy, T.~B. (2008).
\newblock Parsimonious {G}aussian mixture models.
\newblock {\em Statistics and Computing}, 18(3):285--296.

\bibitem[McNicholas et~al., 2010]{mcnicholas_et_al_2010}
McNicholas, P.~D., Murphy, T.~B., McDaid, A.~F., and Frost, D. (2010).
\newblock Serial and parallel implementations of model-based clustering via parsimonious {G}aussian mixture models.
\newblock {\em Computational Statistics \& Data Analysis}, 54(3):711--723.

\bibitem[McParland et~al., 2017]{mcparland_et_al_2017}
McParland, D., Phillips, C.~M., Brennan, L., Roche, H.~M., and Gormley, I.~C. (2017).
\newblock Clustering high-dimensional mixed data to uncover sub-phenotypes: joint analysis of phenotypic and genotypic data.
\newblock {\em Statistics in Medicine}, 36(28):4548--4569.

\bibitem[Melnykov et~al., 2016]{melnykov_et_al_2016}
Melnykov, V., Melnykov, I., and Michael, S. (2016).
\newblock Semi-supervised model-based clustering with positive and negative constraints.
\newblock {\em Advances in Data Analysis and Classification}, 10(3):327--349.

\bibitem[Neath and Cavanaugh, 2012]{neath_et_al_2012}
Neath, A.~A. and Cavanaugh, J.~E. (2012).
\newblock The {B}ayesian information criterion: background, derivation, and applications.
\newblock {\em Wiley Interdisciplinary Reviews: Computational Statistics}, 4(2):199--203.

\bibitem[Pal and Pal, 1993]{pal_et_al_1993}
Pal, N.~R. and Pal, S.~K. (1993).
\newblock A review on image segmentation techniques.
\newblock {\em Pattern Recognition}, 26(9):1277--1294.

\bibitem[Peterson, 2009]{peterson_et_al_2009}
Peterson, L.~E. (2009).
\newblock K-nearest neighbor.
\newblock {\em Scholarpedia}, 4(2):1883.

\bibitem[Punera and Ghosh, 2008]{punera_et_al_2008}
Punera, K. and Ghosh, J. (2008).
\newblock {C}onsensus-based ensembles of soft clusterings.
\newblock {\em Applied Artificial Intelligence}, 22(7-8):780--810.

\bibitem[{R Core Team}, 2023]{rsoftware}
{R Core Team} (2023).
\newblock {\em R: A language and environment for statistical computing}.
\newblock R Foundation for Statistical Computing, Vienna, Austria.

\bibitem[Raftery and Dean, 2006]{DeanRaftery2006}
Raftery, A.~E. and Dean, N. (2006).
\newblock Variable selection for model-based clustering.
\newblock {\em Journal of the American Statistical Association}, 101(473):168--178.

\bibitem[Rinnan et~al., 2009]{rinnan_et_al_2009}
Rinnan, {\AA}., Van Den~Berg, F., and Engelsen, S.~B. (2009).
\newblock Review of the most common pre-processing techniques for near-infrared spectra.
\newblock {\em TrAC Trends in Analytical Chemistry}, 28(10):1201--1222.

\bibitem[Russell et~al., 2015]{russell_et_al_2015}
Russell, N., Murphy, T.~B., and Raftery, A.~E. (2015).
\newblock Bayesian model averaging in model-based clustering and density estimation.
\newblock {\em arXiv preprint arXiv:1506.09035}.

\bibitem[Schwarz, 1978]{schwarz_1978}
Schwarz, G. (1978).
\newblock {E}stimating the dimension of a model.
\newblock {\em The Annals of Statistics}, 6(2):461--464.

\bibitem[Scrucca et~al., 2016]{mclust_2016}
Scrucca, L., Fop, M., Murphy, T.~B., and Raftery, A.~E. (2016).
\newblock {mclust} 5: clustering, classification and density estimation using {G}aussian finite mixture models.
\newblock {\em The {R} Journal}, 8(1):289--317.

\bibitem[Sokal, 1963]{sokal_1963}
Sokal, R.~R. (1963).
\newblock The principles and practice of numerical taxonomy.
\newblock {\em Taxon}, pages 190--199.

\bibitem[Strehl and Ghosh, 2002]{strehl_et_al_2002}
Strehl, A. and Ghosh, J. (2002).
\newblock Cluster ensembles -- a knowledge reuse framework for combining multiple partitions.
\newblock {\em Journal of Machine Learning Research}, 3(Dec):583--617.

\bibitem[Xu et~al., 2020]{xu_et_al_2020}
Xu, J.-L., Riccioli, C., Herrero-Langreo, A., and Gowen, A.~A. (2020).
\newblock Deep learning classifiers for near infrared spectral imaging: a tutorial.
\newblock {\em Journal of Spectral Imaging}, 9.

\bibitem[Zhai et~al., 2021]{zhai:2021}
Zhai, H., Zhang, H., Li, P., and Zhang, L. (2021).
\newblock Hyperspectral image clustering: Current achievements and future lines.
\newblock {\em IEEE Geoscience and Remote Sensing Magazine}, 9(4):35--67.

\end{thebibliography}
}
\newpage

\appendix

\section{Dimensions of hyperspectral images} 
\label{appendix_ImageDimensions}
As discussed in Section \ref{sec:data}, Table \ref{images_dimension_table} details the dimensionality of the nine hyperspectral images that each contain a single puffed cereal grain. The single image that contains multiple puffed cereal grains of potentially different types is of dimension 241 $\times$ 181 i.e., it has 43,621 pixels.

\begin{table}[h!]
    \centering        \caption{The dimensionality of the motivating hyperspectral images of puffed cereals.}
    \begin{tabular}{     |P{6cm}||P{1cm}|P{1cm}|P{1cm}|P{1cm}|}
     \hline
     \textbf{Puffed cereal type} & \textbf{$l$} & \textbf{$S_l$} & \textbf{$T_l$} & \textbf{$S_l \times T_l$}\\
     \hline\hline
     \textbf{Wheat}   & 1 & 67 & 53 & 3551\\
     \hline 
    \textbf{Wheat}   & 2 & 62 & 64 & 3968\\
     \hline      \textbf{Wheat}   & 3 & 65 & 55 & 3575\\
     \hline      \textbf{Corn}   & 4 & 61 & 34 & 2074\\
     \hline 
    \textbf{Corn}   & 5 & 39 & 69 & 2691\\
     \hline      \textbf{Corn}   & 6 & 56 & 54 & 3024\\
     \hline      \textbf{Rice}   & 7 & 56 & 49 & 2744\\
     \hline 
    \textbf{Rice}   & 8 & 78 & 54 & 4212\\
     \hline      \textbf{Rice}   & 9 & 50 & 44 & 2200\\
     \hline 
    \end{tabular}
    \label{images_dimension_table}
\end{table}

\section{The AECM algorithm for constrained-PGMM} \label{appendix_AECM}
The constrained parsimonious Gaussian mixture model (constrained-PGMM) is a finite mixture of $G$ factor analysis models such that each mixture component $g$ has a multivariate Gaussian distribution with mean $\bm{\mu}_{g}$, covariance $\bm{\Sigma}_g = \bm{\Lambda}_g\bm{\Lambda}_g^{\top}+\bm{\Psi}_g$, and the probability of membership of mixture component $\tau_g$. Denoting the parameter vector $\bm{\Theta} = \{\tau_1, \ldots, \tau_G, \bm{\mu}_1, \ldots, \bm{\mu}_G, \bm{\Lambda}_1, \ldots, \bm{\Lambda}_G,\bm{\Psi}_1, \ldots, \bm{\Psi}_G\}$, under the constrained-PGMM, the observed data log-likelihood is  

\begin{equation}\label{obs_log_likelihood}
    \ell_o(\bm{\Theta}; \mathbf{X}) = \sum_{n=1}^{N} \log \sum_{g=1}^G \tau_g\: \mathcal{N}_p(\mathbf{x}_n ; \bm{\mu}_g, \bm{\Lambda}_g, \bm{\Psi}_g).
\end{equation}

As (\ref{obs_log_likelihood}) is difficult to maximize directly, an AECM algorithm is used to fit the constrained-PGMM. This requires the introduction of a latent component indicator $\mathbf{z}_n = (z_{n1}, \ldots, z_{nG})^{\top}$, for $n = 1, \ldots, N$, where $z_{ng} = 1$ if observation $n$ belongs to cluster $g$ or 
 0 otherwise. The AECM algorithm works in two cycles and allows for a different definition of the complete data in each cycle to estimate the parameters.
 
\subsection{First cycle}
In the first cycle, the component indicators $\mathbf{Z}$ = \{$\mathbf{z}_1$,\ldots, $\mathbf{z}_N$\} are assumed to be the missing data. Further, $\bm{\Theta}$ is partitioned as $(\bm{\Theta}_1, \bm{\Theta}_2)$ where $\bm{\Theta}_1$ denotes the mixing proportions $\tau_g$ and the mean parameters $\bm{\mu}_g$ ($g=1, \ldots, G$), and $\bm{\Theta}_2$ denotes the loadings $\bm{\Lambda}_g$ and variances $\bm{\Psi}_g$  for $g=1, \ldots, G$. The complete data log-likelihood is then
\begin{equation}   \ell_c(\bm{\Theta}; \mathbf{X},\mathbf{Z}) = \sum_{n=1}^{N}\sum_{g=1}^{G} {z}_{ng} [\log \tau_g + \log \: \mathcal{N}_p(\mathbf{x}_n;\bm{\mu}_g,\bm{\Lambda}_g,\bm{\Psi}_g)].
\end{equation}

\subsubsection{E-step:}
The expected complete data log-likelihood is

\begin{eqnarray*}
    Q_{1}(\bm{\Theta}) \!\! & = &  \!\! \sum_{n=1}^{N}\sum_{g=1}^{G} E[{z}_{ng};\mathbf{x}_n]\Bigg[\log\tau_g - \frac{p}{2}\log(2\pi) - \frac{1}{2}\log(|\bm{\Lambda}_g\:\bm{\Lambda}_g^{\top}\:+\:\bm{\Psi}_g|) \\
     & ~ & -\frac{1}{2}[(\mathbf{x}_{n}-\bm{\mu}_{g})^{\top}(\bm{\Lambda}_g\:\bm{\Lambda}_g^{\top}\:+\:\bm{\Psi}_g)^{-1}(\mathbf{x}_{n}-\bm{\mu}_{g})]\Bigg].   
\end{eqnarray*}
The posterior probability ${\hat{z}_{ng}}$ of observation $n$ belonging to cluster $g$ is
\begin{eqnarray}\label{pos_prob_noconstraints_1}
E[{z}_{ng};\mathbf{x}_{n}]  & =  & \hat{{z}}_{ng} \: = \: \Pr({z}_{ng} = 1 ; \mathbf{x}_{n},\hat{\tau}_{g},\hat{\bm{\mu}}_{g},\hat{\bm{\Lambda}}_g,\hat{\bm{\Psi}}_g) 
    = 
\frac{\hat{\tau}_g\:\mathcal{N}_p(\mathbf{x}_{n};\hat{\bm{\mu}}_g,\hat{\bm{\Lambda}}_g\:\hat{\bm{\Lambda}}_g^{\top}\:+\:\hat{\bm{\Psi}}_g)}{\sum_{g^{'} = 1}^{G}\hat{\tau}_{g^{'}}\:\mathcal{N}_p(\mathbf{x}_{n};\hat{\bm{\mu}}_{g^{'}},\hat{\bm{\Lambda}}_{g^{'}}\:\hat{\bm{\Lambda}}_{g^{'}}^{\top}\:+\:\hat{\bm{\Psi}}_{g^{'}})}
\end{eqnarray}
where $\hat{\cdot}$ denotes an initial or current parameter estimate, as relevant. Let $B_1$ be the set of $J$ pixels (indexed by $j$) which must be clustered together, shown in the blue blocks in Figure \ref{cereal_constraints}. This is a positive constraint (denoted $+$). The posterior probability $\hat{z}_{{B_1}g}^{+}$ of the pixels in $B_1$ belonging to cluster $g$ given the positive constraints ($+$) is
\begin{equation}\nonumber
\centering
    \indent\indent \hat{{z}}_{{B_{1}},g}^{+} = \frac{\prod\limits_{\substack{j \in {B_{1}} \\ j = 1}}^J \hat{\tau}_g \mathcal{N}_p(\mathbf{x}_j;\hat{\bm{\mu}}_g,\hat{\bm{\Lambda}}_g\:\hat{\bm{\Lambda}}_g^{\top}\:+\:\hat{\bm{\Psi}}_g)}{\displaystyle\sum_{g^{'}=1}^{G} \prod\limits_{\substack{j \in {B_{1}} \\ j = 1}}^J \hat{\tau}_{g^{'}} \mathcal{N}_p(\mathbf{x}_j;\hat{\bm{\mu}}_{g^{'}},\hat{\bm{\Lambda}}_{g^{'}}\:\hat{\bm{\Lambda}}_{g^{'}}^{\top}\:+\:\hat{\bm{\Psi}}_{g^{'}})}.
\end{equation}
Let $B_2$ be the set of $K$ pixels (indexed by $k$) which must be clustered together, shown in the yellow block of Figure \ref{cereal_constraints}; these are positive constraints (denoted $+$). However, the pixels in $B_1$ should not be clustered together with the pixels in $B_2$; this is a negative constraint (denoted $-$). 
Therefore, the joint posterior probability $\hat{{z}}_{{B_{1}},g}^{+ \atop -}$ of the pixels in $B_1$ belonging to cluster $g$ and the pixels in $B_2$ not belonging to cluster $g$ (i.e., given the positive and negative constraints $+ \atop -$) is
\begin{equation}\label{posterior_probability_pos_neg}
\centering
    \indent\indent \hat{{z}}_{{B_{1}},g}^{+ \atop -} = \frac{\prod\limits_{\substack{j \in {B_{1}} \\ j = 1}}^J \hat{\tau}_g \mathcal{N}_p(\mathbf{x}_j;\hat{\bm{\mu}}_g,\hat{\bm{\Lambda}}_g\:\hat{\bm{\Lambda}}_g^{\top}\:+\:\hat{\bm{\Psi}}_g) \sum\limits_{\substack{f=1 \\ f \neq g}}^G\prod\limits_{\substack{k \in {B_{2}} \\ k = 1}}^K \hat{\tau}_f \mathcal{N}_p(\mathbf{x}_k;\hat{\bm{\mu}}_f,\hat{\bm{\Lambda}}_f\:\hat{\bm{\Lambda}}_f^{\top}\:+\:\hat{\bm{\Psi}}_f)}{\displaystyle\sum_{g^{'}=1}^{G} \prod\limits_{\substack{j \in {B_{1}} \\ j = 1}}^J \hat{\tau}_{g^{'}} \mathcal{N}_p(\mathbf{x}_j;\hat{\bm{\mu}}_{g^{'}},\hat{\bm{\Lambda}}_{g^{'}}\:\hat{\bm{\Lambda}}_{g^{'}}^{\top}\:+\:\hat{\bm{\Psi}}_{g^{'}})\sum\limits_{\substack{f=1 \\ f \neq g^{'}}}^G\prod\limits_{\substack{k \in {B_{2}} \\ k = 1}}^K \hat{\tau}_f \mathcal{N}_p(\mathbf{x}_k;\hat{\bm{\mu}}_f,\hat{\bm{\Lambda}}_f\:\hat{\bm{\Lambda}}_f^{\top}\:+\:\hat{\bm{\Psi}}_f)}.
\end{equation}
Similarly, the joint posterior probability $\hat{{z}}_{{B_{2}},g}^{+ \atop -}$ of the pixels in $B_2$ belonging to cluster $g$ and the pixels in $B_1$ not belonging to cluster $g$ is
\begin{equation}\nonumber
\centering
    \indent\indent \hat{{z}}_{{B_{2}},g}^{+ \atop -} = \frac{\prod\limits_{\substack{k \in {B_{2}} \\ k = 1}}^K \hat{\tau}_g \mathcal{N}_p(\mathbf{x}_k;\hat{\bm{\mu}}_g,\hat{\bm{\Lambda}}_g\:\hat{\bm{\Lambda}}_g^{\top}\:+\:\hat{\bm{\Psi}}_g) \sum\limits_{\substack{f=1 \\ f \neq g}}^G\prod\limits_{\substack{j \in {B_{1}} \\ j = 1}}^J \hat{\tau}_f \mathcal{N}_p(\mathbf{x}_j;\hat{\bm{\mu}}_f,\hat{\bm{\Lambda}}_f\:\hat{\bm{\Lambda}}_f^{\top}\:+\:\hat{\bm{\Psi}}_f)}{\displaystyle\sum_{g^{'}=1}^{G} \prod\limits_{\substack{k \in {B_{2}} \\ k = 1}}^K \hat{\tau}_{g^{'}} \mathcal{N}_p(\mathbf{x}_k;\hat{\bm{\mu}}_{g^{'}},\hat{\bm{\Lambda}}_{g^{'}}\:\hat{\bm{\Lambda}}_{g^{'}}^{\top}\:+\:\hat{\bm{\Psi}}_{g^{'}})\sum\limits_{\substack{f=1 \\ f \neq g^{'}}}^G\prod\limits_{\substack{j \in {B_{1}} \\ j = 1}}^J \hat{\tau}_f \mathcal{N}_p(\mathbf{x}_j;\hat{\bm{\mu}}_f,\hat{\bm{\Lambda}}_f\:\hat{\bm{\Lambda}}_f^{\top}\:+\:\hat{\bm{\Psi}}_f)}.
\end{equation}
The remaining pixels in Figure \ref{cereal_constraints} which are not in the highlighted blocks are considered as blocks with only one pixel and no negative constraints.

\subsubsection{CM-step:}
The expected complete data log-likelihood can be rewritten as
\begin{equation}\label{exp_comp_data_log_lik}
\begin{split}
    Q_{1}(\bm{\Theta}) = \sum_{n=1}^{N}\sum_{g=1}^{G} \hat{z}_{ng}^{+ \atop -}\Bigg[\log\tau_g - \frac{p}{2}\log(2\pi) - \frac{1}{2}\log(|\bm{\Lambda}_g\:\bm{\Lambda}_g^{\top}\:+\:\bm{\Psi}_g|)\\ -\frac{1}{2}[(\mathbf{x}_{n}-\bm{\mu}_{g})^{\top}(\bm{\Lambda}_g\:\bm{\Lambda}_g^{\top}\:+\:\bm{\Psi}_g)^{-1}(\mathbf{x}_{n}-\bm{\mu}_{g})]\Bigg],
\end{split}
\end{equation}
where $\hat{z}_{ng}^{+ \atop -}$ indicates in general constrained observations and unconstrained observations, for ease of notation.
The model parameters $\bm{\Theta}_1$ are updated with $\bm{\Theta}_2$ held fixed. Differentiating (\ref{exp_comp_data_log_lik}) with respect to $\bm{\mu}_g$,
\begin{equation}\nonumber
  \indent\indent \hat{\bm{\mu}}_{g} = \frac{\sum_{n=1}^{N} \hat{z}_{ng}^{+ \atop -} \mathbf{x}_n}{\sum_{n=1}^{N} \hat{z}_{ng}^{+ \atop -}}. 
\end{equation}
Differentiating (\ref{exp_comp_data_log_lik}) with respect to $\tau_g$,
\begin{equation}\nonumber
    \indent\indent \hat{\tau}_{g} = \frac{\sum_{n=1}^{N} \hat{z}_{ng}^{+ \atop -}}{N}.
\end{equation}

\subsection{Second cycle}
In the second cycle, both component indicators $\mathbf{Z} = \{\mathbf{z}_1, \ldots, \mathbf{z}_n\}$ and latent factors $\mathbf{U} = \{\mathbf{u}_1, \ldots, \mathbf{u}_n\}$ are assumed to be the missing data. Therefore, the complete data log-likelihood is
\begin{equation}\nonumber    \ell_c(\bm{\Theta};\mathbf{X},\mathbf{Z},\mathbf{U}) = \sum_{n=1}^{N}\sum_{g=1}^{G} z_{ng} [\log \tau_g + \log f(\mathbf{x}_n;\bm{\mu}_g,\bm{\Lambda}_g,\bm{\Psi}_g,\mathbf{u}_n) + \log f(\mathbf{u}_n)],
\end{equation}
where $f(\mathbf{x}_n;\bm{\mu}_g,\bm{\Lambda}_g,\bm{\Psi}_g,\mathbf{u}_n) \sim \mathcal{N}_p(\bm{\mu}_{g} + \bm{\Lambda}_{g} \mathbf{u}_{n}, \bm{\Psi}_g)$ and $f(\mathbf{u}_n) \sim \mathcal{N}_q(0,I)$.
\newline
The complete data log-likelihood can be rewritten as
\begin{equation}\nonumber
\begin{split}
    \ell_{c}(\bm{\Theta};\mathbf{X},\mathbf{Z},\mathbf{U}) = C + \sum_{g=1}^{G}\Bigg[n_g \log \tau_{g} - \frac{n_g}{2}\log (|\bm{\Psi}_g|)
    -\frac{1}{2}\:\Bigg[\sum_{n=1}^{N} z_{ng}[(\mathbf{x}_n-\bm{\mu}_{g})^{\top}\bm{\Psi}_{g}^{-1}(\mathbf{x}_n-\bm{\mu}_{g})]\Bigg]\\ +  \sum_{n=1}^{N} z_{ng}[(\mathbf{x}_n-\bm{\mu}_{g})^{\top}\bm{\Psi}_{g}^{-1}(\bm{\Lambda}_{g}\mathbf{u}_{n})]  -\frac{1}{2} \mathrm{tr}\{\bm{\Lambda}_{g}^{\top}\bm{\Psi}_{g}^{-1}\bm{\Lambda}_{g}\sum_{n=1}^{N}z_{ng}\mathbf{u}_{n}\mathbf{u}_{n}^{\top}\}\Bigg],    
\end{split}
\end{equation}
where $C$ is the constant and $n_g = \sum_{n=1}^{N} z_{ng}$.

\subsubsection{E-step:}
The expected complete data log-likelihood is
\begin{equation}
    \begin{split}
        Q_{2}(\bm{\Theta}) = C + \sum_{g=1}^{G}\Bigg[n_g \log \tau_{g} - \frac{n_g}{2}\log (|\bm{\Psi}_g|) - \frac{n_g}{2} \mathrm{tr}\{\bm{\Psi}_g^{-1} \mathbf{S}_g\}\\
 +  \sum_{n=1}^{N} E[z_{ng};\mathbf{x}_n][(\mathbf{x}_n-\bm{\mu}_{g})^{\top}\bm{\Psi}_{g}^{-1}(\bm{\Lambda}_{g}E[\mathbf{u}_{n};\mathbf{x}_n])] \\ -\frac{1}{2} \mathrm{tr}\{\bm{\Lambda}_{g}^{\top}\bm{\Psi}_{g}^{-1}\bm{\Lambda}_{g}\sum_{n=1}^{N}E[z_{ng};\mathbf{x}_n]E[\mathbf{u}_{n}\mathbf{u}_{n}^{\top};\mathbf{x}_n]\}\Bigg], 
    \end{split}
\end{equation}
where $\mathbf{S}_{g} = \frac{1}{n_g}\sum_{n=1}^{N}{z}_{ng}(\mathbf{x}_{n}-{\bm{\mu}}_{g})(\mathbf{x}_{n}-{\bm{\mu}}_{g})^{\top}$.

The expected value $E[z_{ng};\mathbf{x}_n] = \hat{z}_{ng}$ is computed with $\bm{\mu}_g = \hat{\bm{\mu}}_g$ and $\tau_g = \hat{\tau}_g$ updated in the CM-step of the first cycle with the known positive and negative constraints. The joint posterior probability $\hat{{z}}_{{B_{1}},g}^{+ \atop -}$ of the pixels in $B_1$ belonging to cluster $g$, given the positive and negative constraints $+ \atop -$ is
\begin{equation}\label{pos_neg_b1}
\centering
    \indent\indent \hat{{z}}_{{B_{1}},g}^{+ \atop -} = \frac{\prod\limits_{\substack{j \in {B_{1}} \\ j = 1}}^J \hat{\tau}_g \mathcal{N}_p(\mathbf{x}_j;\hat{\bm{\mu}}_g,\hat{\bm{\Lambda}}_g\:\hat{\bm{\Lambda}}_g^{\top}\:+\:\hat{\bm{\Psi}}_g) \sum\limits_{\substack{f=1 \\ f \neq g}}^G\prod\limits_{\substack{o \in {B_{2}} \\ o = 1}}^O \hat{\tau}_f \mathcal{N}_p(\mathbf{x}_o;\hat{\bm{\mu}}_f,\hat{\bm{\Lambda}}_f\:\hat{\bm{\Lambda}}_f^{\top}\:+\:\hat{\bm{\Psi}}_f)}{\displaystyle\sum_{g^{'}=1}^{G} \prod\limits_{\substack{j \in {B_{1}} \\ j = 1}}^J \hat{\tau}_{g^{'}} \mathcal{N}_p(\mathbf{x}_j;\hat{\bm{\mu}}_{g^{'}},\hat{\bm{\Lambda}}_{g^{'}}\:\hat{\bm{\Lambda}}_{g^{'}}^{\top}\:+\:\hat{\bm{\Psi}}_{g^{'}})\sum\limits_{\substack{f=1 \\ f \neq g^{'}}}^G\prod\limits_{\substack{o \in {B_{2}} \\ o = 1}}^O \hat{\tau}_f \mathcal{N}_p(\mathbf{x}_o;\hat{\bm{\mu}}_f,\hat{\bm{\Lambda}}_f\:\hat{\bm{\Lambda}}_f^{\top}\:+\:\hat{\bm{\Psi}}_f)}.
\end{equation}
Similarly, the computation is repeated for estimating the posterior probabilities of pixels in $B_2$ belonging to cluster $g$ with pixels in $B_1$ as negative constraints. The expected value $E[\mathbf{u}_n;\mathbf{x}_n]$ is computed by considering a joint Gaussian distribution between $\mathbf{x}_n$ and $\mathbf{u}_n$. By Gaussian conditioning formulas,
\begin{equation}\nonumber
    E[\mathbf{u}_{n};\mathbf{x}_{n}] = \hat{\bm{\Lambda}}_{g}^{\top}(\hat{\bm{\Lambda}}_g\:\hat{\bm{\Lambda}}_g^{\top}\:+\:\hat{\bm{\Psi}}_g)^{-1}(\mathbf{x}_{n} - \hat{\bm{\mu}}_{g}).
\end{equation}
Also,
\begin{equation}\nonumber   E[\mathbf{u}_{n}\mathbf{u}_{n}^{\top};\mathbf{x}_{n}] = \mathbf{I}_{q} - \hat{\bm{\beta}}_{g}\hat{\bm{\Lambda}}_{g} + \hat{\bm{\beta}}_{g} (\mathbf{x}_{n}-\hat{\bm{\mu}}_{g})(\mathbf{x}_{n}-\hat{\bm{\mu}}_{g})^{\top} \hat{\bm{\beta}}_{g}^{\top}, 
\end{equation}
where $\hat{\bm{\beta}}_g = \hat{\bm{\Lambda}}_{g}^{\top}(\hat{\bm{\Lambda}}_{g}\hat{\bm{\Lambda}}_{g}^{\top}+\hat{\bm{\Psi}}_{g})^{-1}$.

\subsubsection{CM-step:}
The expected complete log-likelihood can be rewritten as
\begin{equation}\label{eqn22}
\begin{split}
    Q_2(\bm{\Theta}) = C + \sum_{g=1}^{G}n_g^{+ \atop -} \Bigg[\frac{1}{2}\log|\bm{\Psi}_{g}^{-1}| - \frac{1}{2} \mathrm{tr}\{\bm{\Psi}_{g}^{-1}\mathbf{S}_{g}\} + \mathrm{tr}\{\bm{\Psi}_{g}^{-1}\bm{\Lambda}_{g}\hat{\bm{\beta}}_{g}\bm{S}_{g}\}\\ -\frac{1}{2}\mathrm{tr}\{\bm{\Lambda}_{g}^{\top}\bm{\Psi}_{g}^{-1}\bm{\Lambda}_{g}\bm{\Phi}_{g}\}\Bigg],    
\end{split}
\end{equation}
where,
\begin{equation}\nonumber
\centering
\begin{split}
        \bm{\Phi}_{g} = \mathbf{I}_{q} - \hat{\bm{\beta}}_{g}{\bm{\Lambda}}_{g} + \hat{\bm{\beta}}_{g}\mathbf{S}_{g}\hat{\bm{\beta}}_{g}^{\top},\\
        \mathbf{S}_{g} = \frac{1}{n_g^{+ \atop -}}\sum_{n=1}^{N}\hat{z}_{ng}^{+ \atop -}(\mathbf{x}_{n}-\hat{\bm{\mu}}_{g})(\mathbf{x}_{n}-\hat{\bm{\mu}}_{g})^{\top},\\
        n_g^{+ \atop -} = \sum_{n=1}^{N}\hat{z}_{ng}^{+ \atop -}.
\end{split}
\end{equation}
The model parameters $\bm{\Theta}_2$ are updated with $\bm{\Theta}_1$ held fixed. The parameter vector $\bm{\Theta}_2$ denotes the parsimonious covariance matrices $\bm{\Sigma}_g = \bm{\Lambda}_g\bm{\Lambda}_g^{\top}+\bm{\Psi}_g$ ($g=1, \ldots, G$).
\newline
Differentiating \ref{eqn22} with respect to $\bm{\Lambda}_{g}$,
\begin{equation}\nonumber
    \hat{\bm{\Lambda}}_{g} = \mathbf{S}_{g}\hat{\bm{\beta}}_{g}^{\top}\bm{\Phi}_{g}^{-1}.
\end{equation}
Differentiating \ref{eqn22} with respect to $\bm{\Psi}_{g}^{-1}$,
\begin{equation}\nonumber
    \hat{\bm{\Psi}}_{g} = diag[\mathbf{S}_{g} - \hat{\bm{\Lambda}}_{g}\hat{\bm{\beta}}_{g}\mathbf{S}_{g}].
\end{equation}
The algorithm iteratively updates the parameters until convergence and the estimated posterior probabilities $\hat{{z}}_{ng}$ at convergence are used to compute the similarity matrix in the consensus approach.

\newpage
\section{Selected constraints}\label{appendix_selectedConstraints}
\begin{figure}[h!]
     \centering
    \begin{tabular}{c}
    \begin{subfigure}[b]{0.32\textwidth}
         \centering       \includegraphics[width=\textwidth,height=4cm]{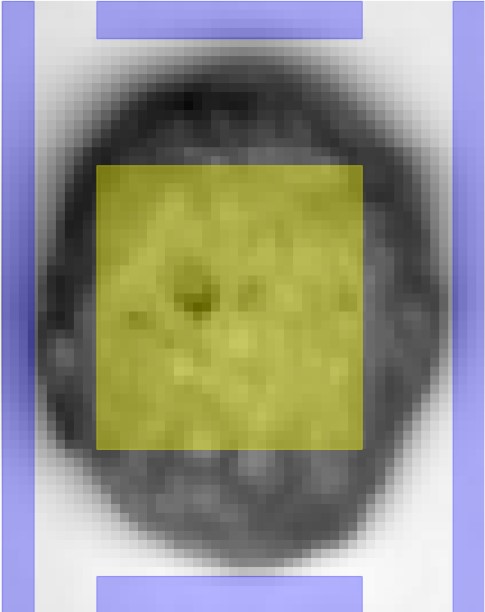}
         \caption{}       \label{fig:Wheat_1_constraints_large}
     \end{subfigure}
     \hfill
     \begin{subfigure}[b]{0.32\textwidth}
         \centering       \includegraphics[width=\textwidth,height=4cm]{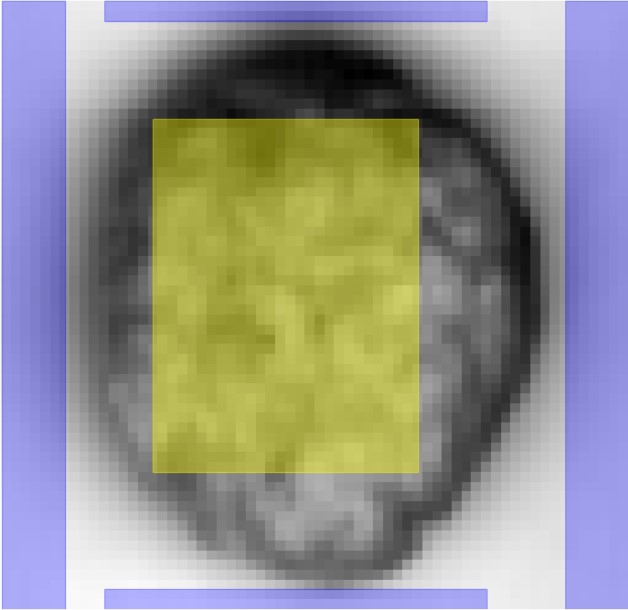}
         \caption{}       \label{fig:Wheat_2_constraints_large}
     \end{subfigure}
     \hfill
     \begin{subfigure}[b]{0.32\textwidth}
         \centering       \includegraphics[width=\textwidth,height=4cm]{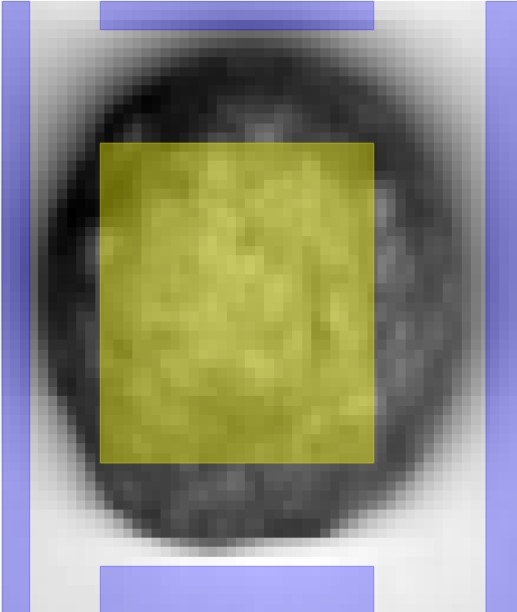}
         \caption{}       \label{fig:Wheat_3_constraints_large}
     \end{subfigure}\\
         \begin{subfigure}[b]{0.32\textwidth}
         \centering       \includegraphics[width=\textwidth,height=4cm]{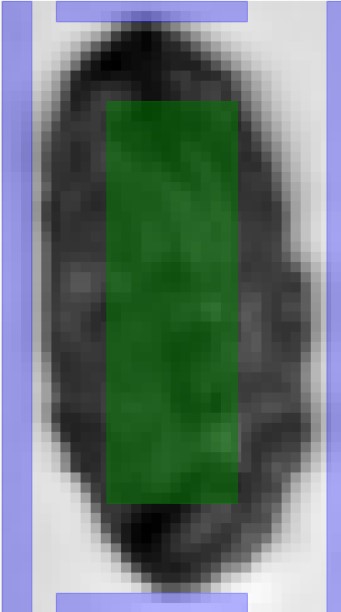}
         \caption{}       \label{fig:Corn_1_constraints_large}
     \end{subfigure}
     \hfill
     \begin{subfigure}[b]{0.32\textwidth}
         \centering       \includegraphics[width=\textwidth,height=4cm]{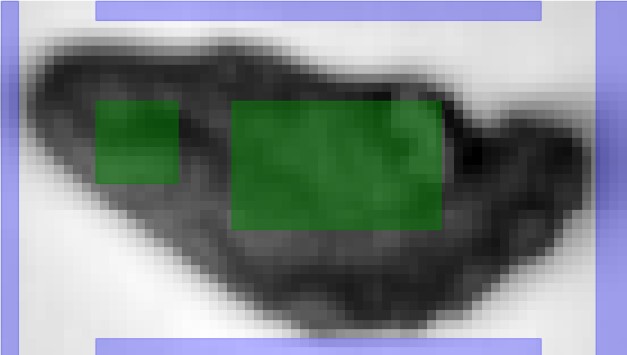}
         \caption{}       \label{fig:Corn_2_constraints_large}
     \end{subfigure}
     \hfill
     \begin{subfigure}[b]{0.32\textwidth}
         \centering       \includegraphics[width=\textwidth,height=4cm]{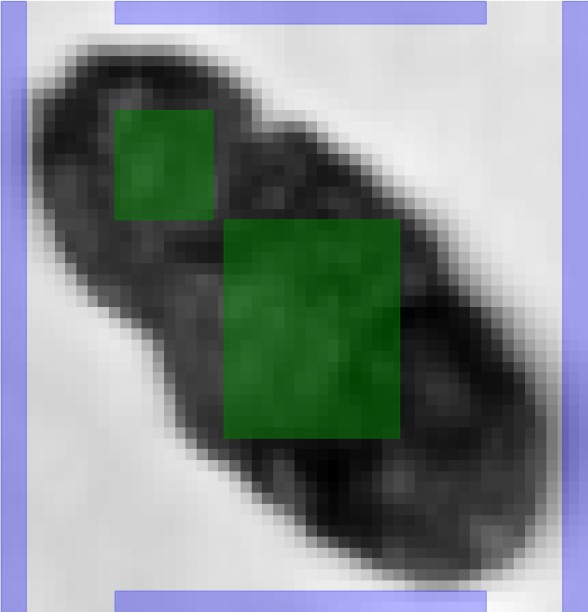}
         \caption{}       \label{fig:Corn_3_constraints_large}
     \end{subfigure}\\
         \begin{subfigure}[b]{0.32\textwidth}
         \centering       \includegraphics[width=\textwidth,height=4cm]{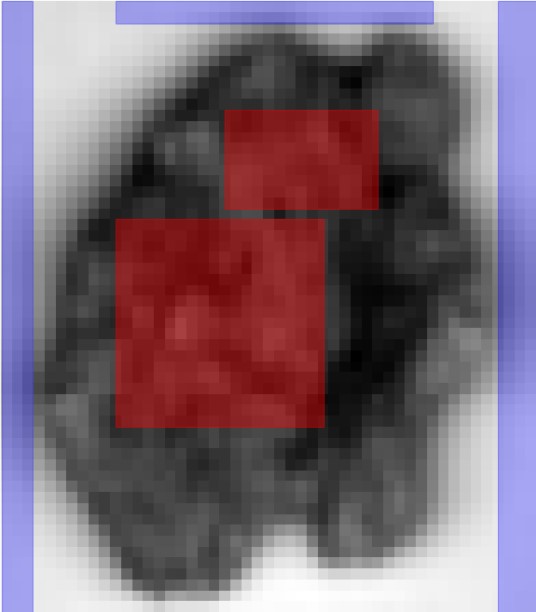}
         \caption{}       \label{fig:Rice_1_constraints_large}
     \end{subfigure}
     \hfill
     \begin{subfigure}[b]{0.32\textwidth}
         \centering       \includegraphics[width=\textwidth,height=4cm]{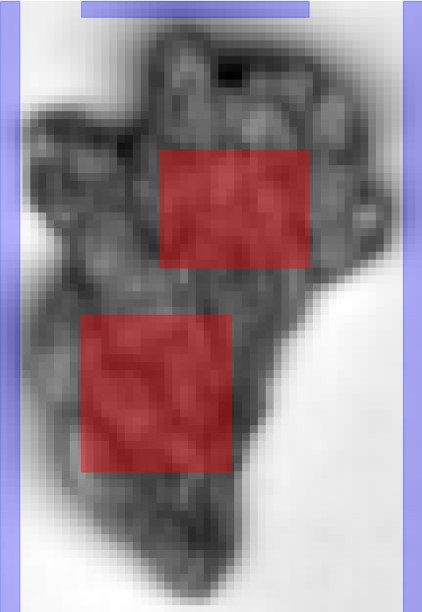}
         \caption{}       \label{fig:Rice_2_constraints_large}
     \end{subfigure}
     \hfill
     \begin{subfigure}[b]{0.32\textwidth}
         \centering       \includegraphics[width=\textwidth,height=4cm]{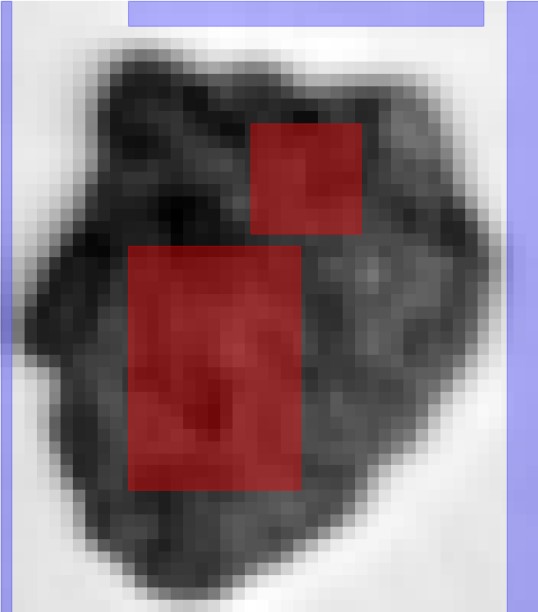}
         \caption{}       \label{fig:Rice_3_constraints_large}
     \end{subfigure}\\
     \end{tabular}
     \caption{Blocks of pixels ($12,215$ pixels of $N = 28,039$ pixels ($43.5\%$)) selected as constraints in three greyscale images for each of the three types of synthetic puffed cereal, wheat (\ref{fig:Wheat_1_constraints_large}, \ref{fig:Wheat_2_constraints_large}, \ref{fig:Wheat_3_constraints_large}), corn (\ref{fig:Corn_1_constraints_large}, \ref{fig:Corn_2_constraints_large}, \ref{fig:Corn_3_constraints_large}), and rice (\ref{fig:Rice_1_constraints_large}, \ref{fig:Rice_2_constraints_large}, \ref{fig:Rice_3_constraints_large}). Of the $12,215$ pixels selected as constraints, $5900$ are in the blue blocks, $3130$ are in the yellow blocks, $1535$ are in the green blocks and $1650$ are in the red blocks.} \label{fig:synthetic_cereal_constraints_large}
\end{figure}

\begin{figure}[h!]
     \centering
    \begin{tabular}{c}
    \begin{subfigure}[b]{0.32\textwidth}
         \centering       \includegraphics[width=\textwidth,height=4cm]{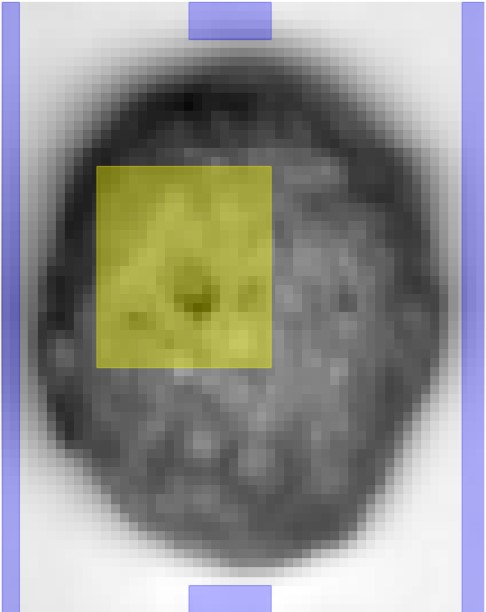}
         \caption{}       \label{fig:Wheat_1_constraints_small}
     \end{subfigure}
     \hfill
     \begin{subfigure}[b]{0.32\textwidth}
         \centering       \includegraphics[width=\textwidth,height=4cm]{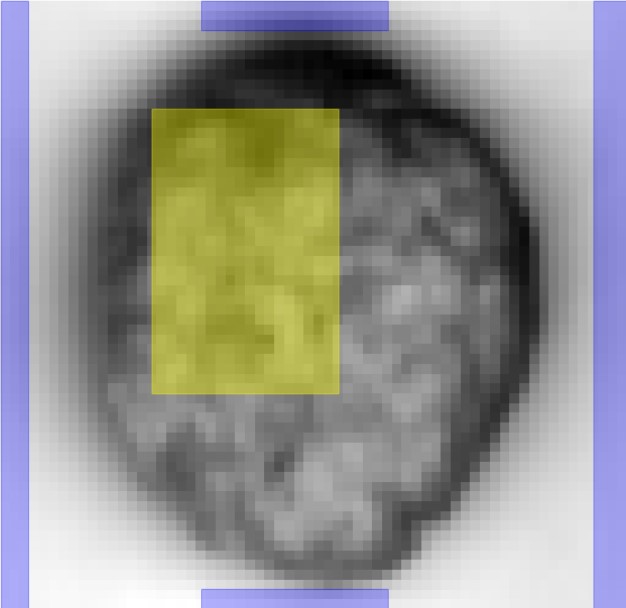}
         \caption{}       \label{fig:Wheat_2_constraints_small}
     \end{subfigure}
     \hfill
     \begin{subfigure}[b]{0.32\textwidth}
         \centering       \includegraphics[width=\textwidth,height=4cm]{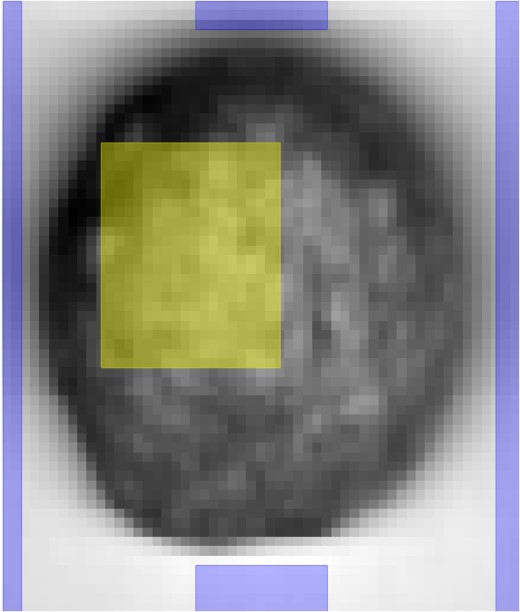}
         \caption{}       \label{fig:Wheat_3_constraints_small}
     \end{subfigure}\\
         \begin{subfigure}[b]{0.32\textwidth}
         \centering       \includegraphics[width=\textwidth,height=4cm]{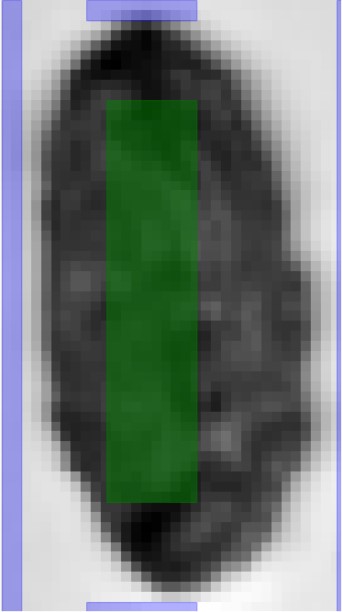}
         \caption{}       \label{fig:Corn_1_constraints_small}
     \end{subfigure}
     \hfill
     \begin{subfigure}[b]{0.32\textwidth}
         \centering       \includegraphics[width=\textwidth,height=4cm]{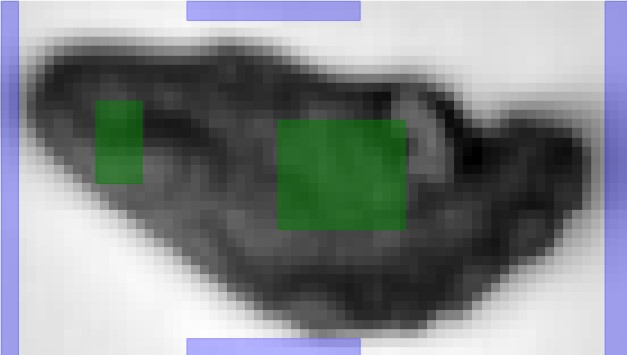}
         \caption{}       \label{fig:Corn_2_constraints_small}
     \end{subfigure}
     \hfill
     \begin{subfigure}[b]{0.32\textwidth}
         \centering       \includegraphics[width=\textwidth,height=4cm]{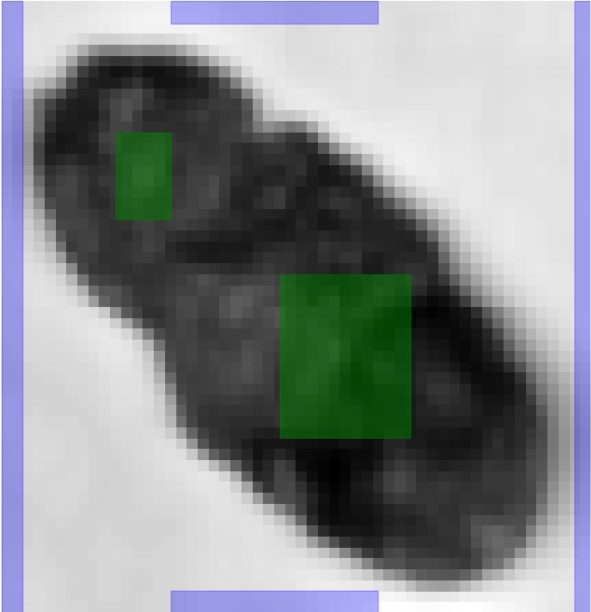}
         \caption{}       \label{fig:Corn_3_constraints_small}
     \end{subfigure}\\
         \begin{subfigure}[b]{0.32\textwidth}
         \centering       \includegraphics[width=\textwidth,height=4cm]{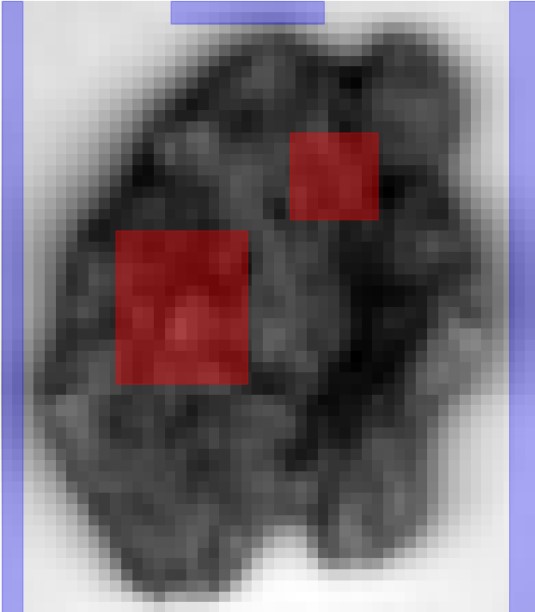}
         \caption{}       \label{fig:Rice_1_constraints_small}
     \end{subfigure}
     \hfill
     \begin{subfigure}[b]{0.32\textwidth}
         \centering       \includegraphics[width=\textwidth,height=4cm]{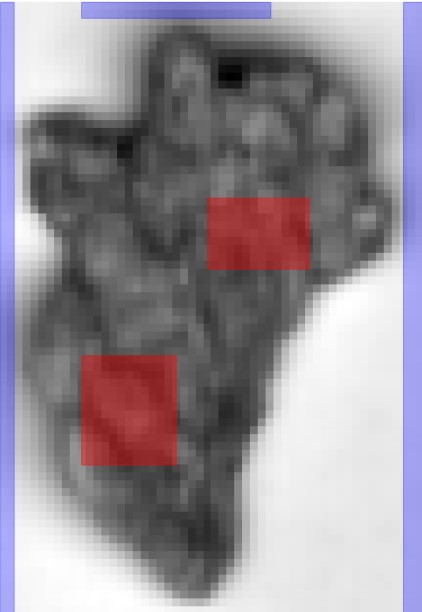}
         \caption{}       \label{fig:Rice_2_constraints_small}
     \end{subfigure}
     \hfill
     \begin{subfigure}[b]{0.32\textwidth}
         \centering       \includegraphics[width=\textwidth,height=4cm]{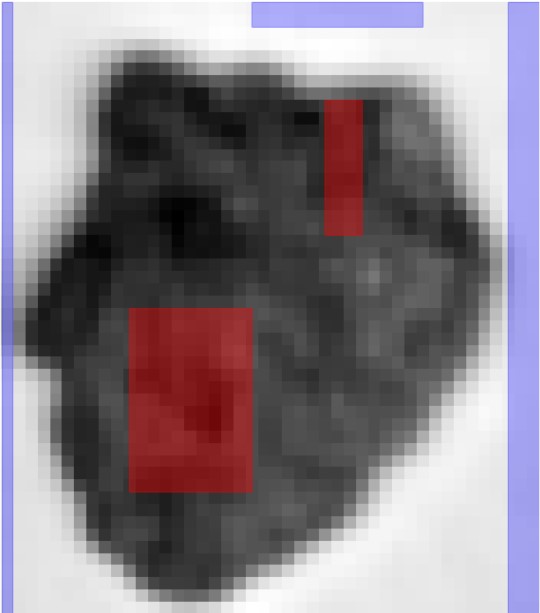}
         \caption{}       \label{fig:Rice_3_constraints_small}
     \end{subfigure}\\
     \end{tabular}
     \caption{Blocks of pixels ($6,951$ pixels of $N = 28,039$ pixels ($24.7\%$)) selected as constraints in three greyscale images for each of the three types of synthetic puffed cereal, wheat (\ref{fig:Wheat_1_constraints_small}, \ref{fig:Wheat_2_constraints_small}, \ref{fig:Wheat_3_constraints_small}), corn (\ref{fig:Corn_1_constraints_small}, \ref{fig:Corn_2_constraints_small}, \ref{fig:Corn_3_constraints_small}), and rice (\ref{fig:Rice_1_constraints_small}, \ref{fig:Rice_2_constraints_small}, \ref{fig:Rice_3_constraints_small}). Of the $6,951$ pixels selected as constraints, $3801$ are in the blue blocks, $1540$ are in the yellow blocks, $783$ are in the green blocks and $827$ are in the red blocks.}     \label{fig:synthetic_cereal_constraints_small}
\end{figure}

As discussed in Section \ref{simstudy}, Figures \ref{fig:synthetic_cereal_constraints_large} and \ref{fig:synthetic_cereal_constraints_small} highlight the $43.5\%$ of pixels and $24.7\%$ of pixels selected as constraints respectively.

\newpage

\newpage
\section{Choice of number of principal components} \label{appendix_pcatable}
As discussed in Section \ref{synthetic_cereal}, Table \ref{pca_table} details the cumulative proportion of the variance explained by the first $5$ principal components of the motivating hyperspectral image dataset.

\begin{table}[h]
    \centering        \caption{Cumulative proportion of variance explained by the principal components of the motivating hyperspectral image dataset on puffed cereals.}
    \begin{tabular}{ 
    |P{6cm}||P{1cm}|P{1cm}|P{1cm}|P{1cm}|P{1cm}|}
     \hline
     \textbf{Number of principal components} & 1 & 2 & 3 & 4 & 5\\
     \hline
     \textbf{Cumulative proportion of variance explained}   & 94.2 & 99.7 & 99.92 & 99.94 & 99.95\\
     \hline        
    \end{tabular}
    \label{pca_table}
\end{table}
 
\newpage
\section{Additional hyperspectral imaging clustering results}\label{appendix_additionalclusterresults}
\begin{figure}[h!]
     \centering
    \begin{tabular}{c}
    \begin{subfigure}[b]{0.32\textwidth}
         \centering
         \includegraphics[width=\textwidth,height=2.1cm]{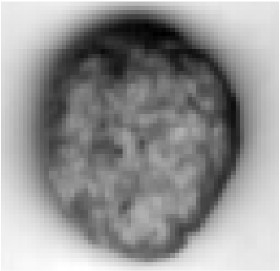}
         \caption{}
         \label{fig:wheat_2_grey}
     \end{subfigure}
     \hfill
     \begin{subfigure}[b]{0.32\textwidth}
         \centering
         \includegraphics[width=\textwidth,height=2.1cm]{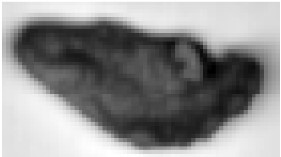}
         \caption{}
         \label{fig:corn_2_grey}
     \end{subfigure}
     \hfill
     \begin{subfigure}[b]{0.32\textwidth}
         \centering
         \includegraphics[width=\textwidth,height=2.1cm]{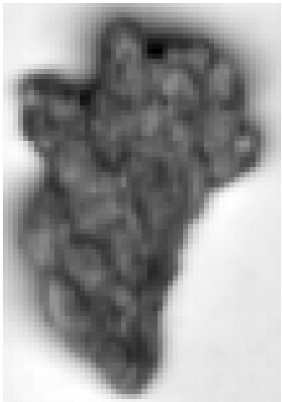}
         \caption{}
         \label{fig:rice_2_grey}
     \end{subfigure}\\
    \begin{subfigure}[b]{0.32\textwidth}
         \centering
         \includegraphics[width=\textwidth,height=2.1cm]{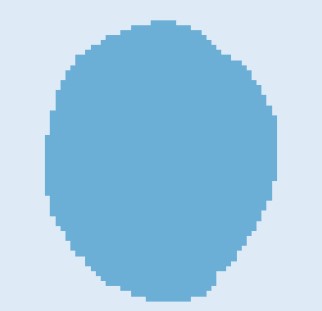}
         \caption{}
         \label{fig:wheat_2_thresh}
     \end{subfigure}
     \hfill
     \begin{subfigure}[b]{0.32\textwidth}
         \centering
         \includegraphics[width=\textwidth,height=2.1cm]{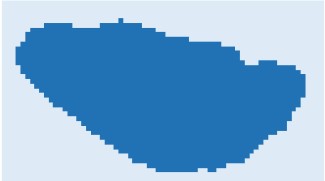}
         \caption{}
         \label{fig:corn_2_thresh}
     \end{subfigure}
     \hfill
     \begin{subfigure}[b]{0.32\textwidth}
         \centering
         \includegraphics[width=\textwidth,height=2.1cm]{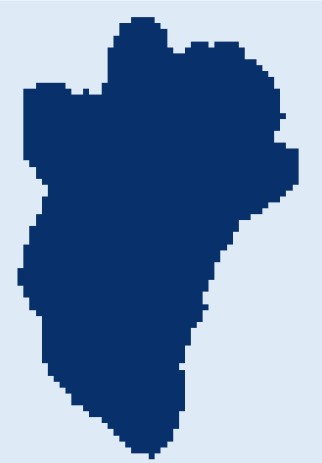}
         \caption{}
         \label{fig:rice_2_thresh}
     \end{subfigure}\\
    \begin{subfigure}[b]{0.32\textwidth}
         \centering
         \includegraphics[width=\textwidth,height=2.1cm]{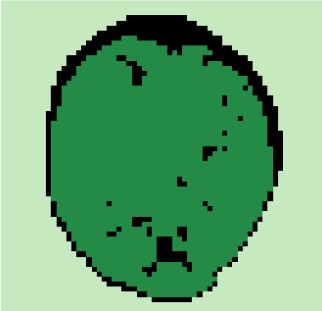}
         \caption{}
         \label{fig:wheat_2_dbscan}
     \end{subfigure}
     \hfill
     \begin{subfigure}[b]{0.32\textwidth}
         \centering
         \includegraphics[width=\textwidth,height=2.1cm]{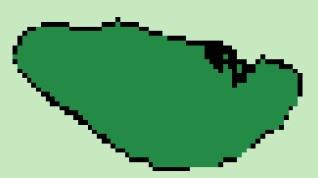}
         \caption{}
         \label{fig:corn_2_dbscan}
     \end{subfigure}
     \hfill
     \begin{subfigure}[b]{0.32\textwidth}
         \centering
         \includegraphics[width=\textwidth,height=2.1cm]{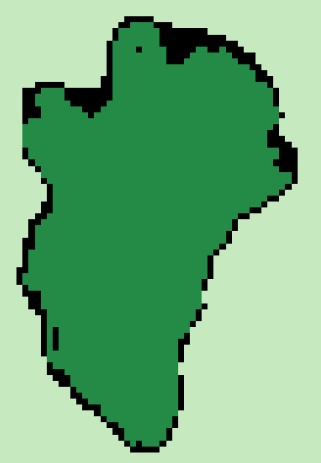}
         \caption{}
         \label{fig:rice_2_dbscan}
     \end{subfigure}\\
    \begin{subfigure}[b]{0.32\textwidth}
         \centering
         \includegraphics[width=\textwidth,height=2.1cm]{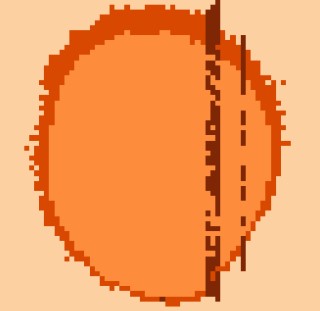}
         \caption{}
         \label{fig:wheat_2_gmm}
     \end{subfigure}
     \hfill
     \begin{subfigure}[b]{0.32\textwidth}
         \centering
         \includegraphics[width=\textwidth,height=2.1cm]{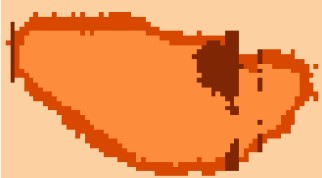}
         \caption{}
         \label{fig:corn_2_gmm}
     \end{subfigure}
     \hfill
     \begin{subfigure}[b]{0.32\textwidth}
         \centering
         \includegraphics[width=\textwidth,height=2.1cm]{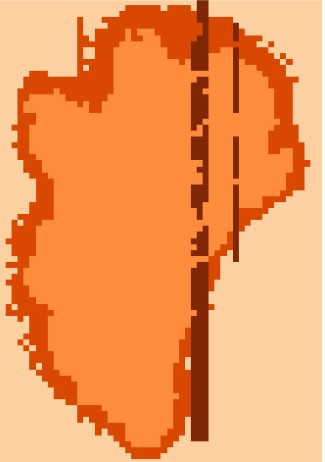}
         \caption{}
         \label{fig:rice_2_gmm}
     \end{subfigure}\\
    \begin{subfigure}[b]{0.32\textwidth}
         \centering
         \includegraphics[width=\textwidth,height=2.1cm]{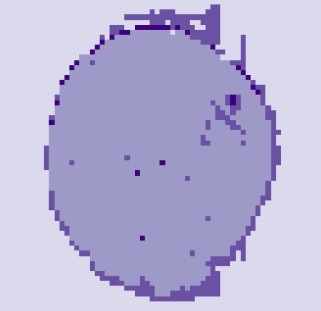}
         \caption{}
         \label{fig:wheat_2_pgmm}
     \end{subfigure}
     \hfill
     \begin{subfigure}[b]{0.32\textwidth}
         \centering
         \includegraphics[width=\textwidth,height=2.1cm]{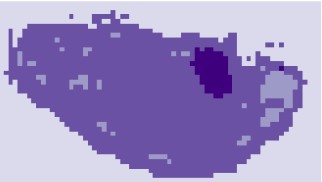}
         \caption{}
         \label{fig:corn_2_pgmm}
     \end{subfigure}
     \hfill
     \begin{subfigure}[b]{0.32\textwidth}
         \centering
         \includegraphics[width=\textwidth,height=2.1cm]{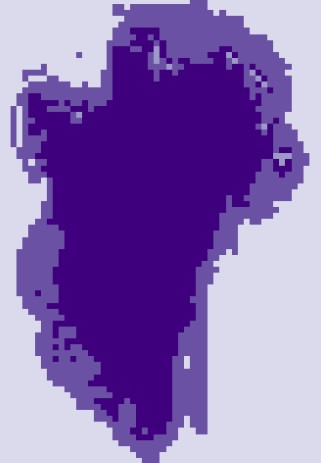}
         \caption{}
         \label{fig:rice_2_pgmm}
     \end{subfigure}\\
    \begin{subfigure}[b]{0.32\textwidth}
         \centering
         \includegraphics[width=\textwidth,height=2.1cm]{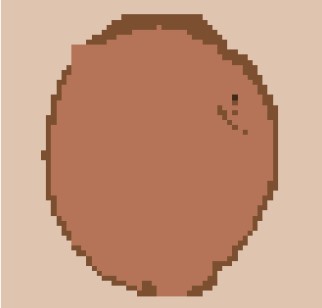}
         \caption{}
         \label{fig:wheat_2_ccpgmm}
     \end{subfigure}
     \hfill
     \begin{subfigure}[b]{0.32\textwidth}
         \centering
         \includegraphics[width=\textwidth,height=2.1cm]{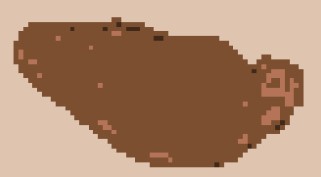}
         \caption{}
         \label{fig:corn_2_ccpgmm}
     \end{subfigure}
     \hfill
     \begin{subfigure}[b]{0.32\textwidth}
         \centering
         \includegraphics[width=\textwidth,height=2.1cm]{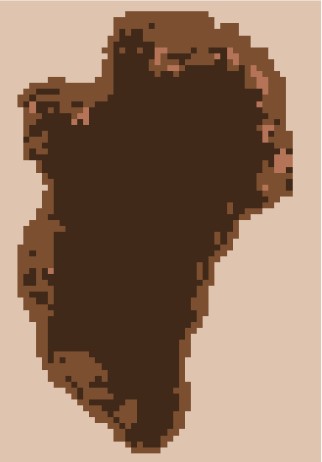}
         \caption{}
         \label{fig:rice_2_ccpgmm}
     \end{subfigure}\\
     \begin{subfigure}[b]{0.32\textwidth}
         \centering
         \includegraphics[width=\textwidth,height=2.1cm]{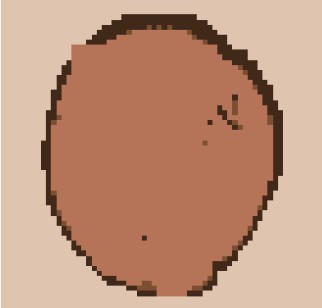}
         \caption{}
         \label{fig:wheat_2_ccpgmm_small}
     \end{subfigure}
     \hfill
     \begin{subfigure}[b]{0.32\textwidth}
         \centering
         \includegraphics[width=\textwidth,height=2.1cm]{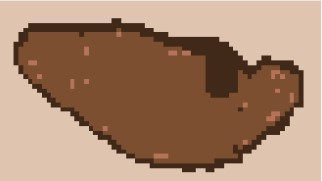}
         \caption{}
         \label{fig:corn_2_ccpgmm_small}
     \end{subfigure}
     \hfill
     \begin{subfigure}[b]{0.32\textwidth}
         \centering
         \includegraphics[width=\textwidth,height=2.1cm]{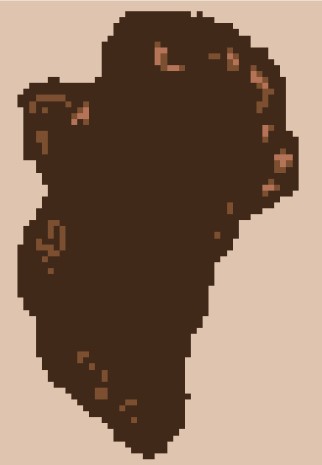}
         \caption{}
         \label{fig:rice_2_ccpgmm_small}
     \end{subfigure}\\
     \end{tabular}
        \caption{Greyscale images and pixel labels for three of the nine hyperspectral images ($l = \{2,5,8\}$) of the cereals using the threshold approach, DBSCAN, GMM, PGMM, ccPGMM with $43.5\%$ pixels as constraints, and ccPGMM with $24.7\%$pixels as constraints respectively for wheat (\ref{fig:wheat_2_grey}, \ref{fig:wheat_2_thresh}, \ref{fig:wheat_2_dbscan}, \ref{fig:wheat_2_gmm}, \ref{fig:wheat_2_pgmm}, \ref{fig:wheat_2_ccpgmm},\ref{fig:wheat_2_ccpgmm_small}), corn (\ref{fig:corn_2_grey}, \ref{fig:corn_2_thresh}, \ref{fig:corn_2_dbscan}, \ref{fig:corn_2_gmm}, \ref{fig:corn_2_pgmm}, \ref{fig:corn_2_ccpgmm}, \ref{fig:corn_2_ccpgmm_small}) and rice (\ref{fig:rice_2_grey}, \ref{fig:rice_2_thresh}, \ref{fig:rice_2_dbscan}, \ref{fig:rice_2_gmm}, \ref{fig:rice_2_pgmm}, \ref{fig:rice_2_ccpgmm}, \ref{fig:rice_2_ccpgmm_small}).}
        \label{fig:results_cereal_2}
\end{figure}

\begin{figure}[h!]
     \centering
    \begin{tabular}{c}
    \begin{subfigure}[b]{0.32\textwidth}
         \centering
         \includegraphics[width=\textwidth,height=2.2cm]{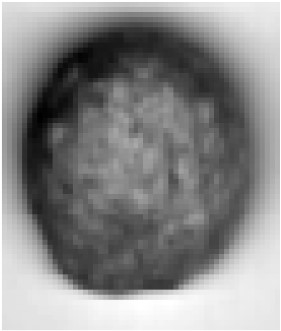}
         \caption{}
         \label{fig:wheat_3_grey}
     \end{subfigure}
     \hfill
     \begin{subfigure}[b]{0.32\textwidth}
         \centering
         \includegraphics[width=\textwidth,height=2.2cm]{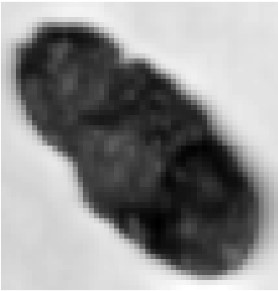}
         \caption{}
         \label{fig:corn_3_grey}
     \end{subfigure}
     \hfill
     \begin{subfigure}[b]{0.32\textwidth}
         \centering
         \includegraphics[width=\textwidth,height=2.2cm]{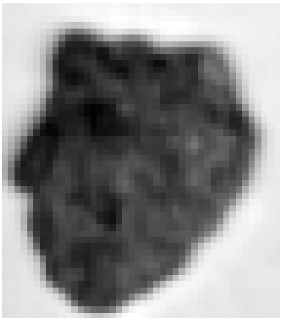}
         \caption{}
         \label{fig:rice_3_grey}
     \end{subfigure}\\
    \begin{subfigure}[b]{0.32\textwidth}
         \centering
         \includegraphics[width=\textwidth,height=2.2cm]{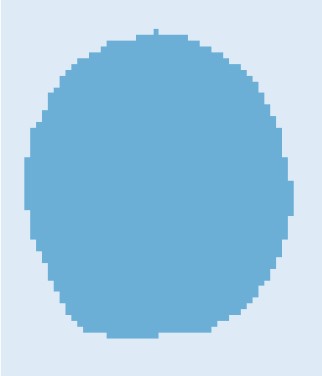}
         \caption{}
         \label{fig:wheat_3_thresh}
     \end{subfigure}
     \hfill
     \begin{subfigure}[b]{0.32\textwidth}
         \centering
         \includegraphics[width=\textwidth,height=2.2cm]{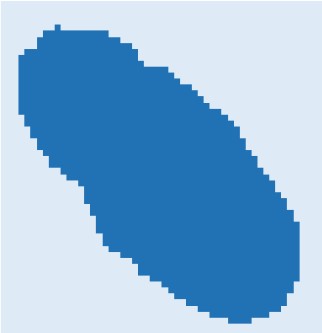}
         \caption{}
         \label{fig:corn_3_thresh}
     \end{subfigure}
     \hfill
     \begin{subfigure}[b]{0.32\textwidth}
         \centering
         \includegraphics[width=\textwidth,height=2.2cm]{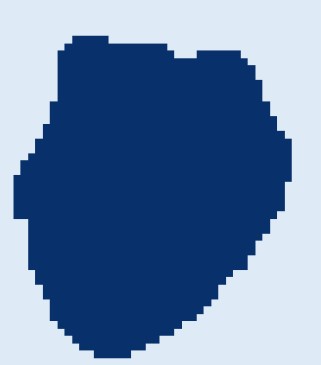}
         \caption{}
         \label{fig:rice_3_thresh}
     \end{subfigure}\\
    \begin{subfigure}[b]{0.32\textwidth}
         \centering
         \includegraphics[width=\textwidth,height=2.2cm]{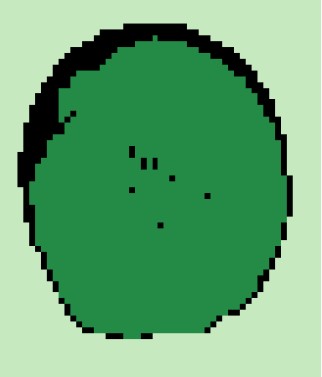}
         \caption{}
         \label{fig:wheat_3_dbscan}
     \end{subfigure}
     \hfill
     \begin{subfigure}[b]{0.32\textwidth}
         \centering
         \includegraphics[width=\textwidth,height=2.2cm]{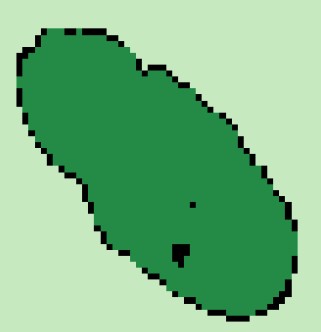}
         \caption{}
         \label{fig:corn_3_dbscan}
     \end{subfigure}
     \hfill
     \begin{subfigure}[b]{0.32\textwidth}
         \centering
         \includegraphics[width=\textwidth,height=2.2cm]{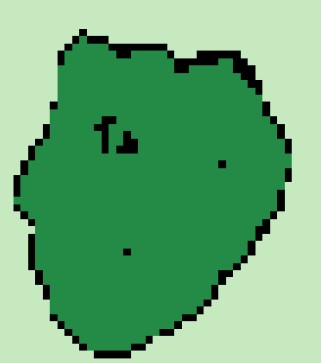}
         \caption{}
         \label{fig:rice_3_dbscan}
     \end{subfigure}\\
    \begin{subfigure}[b]{0.32\textwidth}
         \centering
         \includegraphics[width=\textwidth,height=2.2cm]{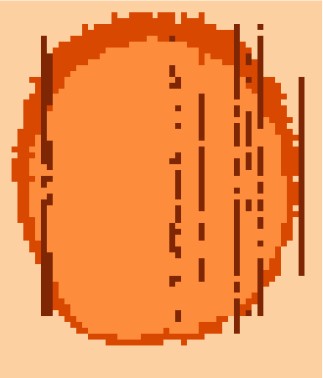}
         \caption{}
         \label{fig:wheat_3_gmm}
     \end{subfigure}
     \hfill
     \begin{subfigure}[b]{0.32\textwidth}
         \centering
         \includegraphics[width=\textwidth,height=2.2cm]{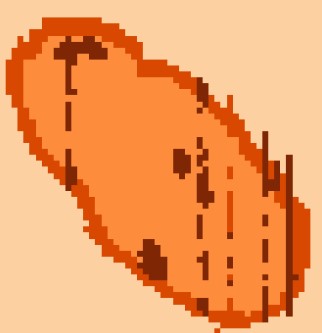}
         \caption{}
         \label{fig:corn_3_gmm}
     \end{subfigure}
     \hfill
     \begin{subfigure}[b]{0.32\textwidth}
         \centering
         \includegraphics[width=\textwidth,height=2.2cm]{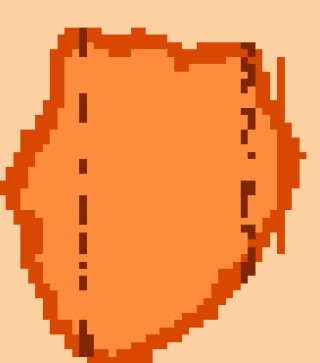}
         \caption{}
         \label{fig:rice_3_gmm}
     \end{subfigure}\\
    \begin{subfigure}[b]{0.32\textwidth}
         \centering
         \includegraphics[width=\textwidth,height=2.2cm]{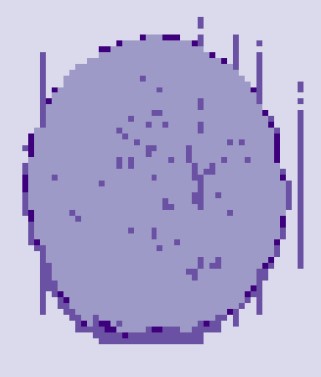}
         \caption{}
         \label{fig:wheat_3_pgmm}
     \end{subfigure}
     \hfill
     \begin{subfigure}[b]{0.32\textwidth}
         \centering
         \includegraphics[width=\textwidth,height=2.2cm]{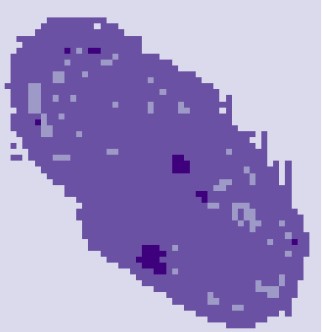}
         \caption{}
         \label{fig:corn_3_pgmm}
     \end{subfigure}
     \hfill
     \begin{subfigure}[b]{0.32\textwidth}
         \centering
         \includegraphics[width=\textwidth,height=2.2cm]{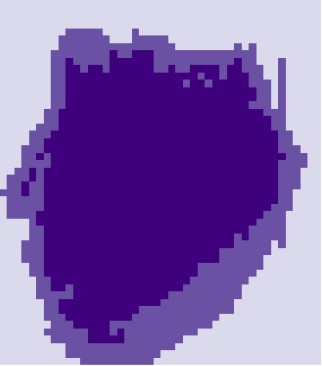}
         \caption{}
         \label{fig:rice_3_pgmm}
     \end{subfigure}\\
    \begin{subfigure}[b]{0.32\textwidth}
         \centering
         \includegraphics[width=\textwidth,height=2.2cm]{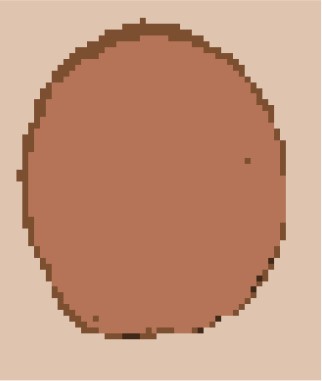}
         \caption{}
         \label{fig:wheat_3_ccpgmm}
     \end{subfigure}
     \hfill
     \begin{subfigure}[b]{0.32\textwidth}
         \centering
         \includegraphics[width=\textwidth,height=2.2cm]{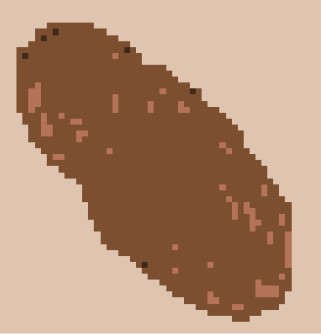}
         \caption{}
         \label{fig:corn_3_ccpgmm}
     \end{subfigure}
     \hfill
     \begin{subfigure}[b]{0.32\textwidth}
         \centering
         \includegraphics[width=\textwidth,height=2.2cm]{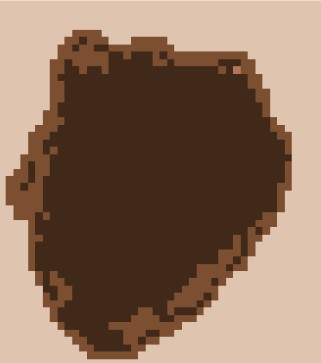}
         \caption{}
         \label{fig:rice_3_ccpgmm}
     \end{subfigure}\\
     \begin{subfigure}[b]{0.32\textwidth}
         \centering
         \includegraphics[width=\textwidth,height=2.2cm]{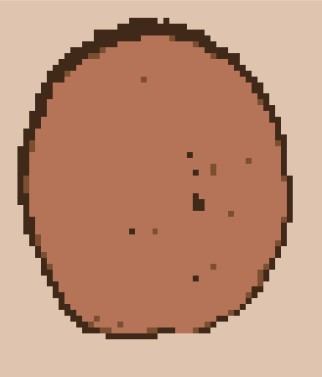}
         \caption{}
         \label{fig:wheat_3_ccpgmm_small}
     \end{subfigure}
     \hfill
     \begin{subfigure}[b]{0.32\textwidth}
         \centering
         \includegraphics[width=\textwidth,height=2.2cm]{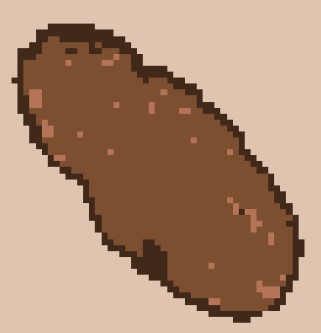}
         \caption{}
         \label{fig:corn_3_ccpgmm_small}
     \end{subfigure}
     \hfill
     \begin{subfigure}[b]{0.32\textwidth}
         \centering
         \includegraphics[width=\textwidth,height=2.2cm]{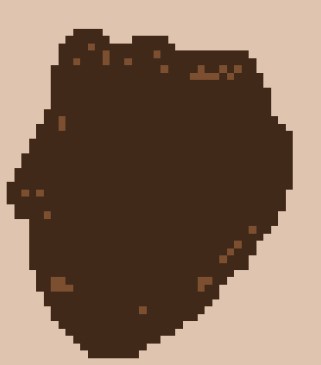}
         \caption{}
         \label{fig:rice_3_ccpgmm_small}
     \end{subfigure}\\
     \end{tabular}
        \caption{Greyscale images and pixel labels for three of the nine hyperspectral images ($l = \{3,6,9\}$) of the cereals using the threshold approach, DBSCAN, GMM, PGMM, ccPGMM with $43.5\%$ pixels as constraints, and ccPGMM with $24.7\%$ pixels as constraints respectively for wheat (\ref{fig:wheat_3_grey}, \ref{fig:wheat_3_thresh}, \ref{fig:wheat_3_dbscan}, \ref{fig:wheat_3_gmm}, \ref{fig:wheat_3_pgmm}, \ref{fig:wheat_3_ccpgmm}, \ref{fig:wheat_3_ccpgmm_small}), corn (\ref{fig:corn_3_grey}, \ref{fig:corn_3_thresh}, \ref{fig:corn_3_dbscan}, \ref{fig:corn_3_gmm}, \ref{fig:corn_3_pgmm}, \ref{fig:corn_3_ccpgmm}, \ref{fig:corn_3_ccpgmm_small}) and rice (\ref{fig:rice_3_grey}, \ref{fig:rice_3_thresh}, \ref{fig:rice_3_dbscan}, \ref{fig:rice_3_gmm}, \ref{fig:rice_3_pgmm}, \ref{fig:rice_3_ccpgmm}, \ref{fig:rice_3_ccpgmm_small}).}
        \label{fig:results_cereal_3}
\end{figure}

As discussed in Section \ref{application}, Figures \ref{fig:results_cereal_2}, and \ref{fig:results_cereal_3} 
shows the greyscale images for the remaining six hyperspectral images, and the respective pixel
labels generated using different clustering approaches.

\newpage
\section{Clustering uncertainty for hyperspectral images}\label{appendix_clusteruncert}
\begin{figure}[h!]
     \centering
    \begin{tabular}{c}
    \begin{subfigure}[b]{0.31\textwidth}
         \centering      \includegraphics[width=\textwidth,height=4cm]{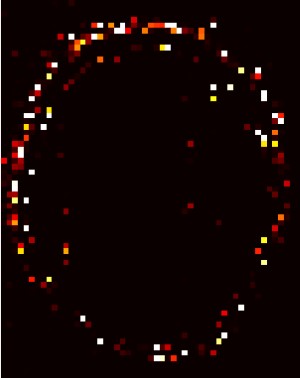}
         \caption{}      \label{fig:wheat_1_gmm_uncertainty}
     \end{subfigure}
     \hfill
     \begin{subfigure}[b]{0.31\textwidth}
         \centering      \includegraphics[width=\textwidth,height=4cm]{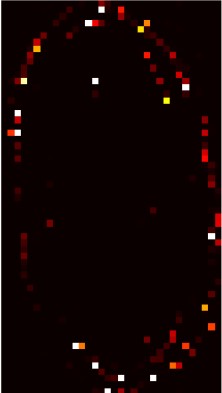}
         \caption{}      \label{fig:corn_1_gmm_uncertainty}
     \end{subfigure}
     \hfill
     \begin{subfigure}[b]{0.31\textwidth}
         \centering      \includegraphics[width=\textwidth,height=4cm]{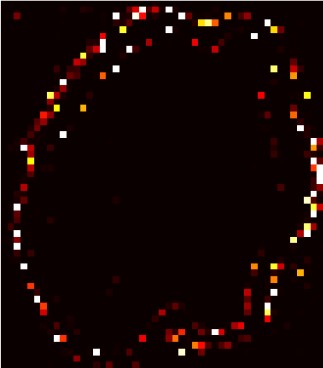}
         \caption{}      \label{fig:rice_1_gmm_uncertainty}
     \end{subfigure}
     \hfill
     \begin{subfigure}[b]{0.07\textwidth}
         \centering      \includegraphics[width=\textwidth,height=4cm]{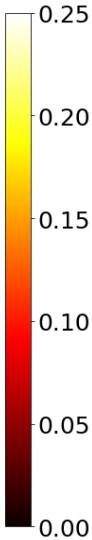}         \captionsetup{labelformat=empty}
         \caption{}
     \end{subfigure}\\
    \begin{subfigure}[b]{0.31\textwidth}
         \centering      \includegraphics[width=\textwidth,height=4cm]{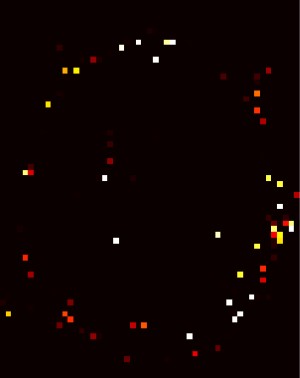}
         \caption{}      \label{fig:wheat_1_pgmm_uncertainty}
     \end{subfigure}
     \hfill
     \begin{subfigure}[b]{0.31\textwidth}
         \centering      \includegraphics[width=\textwidth,height=4cm]{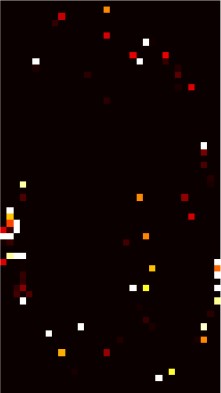}
         \caption{}      \label{fig:corn_1_pgmm_uncertainty}
     \end{subfigure}
     \hfill
     \begin{subfigure}[b]{0.31\textwidth}
         \centering      \includegraphics[width=\textwidth,height=4cm]{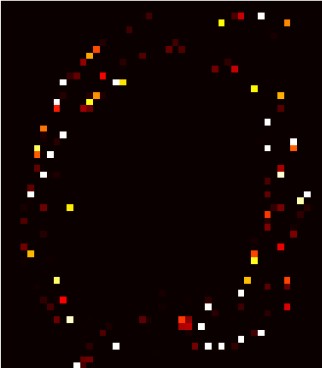}
         \caption{}      \label{fig:rice_1_pgmm_uncertainty}
     \end{subfigure}
     \hfill
     \begin{subfigure}[b]{0.07\textwidth}
         \centering      \includegraphics[width=\textwidth,height=4cm]{Figures/uncertainty_scale.jpg}         \captionsetup{labelformat=empty}
         \caption{}
     \end{subfigure}\\
    \begin{subfigure}[b]{0.31\textwidth}
         \centering      \includegraphics[width=\textwidth,height=4cm]{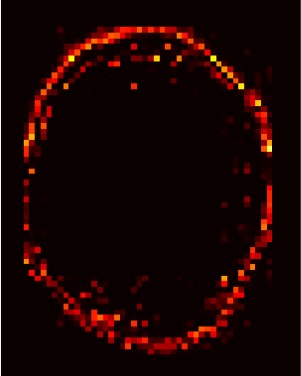}
         \caption{}      \label{fig:wheat_1_ccpgmm_large_uncertainty}
     \end{subfigure}
     \hfill
     \begin{subfigure}[b]{0.31\textwidth}
         \centering      \includegraphics[width=\textwidth,height=4cm]{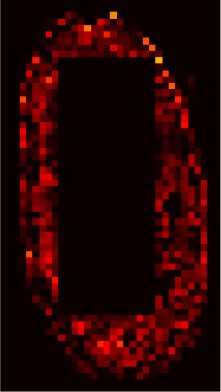}
         \caption{}      \label{fig:corn_1_ccpgmm_large_uncertainty}
     \end{subfigure}
     \hfill
     \begin{subfigure}[b]{0.31\textwidth}
         \centering      \includegraphics[width=\textwidth,height=4cm]{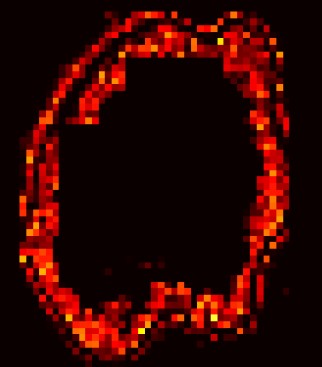}
         \caption{}      \label{fig:rice_1_ccpgmm_large_uncertainty}
     \end{subfigure}
     \hfill
     \begin{subfigure}[b]{0.07\textwidth}
         \centering      \includegraphics[width=\textwidth,height=4cm]{Figures/uncertainty_scale.jpg}         \captionsetup{labelformat=empty}
         \caption{}
     \end{subfigure}\\
    \begin{subfigure}[b]{0.31\textwidth}
         \centering      \includegraphics[width=\textwidth,height=4cm]{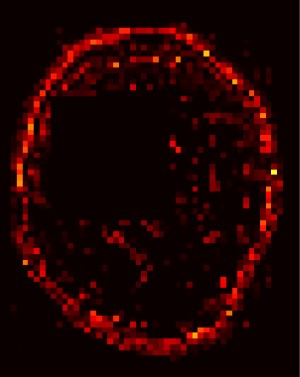}
         \caption{}      \label{fig:wheat_1_ccpgmm_small_uncertainty}
     \end{subfigure}
     \hfill
     \begin{subfigure}[b]{0.31\textwidth}
         \centering      \includegraphics[width=\textwidth,height=4cm]{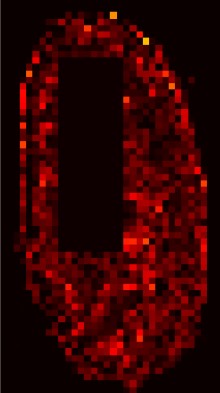}
         \caption{}      \label{fig:corn_1_ccpgmm_small_uncertainty}
     \end{subfigure}
     \hfill
     \begin{subfigure}[b]{0.31\textwidth}
         \centering      \includegraphics[width=\textwidth,height=4cm]{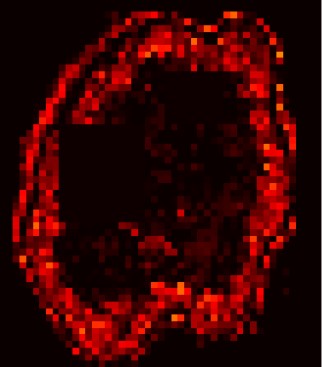}
         \caption{}      \label{fig:rice_1_ccpgmm_small_uncertainty}
     \end{subfigure}
     \hfill
     \begin{subfigure}[b]{0.07\textwidth}
         \centering      \includegraphics[width=\textwidth,height=4cm]{Figures/uncertainty_scale.jpg}         \captionsetup{labelformat=empty}
         \caption{}
     \end{subfigure}\\
     \end{tabular}
        \caption{Uncertainty associated with pixel labels for three of the nine hyperspectral images ($l = \{1,4,7\}$) of the puffed cereals based on the cluster solutions of GMM, PGMM, ccPGMM with $43.5\%$ of pixels as constraints, and ccPGMM with $24.7\%$ of pixels as constraints respectively for wheat (\ref{fig:wheat_1_gmm_uncertainty}, \ref{fig:wheat_1_pgmm_uncertainty}, \ref{fig:wheat_1_ccpgmm_large_uncertainty},\ref{fig:wheat_1_ccpgmm_small_uncertainty}), corn (\ref{fig:corn_1_gmm_uncertainty}, \ref{fig:corn_1_pgmm_uncertainty}, \ref{fig:corn_1_ccpgmm_large_uncertainty}, \ref{fig:corn_1_ccpgmm_small_uncertainty}) and rice (\ref{fig:rice_1_gmm_uncertainty}, \ref{fig:rice_1_pgmm_uncertainty}, \ref{fig:rice_1_ccpgmm_large_uncertainty}, \ref{fig:rice_1_ccpgmm_small_uncertainty}).}
        \label{fig:results_cereal_1_uncertainty}
\end{figure}

\begin{figure}
     \centering
    \begin{tabular}{c}
    \begin{subfigure}[b]{0.31\textwidth}
         \centering      \includegraphics[width=\textwidth,height=4cm]{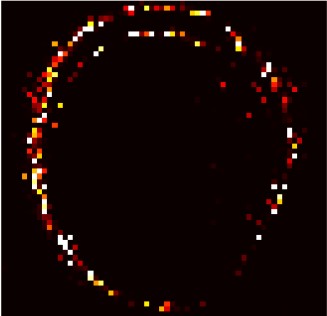}
         \caption{}      \label{fig:wheat_2_gmm_uncertainty}
     \end{subfigure}
     \hfill
     \begin{subfigure}[b]{0.31\textwidth}
         \centering      \includegraphics[width=\textwidth,height=4cm]{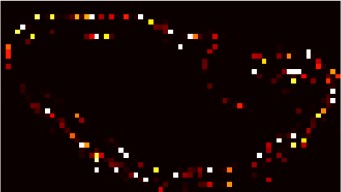}
         \caption{}      \label{fig:corn_2_gmm_uncertainty}
     \end{subfigure}
     \hfill
     \begin{subfigure}[b]{0.31\textwidth}
         \centering      \includegraphics[width=\textwidth,height=4cm]{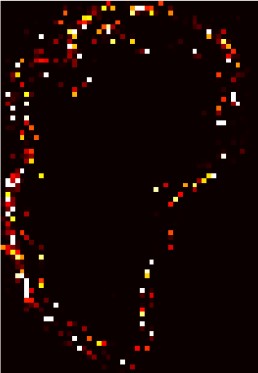}
         \caption{}      \label{fig:rice_2_gmm_uncertainty}
     \end{subfigure}
     \hfill
     \begin{subfigure}[b]{0.07\textwidth}
         \centering      \includegraphics[width=\textwidth,height=4cm]{Figures/uncertainty_scale.jpg}         \captionsetup{labelformat=empty}
         \caption{}
     \end{subfigure}\\
    \begin{subfigure}[b]{0.31\textwidth}
         \centering      \includegraphics[width=\textwidth,height=4cm]{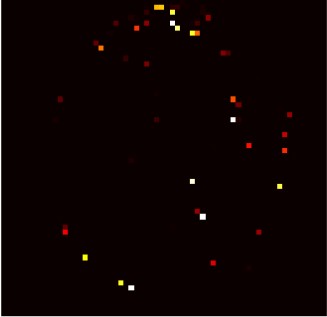}
         \caption{}      \label{fig:wheat_2_pgmm_uncertainty}
     \end{subfigure}
     \hfill
     \begin{subfigure}[b]{0.31\textwidth}
         \centering      \includegraphics[width=\textwidth,height=4cm]{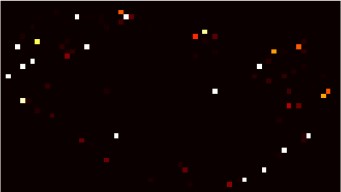}
         \caption{}      \label{fig:corn_2_pgmm_uncertainty}
     \end{subfigure}
     \hfill
     \begin{subfigure}[b]{0.31\textwidth}
         \centering      \includegraphics[width=\textwidth,height=4cm]{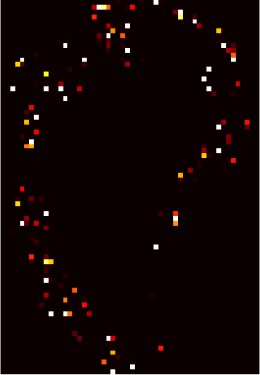}
         \caption{}      \label{fig:rice_2_pgmm_uncertainty}
     \end{subfigure}
     \hfill
     \begin{subfigure}[b]{0.07\textwidth}
         \centering      \includegraphics[width=\textwidth,height=4cm]{Figures/uncertainty_scale.jpg}         \captionsetup{labelformat=empty}
         \caption{}
     \end{subfigure}\\
    \begin{subfigure}[b]{0.31\textwidth}
         \centering      \includegraphics[width=\textwidth,height=4cm]{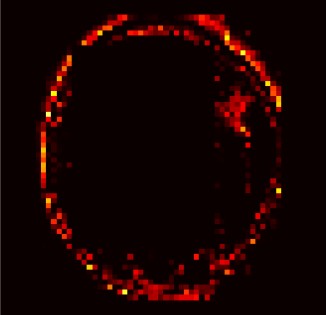}
         \caption{}      \label{fig:wheat_2_ccpgmm_large_uncertainty}
     \end{subfigure}
     \hfill
     \begin{subfigure}[b]{0.31\textwidth}
         \centering      \includegraphics[width=\textwidth,height=4cm]{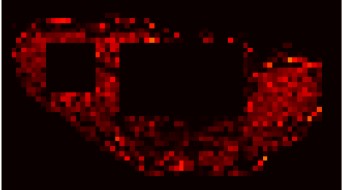}
         \caption{}      \label{fig:corn_2_ccpgmm_large_uncertainty}
     \end{subfigure}
     \hfill
     \begin{subfigure}[b]{0.31\textwidth}
         \centering      \includegraphics[width=\textwidth,height=4cm]{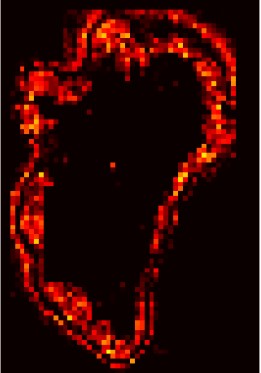}
         \caption{}      \label{fig:rice_2_ccpgmm_large_uncertainty}
     \end{subfigure}
     \hfill
     \begin{subfigure}[b]{0.07\textwidth}
         \centering      \includegraphics[width=\textwidth,height=4cm]{Figures/uncertainty_scale.jpg}         \captionsetup{labelformat=empty}
         \caption{}
     \end{subfigure}\\
    \begin{subfigure}[b]{0.31\textwidth}
         \centering      \includegraphics[width=\textwidth,height=4cm]{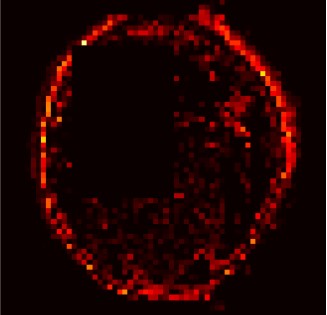}
         \caption{}      \label{fig:wheat_2_ccpgmm_small_uncertainty}
     \end{subfigure}
     \hfill
     \begin{subfigure}[b]{0.31\textwidth}
         \centering      \includegraphics[width=\textwidth,height=4cm]{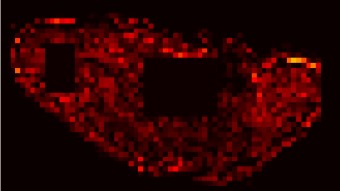}
         \caption{}      \label{fig:corn_2_ccpgmm_small_uncertainty}
     \end{subfigure}
     \hfill
     \begin{subfigure}[b]{0.31\textwidth}
         \centering      \includegraphics[width=\textwidth,height=4cm]{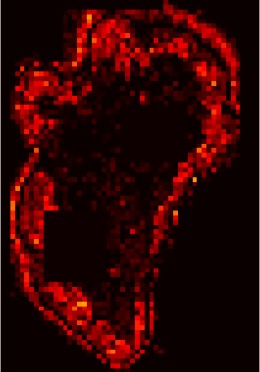}
         \caption{}      \label{fig:rice_2_ccpgmm_small_uncertainty}
     \end{subfigure}
     \hfill
     \begin{subfigure}[b]{0.07\textwidth}
         \centering      \includegraphics[width=\textwidth,height=4cm]{Figures/uncertainty_scale.jpg}         \captionsetup{labelformat=empty}
         \caption{}
     \end{subfigure}\\
     \end{tabular}
        \caption{Uncertainty associated with pixel labels for three of the nine hyperspectral images ($l = \{2,5,8\}$) of the cereals based on the cluster solutions of GMM, PGMM, ccPGMM with $43.5\%$ of pixels as constraints, and ccPGMM with $24.7\%$ of pixels as constraints respectively for wheat (\ref{fig:wheat_2_gmm_uncertainty}, \ref{fig:wheat_2_pgmm_uncertainty}, \ref{fig:wheat_2_ccpgmm_large_uncertainty},\ref{fig:wheat_2_ccpgmm_small_uncertainty}), corn (\ref{fig:corn_2_gmm_uncertainty}, \ref{fig:corn_2_pgmm_uncertainty}, \ref{fig:corn_2_ccpgmm_large_uncertainty}, \ref{fig:corn_2_ccpgmm_small_uncertainty}) and rice (\ref{fig:rice_2_gmm_uncertainty}, \ref{fig:rice_2_pgmm_uncertainty}, \ref{fig:rice_2_ccpgmm_large_uncertainty}, \ref{fig:rice_2_ccpgmm_small_uncertainty}).}
        \label{fig:results_cereal_2_uncertainty}
\end{figure}

\begin{figure}
     \centering
    \begin{tabular}{c}
    \begin{subfigure}[b]{0.31\textwidth}
         \centering      \includegraphics[width=\textwidth,height=4cm]{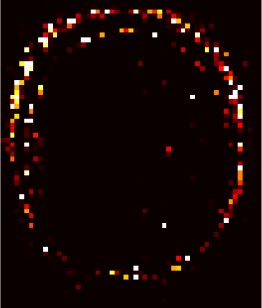}
         \caption{}      \label{fig:wheat_3_gmm_uncertainty}
     \end{subfigure}
     \hfill
     \begin{subfigure}[b]{0.31\textwidth}
         \centering      \includegraphics[width=\textwidth,height=4cm]{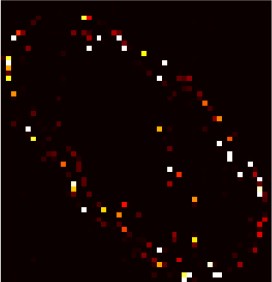}
         \caption{}      \label{fig:corn_3_gmm_uncertainty}
     \end{subfigure}
     \hfill
     \begin{subfigure}[b]{0.31\textwidth}
         \centering      \includegraphics[width=\textwidth,height=4cm]{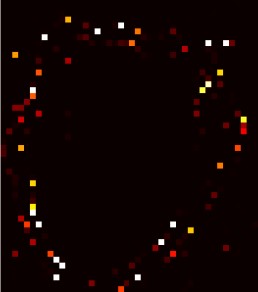}
         \caption{}      \label{fig:rice_3_gmm_uncertainty}
     \end{subfigure}
     \hfill
     \begin{subfigure}[b]{0.07\textwidth}
         \centering      \includegraphics[width=\textwidth,height=4cm]{Figures/uncertainty_scale.jpg}         \captionsetup{labelformat=empty}
         \caption{}
     \end{subfigure}\\
    \begin{subfigure}[b]{0.31\textwidth}
         \centering      \includegraphics[width=\textwidth,height=4cm]{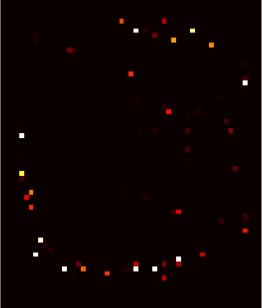}
         \caption{}      \label{fig:wheat_3_pgmm_uncertainty}
     \end{subfigure}
     \hfill
     \begin{subfigure}[b]{0.31\textwidth}
         \centering      \includegraphics[width=\textwidth,height=4cm]{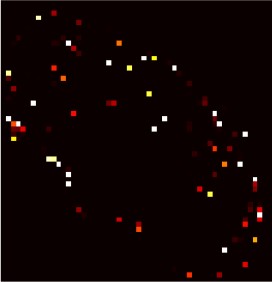}
         \caption{}      \label{fig:corn_3_pgmm_uncertainty}
     \end{subfigure}
     \hfill
     \begin{subfigure}[b]{0.31\textwidth}
         \centering      \includegraphics[width=\textwidth,height=4cm]{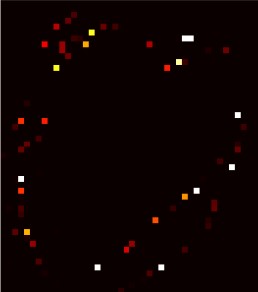}
         \caption{}      \label{fig:rice_3_pgmm_uncertainty}
     \end{subfigure}
     \hfill
     \begin{subfigure}[b]{0.07\textwidth}
         \centering      \includegraphics[width=\textwidth,height=4cm]{Figures/uncertainty_scale.jpg}         \captionsetup{labelformat=empty}
         \caption{}
     \end{subfigure}\\
    \begin{subfigure}[b]{0.31\textwidth}
         \centering      \includegraphics[width=\textwidth,height=4cm]{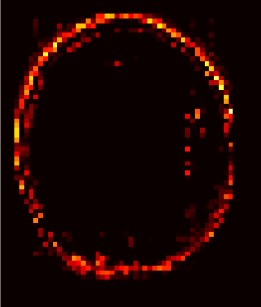}
         \caption{}      \label{fig:wheat_3_ccpgmm_large_uncertainty}
     \end{subfigure}
     \hfill
     \begin{subfigure}[b]{0.31\textwidth}
         \centering      \includegraphics[width=\textwidth,height=4cm]{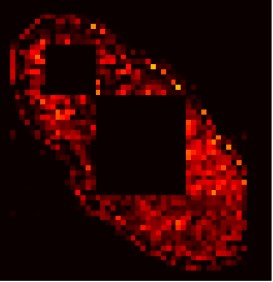}
         \caption{}      \label{fig:corn_3_ccpgmm_large_uncertainty}
     \end{subfigure}
     \hfill
     \begin{subfigure}[b]{0.31\textwidth}
         \centering      \includegraphics[width=\textwidth,height=4cm]{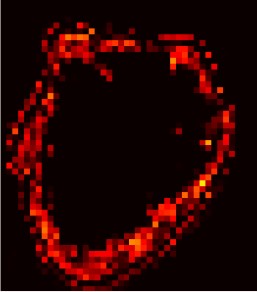}
         \caption{}      \label{fig:rice_3_ccpgmm_large_uncertainty}
     \end{subfigure}
     \hfill
     \begin{subfigure}[b]{0.07\textwidth}
         \centering      \includegraphics[width=\textwidth,height=4cm]{Figures/uncertainty_scale.jpg}         \captionsetup{labelformat=empty}
         \caption{}
     \end{subfigure}\\
    \begin{subfigure}[b]{0.31\textwidth}
         \centering      \includegraphics[width=\textwidth,height=4cm]{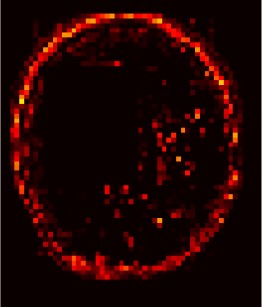}
         \caption{}      \label{fig:wheat_3_ccpgmm_small_uncertainty}
     \end{subfigure}
     \hfill
     \begin{subfigure}[b]{0.31\textwidth}
         \centering      \includegraphics[width=\textwidth,height=4cm]{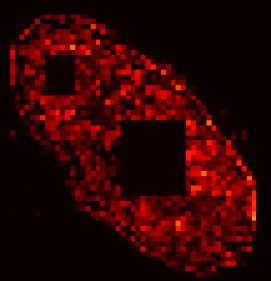}
         \caption{}      \label{fig:corn_3_ccpgmm_small_uncertainty}
     \end{subfigure}
     \hfill
     \begin{subfigure}[b]{0.31\textwidth}
         \centering      \includegraphics[width=\textwidth,height=4cm]{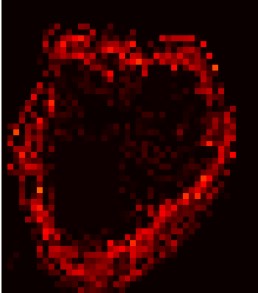}
         \caption{}      \label{fig:rice_3_ccpgmm_small_uncertainty}
     \end{subfigure}
     \hfill
     \begin{subfigure}[b]{0.07\textwidth}
         \centering      \includegraphics[width=\textwidth,height=4cm]{Figures/uncertainty_scale.jpg}         \captionsetup{labelformat=empty}
         \caption{}
     \end{subfigure}\\
     \end{tabular}
        \caption{Uncertainty associated with pixel labels for three of the nine hyperspectral images ($l = \{3,6,9\}$) of the cereals based on the cluster solutions of GMM, PGMM, ccPGMM with $43.5\%$ of pixels as constraints, and ccPGMM with $24.7\%$ of pixels as constraints respectively for wheat (\ref{fig:wheat_3_gmm_uncertainty}, \ref{fig:wheat_3_pgmm_uncertainty}, \ref{fig:wheat_3_ccpgmm_large_uncertainty},\ref{fig:wheat_3_ccpgmm_small_uncertainty}), corn (\ref{fig:corn_3_gmm_uncertainty}, \ref{fig:corn_3_pgmm_uncertainty}, \ref{fig:corn_3_ccpgmm_large_uncertainty}, \ref{fig:corn_3_ccpgmm_small_uncertainty}) and rice (\ref{fig:rice_3_gmm_uncertainty}, \ref{fig:rice_3_pgmm_uncertainty}, \ref{fig:rice_3_ccpgmm_large_uncertainty}, \ref{fig:rice_3_ccpgmm_small_uncertainty}).}
        \label{fig:results_cereal_3_uncertainty}
\end{figure}

As discussed in Section \ref{application}, Figures \ref{fig:results_cereal_1_uncertainty}, \ref{fig:results_cereal_2_uncertainty}, and \ref{fig:results_cereal_3_uncertainty} illustrate the uncertainty associated with the pixel labels for the nine hyperspectral images of the puffed cereals based on the cluster solutions of GMM, PGMM, ccPGMM with $43.5\%$ pixels as constraints and ccPGMM with $24.7\%$ pixels as constraints. 

\begin{figure}[t]
     \centering
    \begin{tabular}{c}
    \begin{subfigure}[b]{0.32\textwidth}
         \centering      \includegraphics[width=\textwidth,height=7cm]{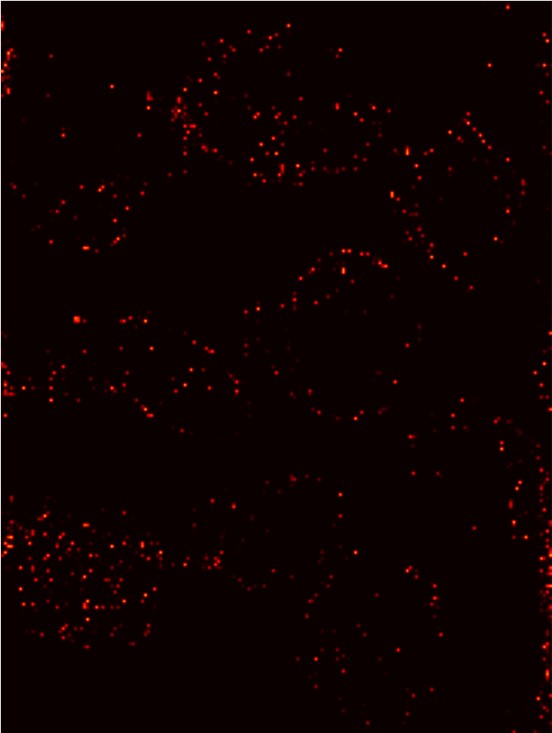}
         \caption{}      \label{fig:Mixture_Puffed_Cereal_PGMM_Classification_Uncertainty}
     \end{subfigure}
     \hfill
     \begin{subfigure}[b]{0.32\textwidth}
         \centering      \includegraphics[width=\textwidth,height=7cm]{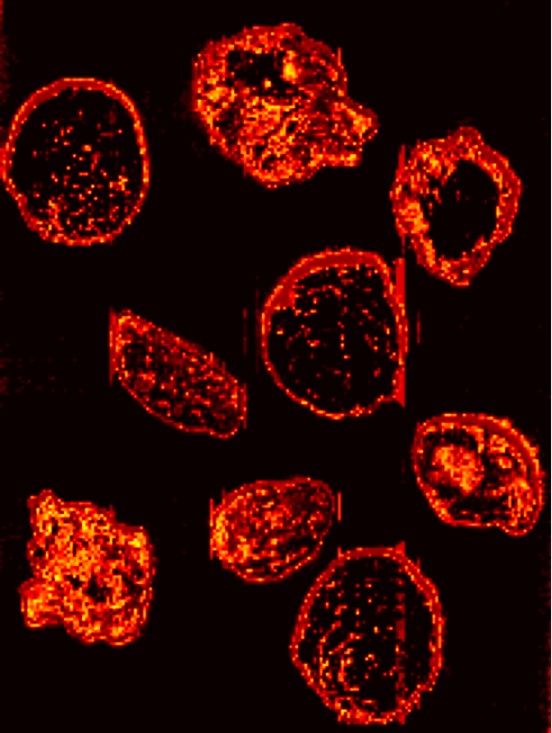}
         \caption{}      \label{fig:Mixture_Puffed_Cereal_ccPGMM_LargeConstraints_Classification_Uncertainty}
     \end{subfigure}
     \hfill
     \begin{subfigure}[b]{0.32\textwidth}
         \centering      \includegraphics[width=\textwidth,height=7cm]{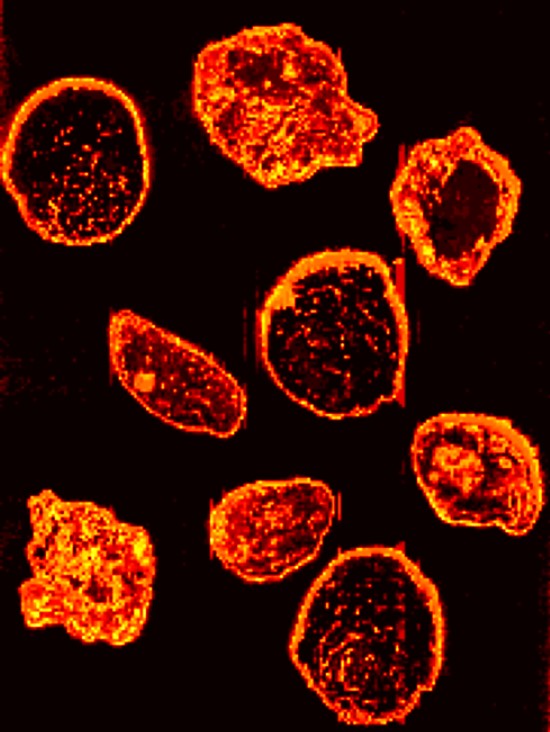}
         \caption{}      \label{fig:Mixture_Puffed_Cereal_ccPGMM_SmallConstraints_Classification_Uncertainty}
     \end{subfigure}
    \hfill
     \begin{subfigure}[b]{0.07\textwidth}
         \centering      \includegraphics[width=\textwidth,height=7cm]{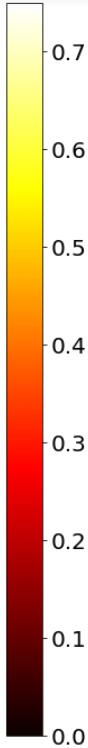}         \captionsetup{labelformat=empty}
         \caption{}      \label{fig:uncertainty_scale_mixtureimage}
     \end{subfigure}\\
     \end{tabular}
        \caption{Uncertainty associated with classification labels for the pixels in the multi-grain hyperspectral image under PGMM (\ref{fig:Mixture_Puffed_Cereal_PGMM_Classification_Uncertainty}) , ccPGMM with $43.5\%$ of pixels as constraints (\ref{fig:Mixture_Puffed_Cereal_ccPGMM_LargeConstraints_Classification_Uncertainty}), and  ccPGMM with $24.7\%$ of pixels as constraints (\ref{fig:Mixture_Puffed_Cereal_ccPGMM_SmallConstraints_Classification_Uncertainty}).}        \label{fig:Mixture_Puffed_Cereal_Classification_Uncertainty}
\end{figure}

Also, as discussed in Section \ref{Mixture_puffed_cereal_classification}, Figure \ref{fig:Mixture_Puffed_Cereal_Classification_Uncertainty} illustrates the uncertainty associated with the classification labels for pixels of the multi-grain image under PGMM, ccPGMM with $43.5\%$ pixels as constraints, and ccPGMM with $24.7\%$ pixels as constraints.

\end{document}